\documentclass[a4paper,fleqn,usenatbib]{mnras}

\usepackage{newtxtext,newtxmath}

\usepackage[T1]{fontenc}
\usepackage{ae,aecompl}

\usepackage{graphicx}	% Including figure files
\usepackage{amsmath}	% Advanced maths commands
\usepackage{amssymb}	% Extra maths symbols
\usepackage{bm}		% Bold maths symbols, including upright Greek
\usepackage{hyperref}
\usepackage{subfigure}
\usepackage{romannum}
\usepackage{float}

\newcommand{\diff}{{\rm d}}
\title[Advective accretion disc in TDEs]{Advective accretion disc-corona model with fallback for tidal disruption events}

\author[Mageshwaran \& Bhattacharyya]{
T. Mageshwaran$^{1,2}$\thanks{E-mail: tmageshwaran2013@gmail.com}
and Sudip Bhattacharyya$^{2}$
\\
$^{1}$Department of Space Science and Astronomy, Chungbuk National University, 12 Gaeshin-dong, Heungduk-gu, Cheongju 361-763, Korea \\
$^{2}$Department of Astronomy and Astrophysics, Tata Institute of Fundamental Research, Mumbai 400005, India
}

\date{Accepted 2022 November 08. Received 2022 November 03; in original form 2021 June 23}

\pubyear{2022}

\begin{document}
\label{firstpage}
\pagerange{\pageref{firstpage}--\pageref{lastpage}}
\maketitle

% Abstract of the paper
%This is a simple template for authors to write new MNRAS papers. The abstract should briefly describe the aims, methods, and main results of the paper. It should be a single paragraph not more than 250 words (200 words for Letters).No references should appear in the abstract.

\begin{abstract}

Tidal disruption events (TDEs) show a correlation between the UV to X-ray spectral index and the Eddington ratio, with non-thermal X-ray emission at the low Eddington ratio. We consider the corona surrounding the accretion disc as a non-thermal X-ray source. We construct a time-dependent and non-relativistic advective accretion disc-corona model for TDEs. The infalling debris is assumed to form a seed disc in time $t_c$, that evolves due to the mass gain from the infalling debris at the constant outer radius with a mass fallback rate $\dot{M}_{\rm fb}$ and the mass loss through accretion onto the black hole. The viscous stress in our model depends on gas ($P_g$) and total ($P_t$) pressures as $\tau_{r\phi} \propto P_g^{1-\mu} P_t^{\mu}$, where $\mu$ is a constant. We find that the mass accretion rate $\dot{M}_a$ evolves from Eddington to sub-Eddington accretion with a late-time evolution close to $t^{-5/3}$, where $t$ is the time. We find that the bolometric disc luminosity follows a late-time evolution close to $t^{-5/3}$. The ratio of total X-ray luminosity from corona to bolometric disc luminosity increases with $\mu$ and increases at late times for $\mu \neq 1$. We obtain the  X-ray blackbody temperature of the disc that agrees with the temperature from X-ray observations ($\sim~10^5~{\rm K}$). We find the radiative efficiency of the disc increases with time and decreases for a disc when the corona is included. We have neglected the outflow, and our model is more applicable for near-to-sub-Eddington accretion and when $\dot{M}_{\rm fb}$ is sub-Eddington.
\end{abstract}

% Select between one and six entries from the list of approved keywords.
% Don't make up new ones.

%%%%%%%%%%%%%%%%%%%%%%%%%%%%%%%%%%%%%%%%%%%%%%%%%%

%%%%%%%%%%%%%%%%% BODY OF PAPER %%%%%%%%%%%%%%%%%%

\section{Introduction}

A star can be tidally stripped into pieces near a black hole as the tidal force exceeds the star's self-gravity and this phenomenon is called a tidal disruption event (TDE) \citep{1976MNRAS.176..633F, 1988Natur.333..523R}. In a simplified fallback model, an impulse approximation is assumed where the star is frozen until it reaches the pericenter where a short duration impulse of tidal potential disrupts the star \citep{2009MNRAS.392..332L}. The bound debris follows a Keplerian orbit and returns to the pericenter with a mass fallback rate that evolves as $\dot{M}_{\rm fb} \propto t^{-5/3}$ at late time $t$ . At the initial time, the mass fallback rate depends on the stellar density profile and the distribution of mass per orbital energy after a disruption. The returning debris interacts with the outflowing debris and this stream interaction results in a formation of an accretion disc \citep{1994ApJ...422..508K,2016MNRAS.461.3760H}. The nature of the formed disc depends on the stream-stream interactions, the thermal radiative efficiency of the debris, viscous dynamics within the debris, and on the pericenter and the eccentricity of the initial stellar orbit; this can result in a circular or an elliptical disc with or without still infalling debris \citep{2020A&A...642A.111C}. If the energy and angular momentum of infalling debris is lost on a timescale smaller than the orbital time of the debris, the mass accretion rate follows the mass fallback rate ($\dot{M}_{\rm fb}$) and the bolometric luminosity is $L_b \propto \dot{M}_{\rm fb} c^2$, where $c$ is the light speed \citep{2002ApJ...576..753L}. However, the accretion of matter to the black hole depends on the viscous dynamics in the accretion disc and the pressure which can be dominated by either radiation or gas pressure. 

TDE accretion disc provides an excellent opportunity to study the evolution of the accretion phenomenon around supermassive black holes, which in general take millions of years for active galactic nuclei (AGN). TDE disc may evolve through super and sub-Eddington phases and the accretion models constructed in literature are at individual phases. The sub-Eddington disc with gas pressure and without fallback has been modelled analytically using a self-similar formulation for a non-relativistic disc by \citet{1990ApJ...351...38C} and numerically for a relativistic disc by \citet{2019MNRAS.489..132M}. \citet{2009MNRAS.400.2070S} has constructed a steady slim disc accretion model with an adiabatic and spherical outflow whereas \citet{2021NewA...8301491M} has constructed a time-dependent and self-similar model with mass infall to the disc for both sub and super-Eddington with outflow phases, where the mass outflow rate is a function of time only. \citet{2020MNRAS.496.1784M} have constructed a relativistic thin disc model with a mass fallback at the outer radius for full and partial TDEs, and showed that the late time luminosity decline is higher than the luminosity obtained using $L \propto \dot{M}_{\rm fb}$. The transition of a disc from one phase to another is usually obtained by equating the mass accretion rate \citep{2014ApJ...784...87S} and results in a non-smooth transition in disc evolving parameters such as surface density and luminosity. Here, we aim to construct an advective accretion model with a fallback that shows a smooth transition from Eddington to the sub-Eddington phase.

\citet{2019ApJ...883...76R} compared the correlations between the UV to X-ray spectral index ($\alpha_{\rm OX}$) and the Eddington ratio ($L / L_E$) in AGN and X-ray binaries, and the AGN and X-ray binaries observations show a remarkable similarity to accretion state transitions. They concluded that the dynamics of black hole accretion ﬂows directly scale across the various black hole masses and the different accretion states. \citet{2020MNRAS.497L...1W} studied the correlation for seven TDEs and showed that the $\alpha_{\rm OX} - L / L_E$ correlation for TDE sources is similar to that observed by \citet{2019ApJ...883...76R} for AGN and X-ray binaries. They found that the X-ray emission is dominated by the power-law spectrum at a low Eddington ratio and by the soft disc spectrum at a high Eddington ratio such that the power-law X-ray emission is suppressed. The spectral state transition occurs around the Eddington ratio $\sim 0.03$. \citet{2016MNRAS.463.3813H} found that an absorbed blackbody plus power-law model shows a good fit to the X-ray spectrum of ASAS-SN 15oi. \citet{2017ApJ...838..149A} have used an absorbed power-law model to fit thirteen TDE X-ray spectra to study the time evolution of TDE X-ray luminosity. \citet{2017A&A...598A..29S} using the X-ray and UV observations of the source XMMSL1 J074008.2-853927 have shown that a power-law emission dominates over the disc emission above 2 keV and the source has both thermal and non-thermal components. The power-law emission indicates the presence of the non-thermal and hotter medium surrounding the standard accretion disc. Here, we assume that this hot medium is a corona above the disc and the energy transported from the disc to the corona is Compton scattered.

The effective temperature in a steady thin disc scales as $T_{\rm eff} \propto M_{\bullet}^{-1/4} \dot{m}^{1/4} (r/r_g)^{-3/4}$, where $M_{\bullet}$ is the black hole mass, $\dot{m}$ is the accretion rate normalized to Eddington rate, $r$ is the radial variable and $r_g$ is the gravitational radii. The disc thermal emission peaks in UV for AGNs, and in soft X-ray for X-ray binaries, whereas the Comptonized coronal emission dominates the hard X-rays. The X-ray spectrum is either dominated by thermal emission from the disc or non-thermal and power-law emission from the corona. The presence of corona plays a part in cooling the accretion flow and explaining the excess hard X-ray emission. We show that the impact on the disc bolometric luminosity and spectrum due to the inclusion of the corona is insignificant when the luminosity is high but is significant when the luminosity is sub-Eddington. 

In this paper, we construct a non-relativistic and time-dependent advective accretion disc-corona model for TDEs with fallback. We include the energy loss to the corona in the energy conservation equation. We use a viscosity that is a combination of total pressure and gas pressure. We include the gravitational and Doppler redshift in calculating the spectral luminosity. The infalling debris is assumed to form a seed accretion disc that evolves due to mass gain through the mass fallback and mass loss through the viscous accretion onto the black hole with energy loss to the corona. The outer radius in our model is constant, where the mass accretion rate is equal to the mass fallback rate. The outer radius of an accretion disc with mass fallback may evolve with time, but we have limited our modelling to a constant outer radius, which simplifies our numerical calculation and shows reasonable solutions. We do not impose a global angular momentum conservation where the disc's outer radius evolves with time but truncate our disc to a constant outer radius assuming the infalling matter loses its angular momentum to the external infalling debris whose evolution is not considered in this paper. We consider the accretion dynamics after the disc has formed and the time at which matter starts crossing the innermost stable circular orbit (ISCO) is the beginning of the accretion to the black hole. The beginning time is obtained using the initial condition and we find that the increase in the contribution of gas pressure to the viscous stress delays the beginning of disc accretion. The presence of corona affects the bolometric disc and spectral luminosity at the late times in the sub-Eddington phase whereas the mass accretion rate increases in presence of corona at initial times and shows a weak rise at late times. Our time-dependent accretion model shows a disc evolution from Eddington to the sub-Eddington phase. We also estimate the evolution of coronal properties such as electron temperature, optical depth, and Compton $y$ parameter using a two-temperature plasma model in the corona where electrons cool via bremsstrahlung, synchrotron, and Compton cooling.

In the super-Eddington phase, outflows are driven by the strong radiation force at the disc surface. The dynamics of a time-dependent disc with an outflow are complex and the presence of corona surrounding such a disc is uncertain. \citet{2020MNRAS.497L...1W} showed that the non-thermal X-ray emission is suppressed at high luminosity and the spectrum is dominated by the disc spectrum. We also show that the disc luminosity declines significantly due to corona at low luminosity. This suggests that the non-thermal emission is negligible if the disc has a radiatively driven outflow. We have neglected the outflow in our disc-corona model, and this model is more applicable for near to sub-Eddington accretion. We have included the mass fallback rate in our formulation and show that the late-time mass accretion rate nearly follows the mass fallback rate. The obtained mass accretion rate is super-Eddington if the mass fallback rate is super-Eddington. We emphasise that our model is suitable when the mass fallback rate has dropped to sub-Eddington or when the peak mass fallback rate is already sub-Eddington.

In section \ref{mfb}, we present the mass fallback rate of the disrupted debris constructed using an impulse approximation. In section \ref{dcm}, we present the advective disc-corona model with fallback where the basic assumptions and conditions are discussed. In section \ref{result}, we present the results of our accretion model and the time-evolution of accretion rate and luminosity. We discuss our results in section \ref{discuss} and present the summary in section \ref{summary}.
  
\section{Mass fallback rate}
\label{mfb}

We assume that the star is on a nearly parabolic orbit and is tidally disrupted for the pericenter $r_p \leq r_t$, where the tidal radius $r_t \simeq (M_{\bullet}/M_{\star})^{1/3} R_{\star}$. The specific energy of the disrupted debris is governed by the variation of the black hole potential across the star and the tidal spin-up of the star as a result of the tidal interaction. The dynamics of tidal interaction that results in the stellar spin-up are complex and depend on the stellar structure \citep{1992ApJ...385..604K}. A numerical simulation by \citet{2001ApJ...549..948A} showed that the tidal interaction can spin up the rotating velocity of the star close to the angular velocity of the stellar orbit at the pericenter. By including the tidal interaction, \citet{2002ApJ...576..753L} formulated the disrupted debris energy given by $E_{\rm d}= -k G M_{\bullet} \Delta R/ r_t^2$, where $k=1$ for no tidal spin up and $k=3$ for tidal spin up. The disrupted debris follows a Keplerian orbit and the time period of innermost debris is given by \citep{1988Natur.333..523R}

\begin{equation}
t_m= 40.8~{\rm days}~ M_6^{1/2} m^{1/5} k^{-3/2},
\label{tmt}
\end{equation}

\noindent where $M_6= M_{\bullet}/[10^6 M_{\odot}]$, $m=M_{\star}/M_{\odot}$ and the radius of star $R_{\star}=R_{\odot} m^{0.8}$ \citep{1994sse..book.....K}. Following the impulse approximation, the mass fallback rate is given by  \citep{2009MNRAS.392..332L}

\begin{equation}
\dot{M}_{\rm fb}= \frac{4 \pi  b}{3} \frac{M_{\star}}{t_m} \tau^{-5/3} \int_{x}^{1} \theta^{u}(x') x' \, \diff x',
\label{mfbn}
\end{equation}

\noindent where $\theta(x)$ is the solution of Lane-Emden equation with a polytropic index $\Gamma= 1+ 1/u$, $b$ is the ratio of central to mean density of the star \citep{1943ApJ....97..255C} and $x=\Delta R / R_{\star}= \tau^{-2/3}$ with $\tau = t/t_m$. The integral is nearly constant at late times that results in the mass fallback rate $\dot{M}_{\rm fb} \propto t^{-5/3}$. We consider the tidal spin up in our model by taking $k=3$ and the polytropic index $\Gamma= 5/3$ which results in $u=3/2$. 

The specific angular momentum at the pericenter for a parabolic stellar orbit is $J= \sqrt{2 G M_{\bullet} r_t}$. The specific angular momentum for a circular orbit is $J_c= r^2 \omega= \sqrt{G M_{\bullet} r_c}$, where the Keplerian angular frequency is $\omega=\sqrt{G M_{\bullet}/r_c^3}$, and $r_c$ is the circularization radius. The angular momentum conservation of debris results in $r_c= 2 r_t$ \citep{1999ApJ...514..180U,2009MNRAS.400.2070S}, and is used as the outer radius in the steady accretion model of \citet{2009MNRAS.400.2070S} and in the time-dependent accretion model by \citet{2011ApJ...736..126M}. \citet{2016MNRAS.461.3760H} showed through the numerical simulation that the circularization radius is close to the $r_c$. To avoid an uncertainty in the outer radius, we consider it to be $r_c= 2 r_t$ in our calculation.

\section{Disc accretion model}
\label{dcm}

Here, we develop the advective accretion model for a TDE disc. In our accretion model, we consider the equations in the cylindrical coordinate and the vertical flow is assumed to be zero. We employ the vertically integrated mass and momentum conservation equations. The mass accretion rate is given by

\begin{equation}
\frac{\partial \Sigma}{\partial t} = \frac{1}{2 \pi r} \frac{\partial \dot{M}}{\partial r},
\label{mcons}
\end{equation} 

\noindent where $\Sigma$ is the surface density and $\dot{M}$ is the mass accretion rate that depends on radial velocity $v_r$ as $\dot{M} = -2 \pi r \Sigma v_r$. We assume the angular velocity to be Keplerian given by $v_{\phi} = r \Omega_K $, where $\Omega_K = \sqrt{G M_{\bullet}/r^3}$, such that the angular momentum conservation results in 

\begin{equation}
\dot{M} = 6 \pi \sqrt{r} \frac{\partial}{\partial r}(\sqrt{r} \nu \Sigma),
\label{mdot}
\end{equation}

\noindent where $\nu$ is the viscosity. In the standard thin disc model, the disc height is estimated using the vertical conservation equation with the assumption $H / r \ll 1$ and is given by $c_s^2 = H^2 \Omega_K^2$ which is used by \citet{2009MNRAS.400.2070S} for a steady state slim disc and they showed that $H \sim r$ for near to super-Eddington accretion. \citet{2002ApJ...576..908J} showed that vertically averaged disc scale height is $ c_s^2 = C_1 H^2 \Omega_K^2$, where $C_1 < 1$. The $C_1$ represents the correction due to the large disc scale height and the determination of $C_1$ requires the vertical structure of density. We use their scale-height formulation and consider $C_1$ to be a free parameter. The viscous heating is given by \citep{2002apa..book.....F}

\begin{equation}
Q^{+} = \frac{9}{4} \nu \Sigma \Omega_K^2,
\label{vis}
\end{equation}

\noindent and the advection flux $Q_{\rm adv}$ is given by 

\begin{align}
Q_{\rm adv} &= C_v \Sigma T \left[\frac{1}{T} \frac{\partial T}{\partial t} + \frac{v_r}{T} \frac{\partial T}{\partial r} - (\Gamma_3 -1) \left\{\frac{1}{\rho} \frac{\partial \rho}{\partial t} + \frac{v_r}{\rho} \frac{\partial \rho}{\partial r} \right\} \right], \label{qadv}\\
C_v &= \frac{4 - 3 \beta_g}{\Gamma_3-1} \frac{\bar{P}}{\Sigma T},~~{\rm and}~~
\Gamma_3-1 = \frac{(4 -3 \beta_g)(\gamma_g -1)}{\beta_g + 12 (1-\beta_g)(\gamma_g-1)},
\end{align}

\noindent where $T$ is the temperature, $\rho = \Sigma / (2 H)$ is the density, vertically integrated pressure $\bar{P} \approx 2 P H$, $\beta_g$ is the ratio of gas to total pressure and $\gamma_g$ is the ratio of specific heats. 

The corona above the accretion disc is heated via magnetic reconnections and a fraction of energy generated via viscous heating is transported to the corona by the magnetic fields. The amount of energy flux escapes to the corona from the disc depends on the vertical transport of magnetic flux tubes. In this paper, we follow the prescription of \citet{2003MNRAS.341.1051M} where the standard conservation equations of an optically thick disc are self-consistently coupled with the corona, and the fraction of energy dissipated to the corona is given by $f$ which at any radius $r$ is given by  

\begin{equation}
f = \frac{Q_{\rm cor}}{Q^{+}},
\label{feq}
\end{equation} 

\noindent where $Q_{\rm cor}$ is the coronal energy flux at any $r$ which is the vertical Poynting flux given by $Q_{\rm cor} = v_D P_{\rm mag}$ \citep{1994ApJ...436..599S}. The magnetic pressure is given by $P_{\rm mag} = B^2/(8 \pi)$ with magnetic field $B$ and the drift velocity $v_D$ is taken to be proportional to the Alfvén speed $v_A$ via an order-unity constant $b_1$ \citep{2019A&A...628A.135A}. The drift velocity is given by $v_D = b_1 \sqrt{2 P_{\rm mag}/\rho}$, where $\rho$ is the density. We assume that the stress is dominated by Maxwell stresses \citep{2015ApJ...808...54M} and thus given by $\tau_{r\phi} = k_0 P_{\rm mag}$, where $k_0$ is of order unity. The magneto-rotational instability (MRI) growth rate depends on the ratio of radiation to gas pressure \citep{2002ApJ...566..148T}. \citet{2001ApJ...553..987B} through linear perturbation theory showed the compressibility of MHD turbulence in the radiation pressure dominated regime which slows the magnetic field growth rate and it attains a saturation point. In the saturation limit, \citet{2003MNRAS.341.1051M} derived the viscous stress to be $ \tau_{r\phi} \propto P_{\rm mag} \propto \sqrt{P_g P_t}$, and a more generalised version of the viscous stress is $\tau_{r\phi} \propto P_{\rm mag} \propto P_g^{1-\mu} P_t^{\mu}$ \citep{2008ApJ...683..389D}. \citet{2009MNRAS.394..207C} considered $\mu = 1/2$ in his steady disc-corona model. We use the generalised version of magnetic pressure that is given by $P_{\rm mag} = \alpha_0 P_g^{1-\mu} P_t^{\mu}$, with $\mu$ constant and the total pressure $P_t= P_r + P_g$, radiation pressure $P_r = a T^4/3$ where $a$ is a radiation constant, and the gas pressure is given by $P_g =  k_B \rho T/(\mu_m m_p)$, where $T$ is the temperature in the disc, $m_p$ is the mass of a proton, $k_B$ is the Boltzmann constant, and $\mu_m$ is the mean molecular weight taken to be ionized solar mean molecular weight of $0.65$.

The vertically integrated viscous stress is given by $\tau_{r\phi} = (\nu \Sigma/ 2 H)  r (\partial \Omega / \partial r)$ \citep{2002apa..book.....F}, such that for a Keplerian velocity 

\begin{equation}
\nu \Sigma = \frac{4 \alpha_s}{3} \frac{H}{\Omega_K} P_g^{1-\mu} P_t^{\mu},
\label{nuvis}
\end{equation}

\noindent where $\alpha_s = k_0 \alpha_0$. Then, the fraction of energy transported $f$ using equations (\ref{feq}) and (\ref{vis}) is given by

\begin{equation}
f(\beta_g) = b_2 \left(\frac{P_g}{P_t}\right)^{\frac{1-\mu}{2}} = b_2~ \beta_g^{\frac{1-\mu}{2}},
\label{fcor}
\end{equation}

\noindent where the constant $b_2 = \sqrt{2}b_1 \alpha_0^{1/2} / (3 k_0) $. The quantities $f(\beta_g),~ \beta_g,~{\rm and}~\mu$ are within the limits $\{0,~1\}$, and thus imposes a constrain $b_2 \in \{0,~1\}$. The $b_2 = 0$ implies no disc energy is transported to the corona and the disc evolves as if the corona is absent. In the case $\mu \neq 1$, $f(\beta_g)$ decreases with a decrease in $\beta_g$ implying that the radiation pressure reduces the cooling of accretion flow by decreasing the fractional energy loss to the corona.  The $\alpha_s$ and $b_2$ are the free parameters and we take their values within $\{0,~1\}$. 

The energy conservation equation is given by $Q^{+} = Q_{\rm adv} + Q_{\rm rad} + Q_{\rm cor}$, where $Q_{\rm rad}$ is the radiative energy loss. The radiation flux is given by $Q_{\rm rad} = 4 \sigma T^4 / (3 \kappa \Sigma)$ with Thomson opacity $\kappa = 0.34~{\rm g~cm^{-2}}$. Using equation (\ref{feq}), we obtain $Q_{\rm adv} + Q_{\rm rad} = [1-f(\beta_g)]Q^{+} $. 

The advection flux in the steady state is approximated as $Q_{\rm adv} = [\dot{M}/(2 \pi r^2)] c_s^2 \xi $, where $\xi$ which is a function of log radial variation of entropy is of the order of unity \citep{2002apa..book.....F}. Thus, $Q_{\rm adv} / Q^{+} = [2 C_1 / (9 \pi)] \dot{M}/(\nu \Sigma) (H/r)^2 $ and can be approximated to $Q_{\rm adv}  = s(r) (H/r)^2 Q^{+}$, where $s(r) = [4 c_1 / 3] \partial [\ln (\sqrt{r} \nu \Sigma)])/\partial[\ln r]$. The $s(r)$ depends on the angular momentum distribution throughout the disc, but in general, it is close to the unity \citep{1995ApJ...438L..37A}. \citet{2011ApJ...736..126M} considered it to be a unity whereas we consider it to be a constant parameter denoted by $k_1$ such that $Q_{\rm adv}  = k_1 (H/r)^2 Q^{+}$. In presence of an energy loss to the corona, the amount of energy available for advection and radiation decreases by $1-f(\beta_g)$, and thus, we approximate the advection energy loss as

\begin{equation}
Q_{\rm adv}  = k_1 \left(\frac{H}{r}\right)^2 [1-f(\beta_g)] Q^{+},
\label{qadv1}
\end{equation}

\noindent which results in radiation energy loss given by 

\begin{equation}
Q_{\rm rad} = [1-f(\beta_g)] \left[1-k_1 \left(\frac{H}{r}\right)^2\right]Q^{+}.
\label{qrad}
\end{equation}

\noindent In the absence of corona, the radiative flux is given by $Q_{\rm rad} = [1-k_1 (H/r)^2]Q^{+}$ and thus implies that the total cooling flux of accretion flow in presence of corona is $[1-f(\beta_g)]^{-1} 4 \sigma T^4 / (3 \kappa \Sigma)$ which is similar to the cooling rate approximated by \citet{2002ApJ...576..908J} for a constant $f(\beta_g)$.

The conservation equations (\ref{mcons}) and (\ref{mdot}) results in a second order radial derivative in surface density and temperature. We transform the conservation equations from $\Sigma$ and $T$ to $\mathcal{F}$ and $W = \nu \Sigma$. This simplifies the non-derivative and the second-order derivative terms in the conservation equations. This transformation is also useful in estimating the boundary values with the assumption and the conditions we employ. We perform a transformation using $\mathcal{F} = \left[(\mu_m m_p/k_B C_1) (a^2/9)\right]^{1/2} T^{7/2} \Sigma^{-1} \Omega_K^{-1}$, such that the total pressure ($p_t$) and gas pressure ($p_g$) are given by

\begin{eqnarray}
p_t =& \frac{1}{2} \left(\frac{k_B C_1}{\mu_m m_p}\right)^{4/7} \left(\frac{9}{a^2}\right)^{1/14} \Sigma^{8/7} \Omega_K^{8/7} \mathcal{F}^{1/7} \left[\mathcal{F} + \sqrt{\mathcal{F}^2 + 1}\right] \\
p_g =& \frac{1}{2} \left(\frac{k_B C_1}{\mu_m m_p}\right)^{4/7} \left(\frac{9}{a^2}\right)^{1/14} \Sigma^{8/7} \Omega_K^{8/7} \mathcal{F}^{1/7} \left[\mathcal{F} + \sqrt{\mathcal{F}^2 + 1}\right]^{-1}.
\end{eqnarray}

\noindent The ratio of gas to total pressure $\beta_g = \left[\mathcal{F} + \sqrt{\mathcal{F}^2 + 1}\right]^{-2} $, and the viscosity using equation (\ref{nuvis}) is given by 

\begin{equation}
\nu \Sigma = \chi \Sigma^{9/7} \Omega_K^{-5/7} \mathcal{F}^{2/7} \left[\mathcal{F} + \sqrt{\mathcal{F}^2 + 1}\right]^{2 \mu},
\label{nusig}
\end{equation}

\noindent where $\chi = (2\alpha_s / 3 C_1) (k_B C_1 / \mu_m  m_p)^{8/7} (9/a^2)^{1/7}$. The sound speed $c_s = \sqrt{p/\rho}$ is then given by 

\begin{equation}
c_s^2 = \eta_1 \Sigma^{2/7}\Omega_K^{2/7} \mathcal{F}^{2/7} \left[\mathcal{F} + \sqrt{\mathcal{F}^2 + 1}\right]^{2},
\end{equation}

\noindent where $\eta_1 = (1/C_1) (k_B C_1 / \mu_m  m_p)^{8/7} (9/a^2)^{1/7} $. The radiation flux is given by $Q_{\rm rad} = \eta_2 \Sigma^{1/7} \Omega_K^{8/7} \mathcal{F}^{8/7}$, where $\eta_2 = [4 \sigma/(3 \kappa)] [k_B C_1/ \mu_m m_p ]^{4/7} (9 / a^2)^{4/7}$. 

We assume that the debris circularizes to form a seed disc with outer radius $r_{\rm out} = r_c$ and the mass is added to the disc at $r_{\rm out}$ by the later infalling debris. The rate at which the mass is added at the outer radius is the mass fallback rate $\dot{M}_{\rm fb}$. If the mass accretion rate is smaller or higher than the mass fallback rate, the disc's outer radius will increase or decrease accordingly. In this paper, we limit our calculation to the constant outer radius and assume that mass added by the infalling debris is accreted inward. To calculate the boundary conditions at $r_{\rm out}$, we assume that the mass accretion rate at the outer radius is equal to the mass fallback rate \citep{2011ApJ...736..126M}. In the steady limit and for no-corona energy loss case ($b_2 = 0$), the $Q_{\rm adv} = k_1 (H/r)^2 Q^{+}$ with $Q_{\rm adv} = \dot{M} c_s^2/(2 \pi r^2)$, and $\dot{M} = 6 \pi \sqrt{r}~\diff(\sqrt{r} \nu\Sigma)/\diff r$ \citep{2002apa..book.....F}. Using equation (\ref{vis}), we get $\displaystyle{\nu \Sigma \propto r^{\frac{3}{4}\frac{k_1}{C_1} - \frac{1}{2}}}$. For $k_1 = (2/3)C_1$, $\nu \Sigma$ is a constant such that $\diff(\nu\Sigma)/\diff r = 0$. We assume this result in our calculation to get the boundary condition at the outer radius, and thus $\partial (\nu \Sigma)/\partial r|_{r_{\rm out}} = 0$. Then, the equation (\ref{mdot}) at $r_{\rm out}$ results in $\dot{M}_{\rm fb} = 3 \pi \nu \Sigma$ which results in

\begin{equation}
\Sigma_{\rm out}(t) = \left[\frac{\dot{M}_{\rm fb}(t)}{3 \pi \chi}\right]^{7/9} \Omega_K^{5/9}(r_{\rm out}) \mathcal{F}^{-2/9} \left[\mathcal{F} + \sqrt{\mathcal{F}^2 + 1}\right]^{-14 \mu /9}.
\label{sigout}
\end{equation} 

\noindent Using this in equation (\ref{qrad}), the energy conservation equation results in

\begin{multline}
1 -\frac{k_1 \eta_1}{C_1}\left[\frac{\dot{M}_{\rm fb}(t)}{3 \pi \chi}\right]^{2/9} \frac{\Omega_K^{-14/9}(r_{\rm out})}{r_{\rm out}^2} \mathcal{F}^{2/9} \left[\mathcal{F} + \sqrt{\mathcal{F}^2 + 1}\right]^{2-4 \mu /9} - \\ \frac{1}{1-f(\beta_g)} \frac{4 \eta_2}{9 \chi} \left[\frac{\dot{M}_{\rm fb}(t)}{3 \pi \chi}\right]^{-8/9} \Omega_K^{-7/9}(r_{\rm out}) \mathcal{F}^{10/9} \left[\mathcal{F} + \sqrt{\mathcal{F}^2 + 1}\right]^{-2\mu /9} \\ = 0,
\end{multline}

\noindent which results in $\mathcal{F}_{\rm out}(t)$ and thus we can obtain $\Sigma_{\rm out}(t)$ and $T_{\rm out}(t)$ {which are the boundary conditions in our calculations. We do not assume any boundary conditions at the inner radius which is taken to be the innermost stable circular orbit (ISCO), and we let all the variables evolve according to their equations.

We perform another transformation by assuming $\nu \Sigma = W$, such that mass conservation equation (\ref{mcons}) is given by

\begin{equation}
\frac{7}{9} \frac{1}{W} \frac{\partial W}{\partial t} - \frac{2}{9}\left[1+ \frac{7 \mu \mathcal{F}}{\sqrt{\mathcal{F}^2+1}}\right] \frac{1}{\mathcal{F}} \frac{\partial \mathcal{F}}{\partial t} = \frac{3}{r \Sigma} \frac{\partial}{\partial r} \left[\sqrt{r} \frac{\partial}{\partial r} (\sqrt{r} W)\right],
\label{twft1}
\end{equation} 

\noindent and the energy conservation using equations (\ref{qadv})and (\ref{qadv1}) is given by 

\begin{multline}
\frac{2}{3} \left(\frac{4}{3}-\Gamma_3\right) \frac{1}{W} \frac{\partial W}{\partial t} + \frac{S(\mathcal{F})}{\mathcal{F}} \frac{\partial \mathcal{F}}{\partial t} + \frac{v_r}{r} \left[\frac{2}{3} \left(\frac{4}{3}-\Gamma_3\right) \frac{r}{W} \frac{\partial W}{\partial r} + \right. \\ \left. \frac{S(\mathcal{F}) r}{\mathcal{F}} \frac{\partial \mathcal{F}}{\partial r} - 2 \left(\frac{4}{3}-\Gamma_3\right)\right] = \frac{9 k_1}{4 C_1} \frac{\Gamma_3 -1}{4 - 3 \beta_g} \frac{W}{r^2 \Sigma} [1-f(\beta_g)],
\label{twft2}
\end{multline}

\noindent where $\Sigma$ and $S(\mathcal{F})$ are given by 

\begin{equation}
\Sigma = \chi^{-7/9} W^{7/9} \Omega_K^{5/9} \mathcal{F}^{-2/9}  \left[\mathcal{F} + \sqrt{\mathcal{F}^2 + 1}\right]^{-14 \mu /9},
\label{sigw}
\end{equation}

\begin{equation}
S(\mathcal{F}) = -\frac{1}{9} - \left(1 + \frac{16 \mu}{9}\right)\frac{\mathcal{F}}{\sqrt{\mathcal{F}^2 + 1}} + \left[\frac{1}{3} + \left(1 + \frac{4 \mu}{3}\right) \frac{\mathcal{F}}{\sqrt{\mathcal{F}^2 + 1}}\right]\Gamma_3.
\label{sft}
\end{equation}

We solve equations (\ref{twft1}) and (\ref{twft2}) with initial and boundary conditions to obtain the disc evolution. We now focus on calculating the initial solution for the disc. We assume that at initial time, the solutions are given by  $W = W_t W_r(r) $ and $\mathcal{F} = \mathcal{F}_t \mathcal{F}_r(r)$, where $W_t$ and $\mathcal{F}_t$ have values equal to $W_{\rm out}$ and $\mathcal{F}_{\rm out}$, such that $W_r(r) $ and $\mathcal{F}_r(r)$ is equal to unity at $r_{\rm out}$. We further assume that the time derivatives represented by $\frac{\partial W_t}{\partial t}$ and $\frac{\partial \mathcal{F}_t}{\partial t}$ are constant at all radius and take the value at the boundary $r_{\rm out}$. Then, the equations (\ref{twft1}) and (\ref{twft2}) became a function of $r$ only and we solved them with boundary values $W_r(r_{\rm out}) =1$, $\partial W_r / \partial r |_{r_{\rm out}} =0 $ and $\mathcal{F}_r (r_{\rm out}) =1$. These assumptions are made for simplification to get the initial solution for $W$ and $\mathcal{F}$ that are consistent with the boundary conditions at the outer radius. This postulation is only at the initial time to get the initial solution and at later times, the solution for $W$ and $\mathcal{F}$ are obtained by solving the coupled differential equations numerically in $\{t,~r\}$ space. We calculate the surface density at initial time given by $\Sigma_i$ using the equation (\ref{sigw}) with the solution $W = W_t W_r(r)$ and $\mathcal{F} = \mathcal{F}_t \mathcal{F}_r(r)$. 

If the moment of disruption is $t=0$, the innermost bound debris return to the pericenter at a time $t=t_m$ and the infalling debris forms an initial accretion disc in time $t_c$ with mass $\displaystyle{M_{\rm i}(t_c) = \int_{t_m}^{t_c} \dot{M}_{\rm fb} \, \diff t}$. With the obtained surface density profile ($\Sigma_i$) at initial time, the disc mass at initial time is given by $\displaystyle{M_d = \int_{r_{\rm in}}^{r_{\rm out}} \Sigma_i 2 \pi r \, \diff r}$. Since, we do not know the initial time $t_c$, we get the initial $\Sigma$ profile and thus the corresponding disc mass $M_d$ at various time $t$. Then, by equating $M_{\rm i}(t_c) = M_d$, we calculate the initial time $t_c$ which is the time corresponding to the matter crossing the ISCO and the beginning of accretion to the black hole. With the obtained initial and boundary solutions of the disc, we solve equations (\ref{twft1}) and (\ref{twft2}) for the disc evolutions. To solve the coupled partial differential equations, we use the NDSolve package in Wolfram Mathematica Version 12\footnote{\url{https://reference.wolfram.com/language/tutorial/NDSolveIntroductoryTutorial.html}} with the method of line and spatial discretization using TensorProductGrid.

We then calculate the effective temperature of disc using $ \sigma T_{\rm eff}^4 = [1- f(\beta_g)] Q^{+} - Q_{\rm adv} $, and thus the luminosity is calculated assuming a black body approximations. The bolometric luminosity from the disc is given by 

\begin{equation}
L_b = \int_{r_{\rm in}}^{r_{\rm out}} \sigma T_{\rm eff}^4 2 \pi r \,  \diff r.
\label{lbol}
\end{equation}

\noindent The energy transported to the corona $Q_{\rm cor} = f(\beta_g)Q^{+}$ is Compton scattered and constitutes the X-ray power-law spectrum. We assume a plane-parallel geometry for the corona and the total X-ray luminosity is given by

\begin{equation}
L_c = \int_{r_{\rm in}}^{r_{\rm out}} Q_{\rm cor} 2 \pi r \,  \diff r.
\end{equation}

Even though we have assumed that the accretion disc is non-relativistic, we consider the effect of relativistic gravitational redshift of photons in Kerr geometry. The gravitational redshift effect and Doppler redshift are included by considering the Lorentz invariant $I_{\nu}/\nu^3$ \citep{1979rpa..book.....R}, such that $I_{\nu}(\nu_{\rm obs}) = g^3 I_{\nu}(\nu_{\rm em})$, where $\nu_{\rm em}$ is the emitted frequency, $g$ is the redshift factor and $I_{\nu}$ is the blackbody intensity. The kinematic and gravitational redshift is taken to be  

\begin{equation}
g = \frac{1}{U^0} = \frac{1-3/x +2j/x^{3/2}}{1 +j/x^{3/2}}, 
\end{equation}

\noindent where $j$ is the black hole spin, $x = r/r_g$ with $r_g$ as the gravitational radii and $U^0$ is the time component of four velocity for a circular orbit. The radiation flux per unit frequency as seen by a distant observer at rest, is given by

\begin{equation}
F_{\nu} (\nu_{\rm obs}) = \int I_{\nu}(\nu_{\rm obs}) \, \diff \Theta, 
\end{equation}

\noindent where $I_{\nu}(\nu_{\rm obs})$ is the intensity at the observer's wavelength and we approximate the differential element $\diff \Theta$ as seen from a distant observer in the Newtonian limit that is given by 

\begin{equation}
\diff \Theta = \frac{2 \pi r \diff r }{D_L^2} \cos\theta_{\rm obs},
\label{dtheta}
\end{equation}

\noindent where $D_L$ is the luminosity distance of the source to the observer. Using Lorentz invariant, the observed flux in a spectral band $\{\nu_l,~\nu_h\}$ where $\nu_l$ and $\nu_h$ are low and high frequencies, is given by

\begin{equation}
F (\nu_{\rm obs}) = \frac{\cos\theta_{\rm obs}}{D_L^2} \int_{r_{\rm in}}^{r_{\rm out}} \int_{\nu_l}^{\nu_h}  g^3 I_{\nu}\left(\frac{\nu_{\rm obs}}{g}\right) \, \diff A \, \diff \nu_{\rm obs}.
\label{fnuobs}
\end{equation} 

\noindent Then, the observed luminosity in a given spectral band is $L (\nu_{\rm obs}) = 4 \pi D_L^2 F (\nu_{\rm obs}) $ and we consider the galaxy to be face on such that $\theta_{\rm obs} = 0^{\circ} $. In the next section, we present the result of our model.

\begin{table}
%\center{
\scalebox{0.93}{
\begin{tabular}{|c|ccc|}
\hline
&&&\\
Cases & $M_{\bullet,6}$ & $m$ & $j$  \\
&&& \\
\hline
&&&\\
\Romannum{1} & 1 & 1 & 0 \\ 
&&&\\
\Romannum{2} & 5 & 1 & 0 \\ 
&&&\\
\Romannum{3} & 10 & 1 & 0 \\ 
&&&\\
\Romannum{4} & 1 & 5 & 0 \\ 
&&&\\
\Romannum{5} & 1 & 10 & 0 \\ 
&&&\\
\Romannum{6} & 1 & 1 & 0.5 \\ 
&&&\\
\Romannum{7} & 1 & 1 & 0.8 \\ 
&&&\\
\hline
\end{tabular}
%}
}
{
\begin{tabular}{|c|c|}
\hline
&\\
Cases & $b_2$   \\
& \\
\hline
&\\
B1 & 0 \\ 
&\\
B2 & 0.1 \\ 
&\\
B3 & 0.5 \\ 
&\\
B4 & 0.9 \\ 
&\\
\hline
\end{tabular}
%}
}
{
\begin{tabular}{|c|c|}
\hline
&\\
Cases & $\mu$   \\
& \\
\hline
&\\
A1 & 1 \\ 
&\\
A2 & 0.9 \\ 
&\\
A3 & 0.8 \\ 
&\\
A4 & 0.7 \\ 
&\\
\hline
\end{tabular}
%}
}
\caption{The free variables in our model are $b_2$, $\mu$, $M_{\bullet,6} = M_{\bullet}/[10^6 M_{\odot}]$, $m = M_{\star}/M_{\odot}$ and $j$ is the black hole spin. We have shown the parameters set taken in our model. The $b_2$ shown in equation (\ref{fcor}) signifies the amount of energy transported from disc to corona and $\mu$ shown below equation (\ref{feq}) is a constant power index that indicates the contribution of gas and total pressure to the viscous stress. See section \ref{result} for more discussion. }
\label{parset}
\end{table}

\begin{table}
%\center{
\scalebox{0.82}{
\begin{tabular}{|c|c|c|c|c|c|c|c|c|}
\hline
&&&&&&&&\\
$b_2$ & $\mu$ & \Romannum{1} & \Romannum{2} & \Romannum{3} & \Romannum{4} & \Romannum{5} & \Romannum{6} & \Romannum{7}    \\
&&&&&&&&\\
\hline
&&&&&&&&\\
B1 & A1 & 3.47 & 7.28 & 10.25 & 6.67 & 7.2 & 3.48 & 3.48 \\
&&&&&&&&\\
B1 & A2 & 4.54 & 8.57 & 11.54 & 8.07 & 10.49  & 4.55 & 4.56 \\
&&&&&&&&\\
B1 & A3 & 6.62 & 11.9 & 15.26 & 12.13 & 15.68 & 6.65 & 6.66 \\
&&&&&&&&\\
B1 & A4 & 10.09 & 18.47 & 23.38 & 17.67 & 20.84 & 10.11 & 10.11\\
&&&&&&&&\\
\hline
&&&&&&&&\\
B2 & A1 & 3.48 & 7.32 & 10.31 & 5.71 & 7.25 & 3.49 & 3.50 \\
&&&&&&&&\\
B2 & A2 & 4.56 & 8.59 & 11.58 & 8.11 & 10.53 & 4.57 & 4.58 \\
&&&&&&&&\\
B2 & A3 & 6.63 & 11.92 & 15.28 & 12.15 & 15.73 & 6.66 & 6.68 \\
&&&&&&&&\\
B2 & A4 & 10.10 & 18.48 & 23.39 & 17.63 & 21.13 & 10.12 & 10.12 \\
&&&&&&&&\\
\hline
&&&&&&&&\\
B3 & A1 & 3.58 & 7.53 & 10.64 & 5.84 & 7.48 & 3.59 & 3.59 \\
&&&&&&&&\\
B3 & A2 & 4.62 & 8.71 & 11.74 & 8.23 & 10.69 & 4.64 & 4.65 \\
&&&&&&&&\\
B3 & A3 & 6.68 & 11.99 & 15.37 & 12.23 & 15.82 & 6.71 & 6.73 \\
&&&&&&&&\\
B3 & A4 & 10.14 & 18.54  & 23.22 & 17.68 & 20.93 & 10.16 & 10.14 \\
&&&&&&&&\\
\hline
&&&&&&&&\\
B4 & A1 & 3.82 & 8.44 & 12.18 & 6.19 & 7.83 & 3.83 & 3.84 \\
&&&&&&&&\\
B4 & A2 & 4.69 & 8.86 & 11.98 & 8.37 & 10.87 & 4.72 & 4.73 \\
&&&&&&&&\\
B4 & A3 & 6.73 & 12.07 & 15.48 & 12.31 & 15.98 & 6.76 & 6.78 \\
&&&&&&&&\\
B4 & A4 & 10.16 & 18.59 & 23.53 & 17.69 & 21.05 & 10.18 & 10.17 \\
&&&&&&&&\\
\hline
\end{tabular}
%}
}
\caption{The time $t_c$ in days for the infalling debris to form a seed disc after disruption ($t = 0$), and the formed disc evolves via a viscous mechanism in the disc and the later debris infall. The parameter set is given in Table \ref{parset}.  See section \ref{result} for more discussion.}
\label{tctm}
\end{table}

\section{Result}
\label{result}

In this section, we present the results of the advection disc-corona model constructed in section \ref{dcm}. First, we discuss the various unknown parameters in the model. In the steady approximation (discussed before equation \ref{sigout}), we have shown $ k_1 = 2 C_1 / 3$. We follow this approximation and for simplicity, we assume $C_1 = 1$ such that $c_s^2 = H^2 \Omega_K^2$ and thus $k_1 = 2/3$. We take $\alpha_s=0.1$ and $\gamma_g = 5/3$ throughout in our calculations. The free variables in our model are $b_2$, $\mu$, $M_{\bullet}$, $M_{\star}$ and $j$. We normalise the black hole mass as $M_{\bullet,6} = M_{\bullet}/[10^6 M_{\odot}]$ and the stellar mass as $m = M_{\star}/M_{\odot}$. Table \ref{parset} shows the parameters set used in this paper. We take the disc inner radius to be innermost stable circular orbit (ISCO) given by $r_{\rm in} = r_g Z(j)$, where $r_g = G M_{\bullet}/c^2 $ and $Z(j)$ is given by 

\begin{equation}
Z(j) =3+Z_2(j)-\sqrt{(3-Z_1(j)) (3+Z_1(j)+2 Z_2(j))}, \label{zjb}
\end{equation}

\noindent where

\begin{subequations}
\begin{align}
Z_1(j) &=1+(1-j^2)^{\frac{1}{3}} \left[(1+j)^{\frac{1}{3}}+(1-j)^{\frac{1}{3}}\right]\\
Z_2(j) &=\sqrt{3 j^2+Z_1(j)^{2}}.
\end{align}
\end{subequations} 

We have obtained an initial solution for the disc following discussion presented below equation (\ref{sft}) for $\mu \geq 0.7$, and below this, we did not get any solutions that satisfy the mass constraint (disc mass at initial time equal to the debris mass infall by that time). This may be due to the assumption of constant $\partial W_t/\partial t$ and $\partial \mathcal{F}_t/\partial t$ at the initial time. We are able to produce the time-dependent advective solution for a disc with fallback for $\mu \geq 0.7$. Thus, our solution implies that the viscous stress is dominated by total pressure, and the seed disc formed is at a very early time (low $t_c/t_m$). Table \ref{tctm} shows the beginning time of accretion $t_c$ for the various parameters given in Table \ref{parset}. The time $t_c$ is measured from the moment of disruption which is taken to be $t= 0$. For a given $b_2$, by comparing case \Romannum{1} for various $\mu$ (cases A1, A2, A3, A4), we can see that the time $t_c$ increases with $\mu$ indicating that the increase in the contribution of gas pressure in the viscous stress delays the beginning of accretion. The time $t_c$ also increases with black hole mass (cases \Romannum{1}, \Romannum{2} and \Romannum{3}) and stellar mass (cases \Romannum{1}, \Romannum{4} and \Romannum{5}). Since the black hole spin ($j$) in the calculation is through the inner radius only, the time $t_c$ shows a weak variation with $j$. The energy loss to the corona does not have a severe effect on time $t_c$ as it shows a small increment with $b_2$ (cases B1, B1, B3, B4).

\begin{figure*}
\begin{center}
\subfigure[]{\includegraphics[scale=0.56]{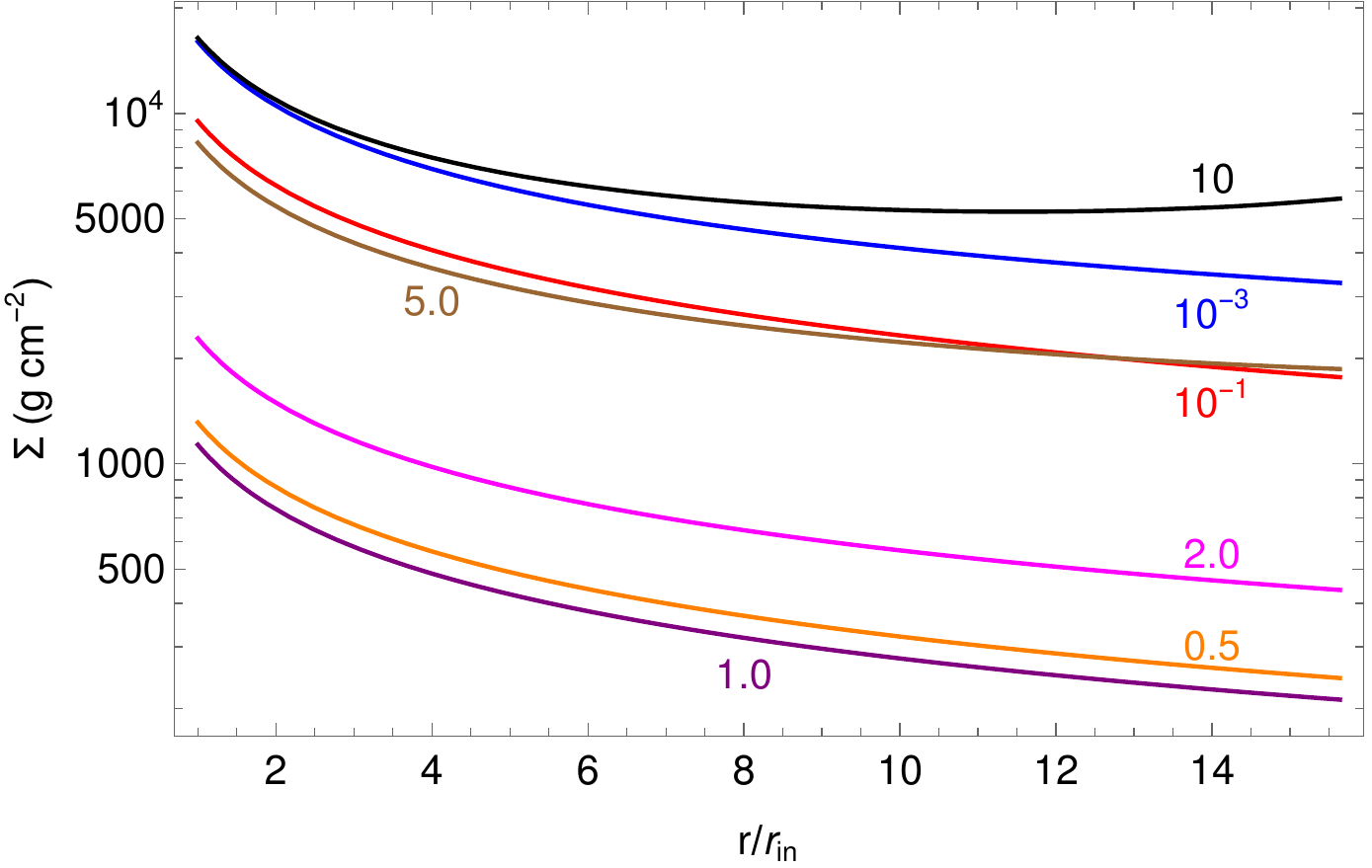}}~~
\subfigure[]{\includegraphics[scale=0.56]{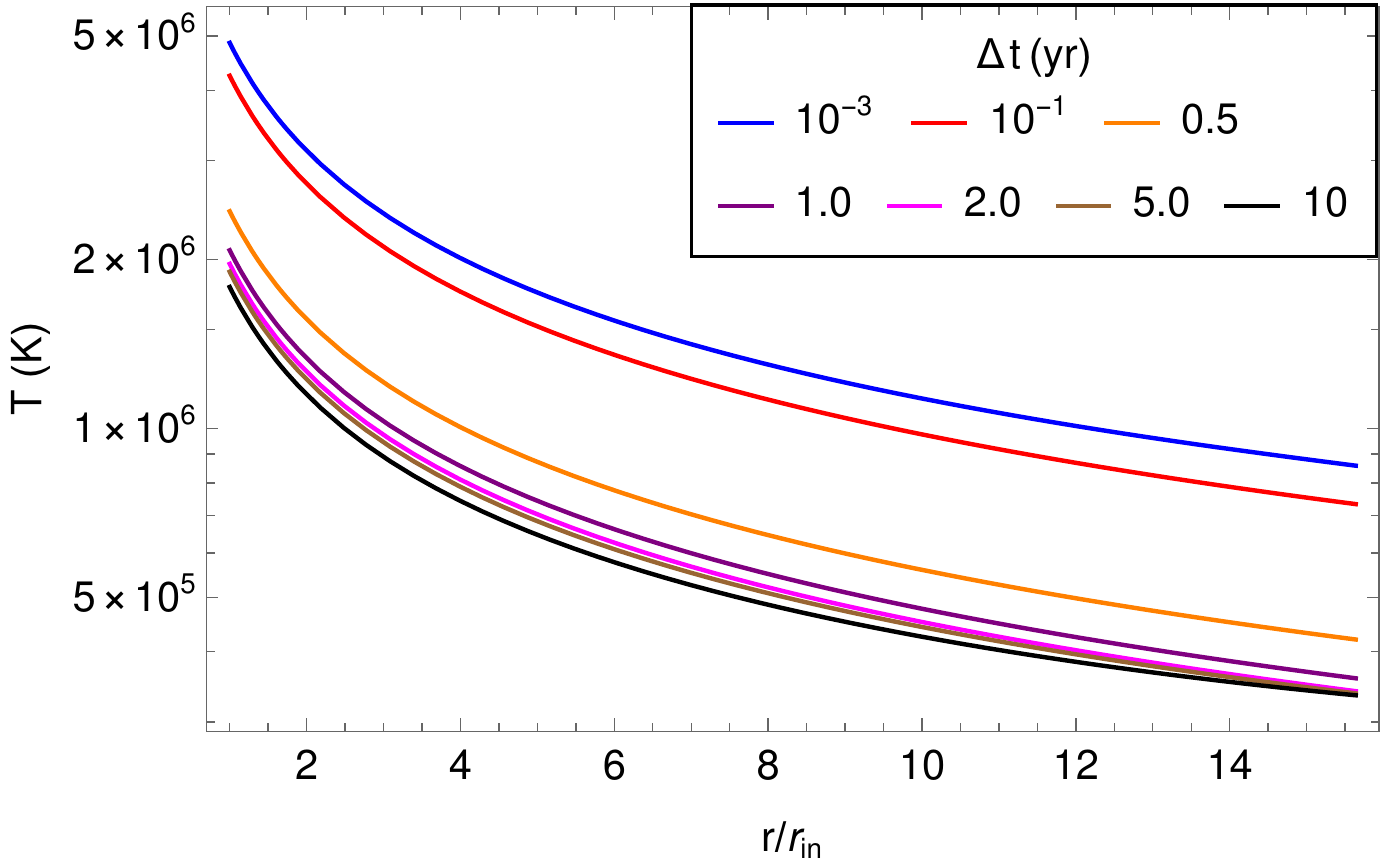}}
\subfigure[]{\includegraphics[scale=0.56]{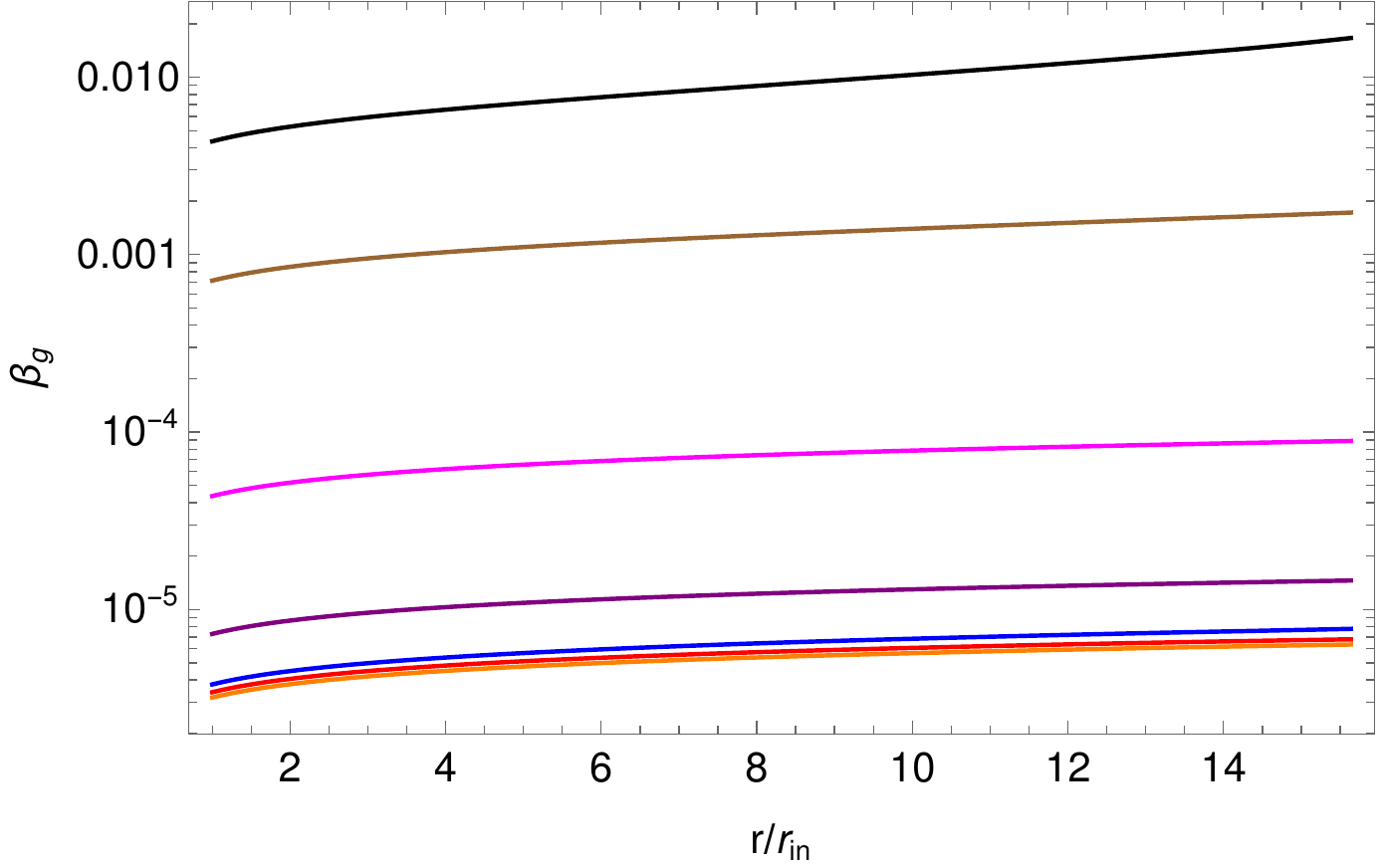}}~~
\subfigure[]{\includegraphics[scale=0.56]{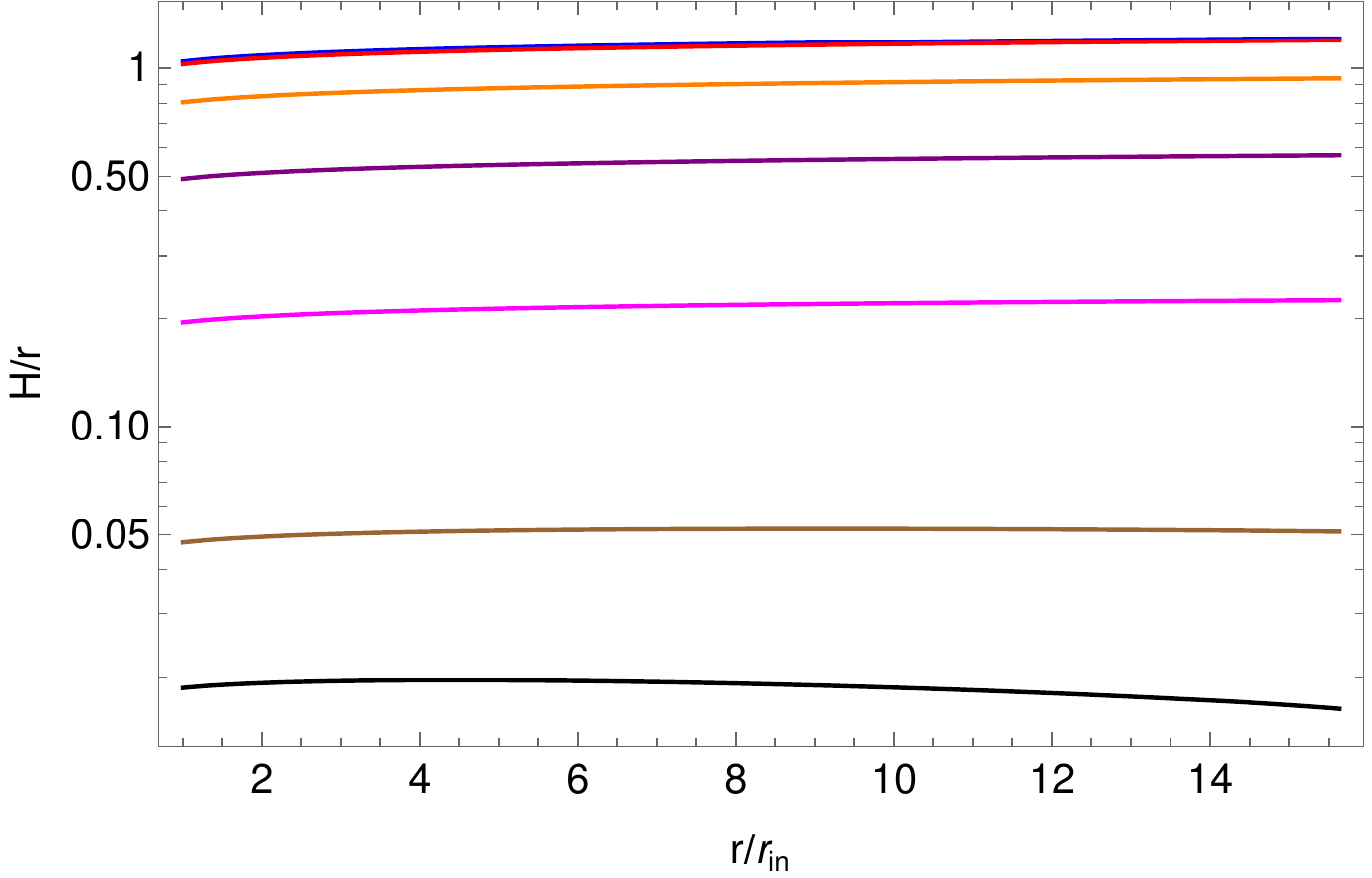}}
\end{center}
\caption{The time evolution of surface density $\Sigma$, disc temperature $T$, the ratio of gas to total pressure $\beta_g$, and disc height are shown for the parameter set B1-A1-\Romannum{1}. The $\beta_g$ increases at late times implying that the gas pressure dominates at late times. The disc height to radius ratio decreases with time and thus the disc tends to be a thin disc at later times. The $\Delta t=0$ corresponds to the beginning of accretion to the black hole and the duration between stellar disruption and the beginning of accretion is the initial time $t_c$ given in Table \ref{tctm}. The coloured number in plot (a) near the curve represents $\Delta t$ in the year. See section \ref{result} for more discussion.}
\label{sigt}
\end{figure*}

From now onwards in the plots, we consider $\Delta t = t-t_c$ such that $\Delta t=0$ corresponds to the matter crossing the ISCO and the beginning of accretion to the black hole. The time evolution of surface density $\Sigma$ and disc temperature $T$ are shown in Fig. \ref{sigt}. The ratio of gas pressure to total pressure $\beta_g$ increases with time and the gas pressure dominates at the late time. Since the radiation pressure dominates at an early time, the total pressure is dominated by radiation pressure, the accretion is near to super Eddington and decreases to sub-Eddington at later times. The disc height to radius ratio decreases with time and the disc tends to a sub-Eddington thin disc phase at later times. 

The time evolution of mass accretion rate at the inner radius $\dot{M}_a$ is shown in Fig. \ref{macc} for case \Romannum{1} with various values of $b_2$ and $\mu$. The fraction of energy transported to corona is $f(\beta_g) = b_2 \beta_g^{(1-\mu)/2}$ and for $\mu = 1$, it is a constant. For $\mu \neq 1$, it is a function of $\beta_g$ and is small when pressure is dominated by radiation pressure ($\beta_g \ll 1$). Thus, variation in mass accretion rate with $b_2$ is higher for $\mu = 1$ at initial time. The accretion rate is close to the fallback rate at late times and with an increase in $b_2$, the mass accretion rate deviates from the fallback rate for higher $\mu$. The increase in $b_2$ corresponding to an increase in energy transport to corona results in a decrease in the time corresponding to deviation from the mass fallback rate. The beginning time $t_c$ for accretion increases with a decrease in $\mu$ which suggests that the initial disc mass is higher for lower $\mu$ as can be seen from figure (d). The disc mass evolves with time that decreases when $\dot{M}_a > \dot{M}_{\rm fb}$ and increases for $\dot{M}_a < \dot{M}_{\rm fb}$.

\begin{figure*}
\begin{center}
\subfigure[]{\includegraphics[scale=0.6]{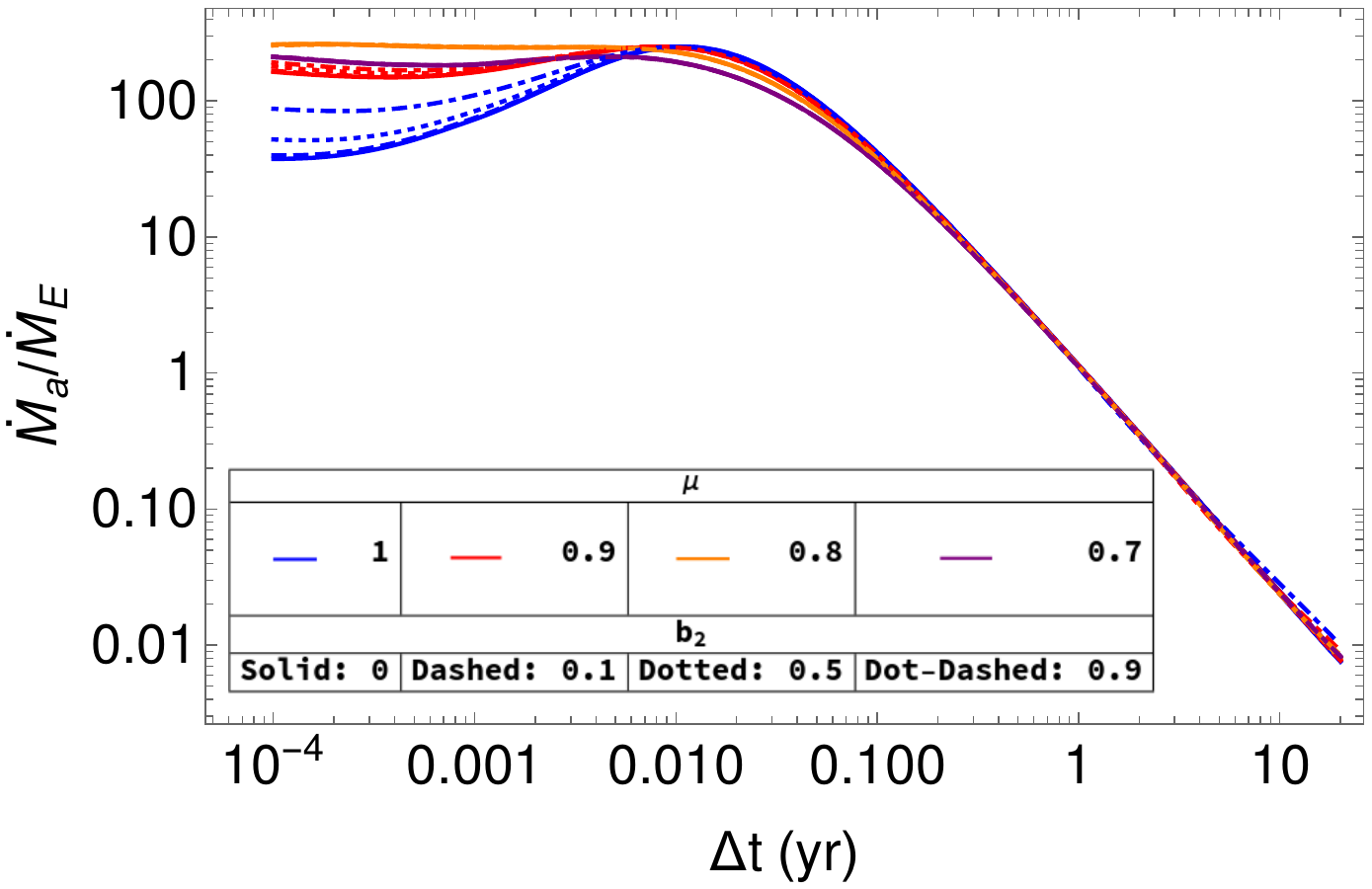}}~~
\subfigure[]{\includegraphics[scale=0.6]{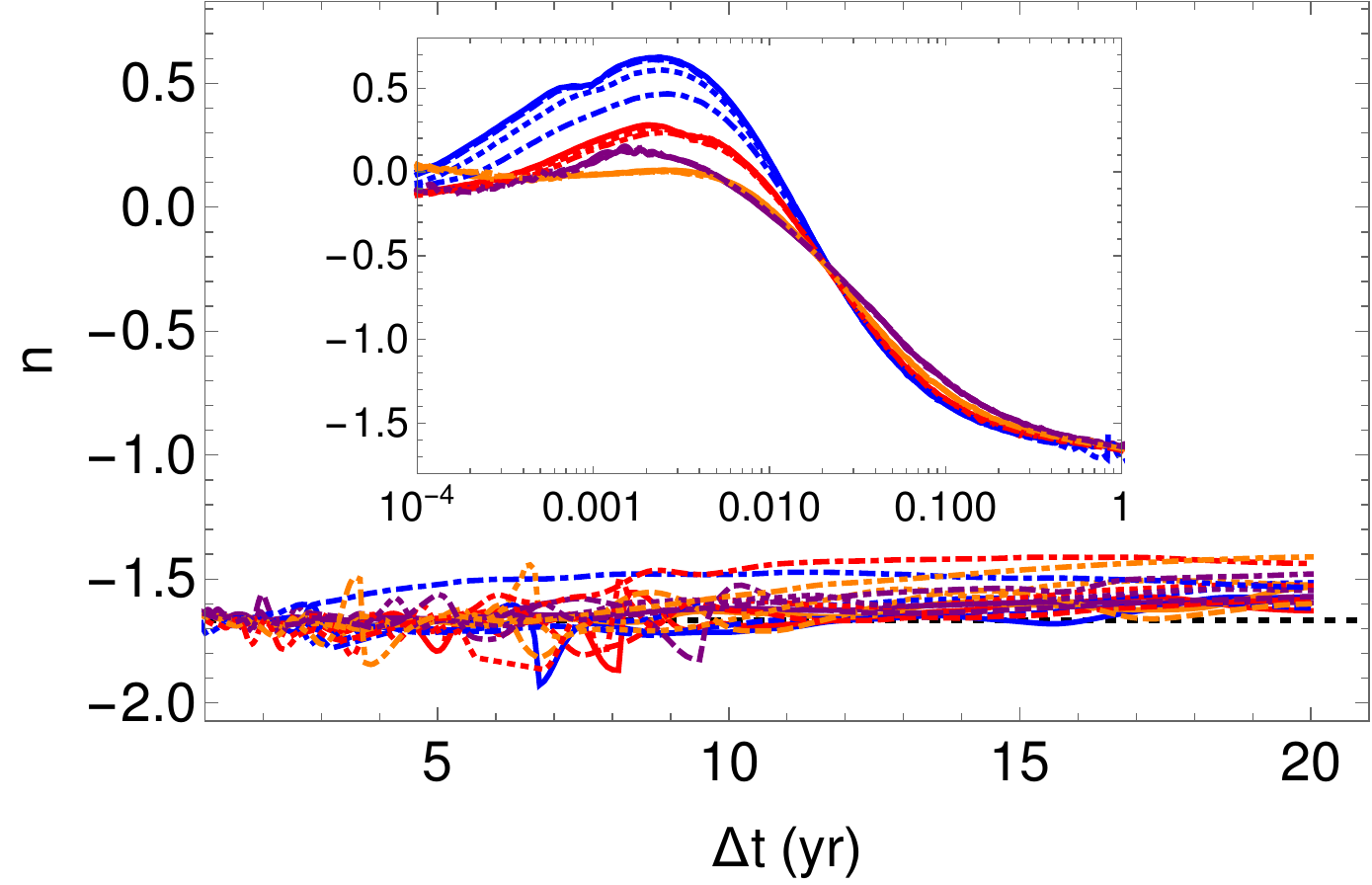}}
\subfigure[]{\includegraphics[scale=0.48]{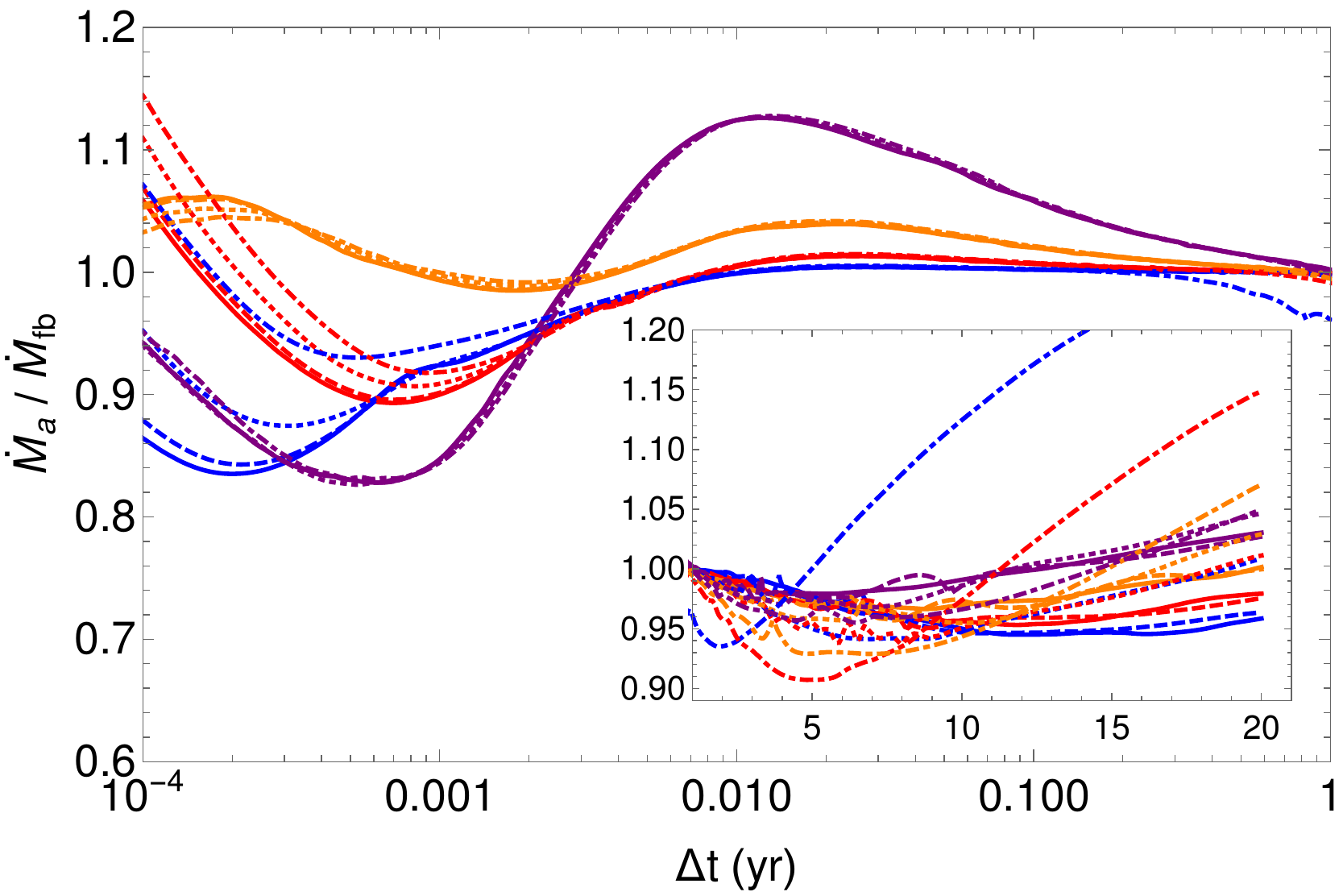}}~~
\subfigure[]{\includegraphics[scale=0.6]{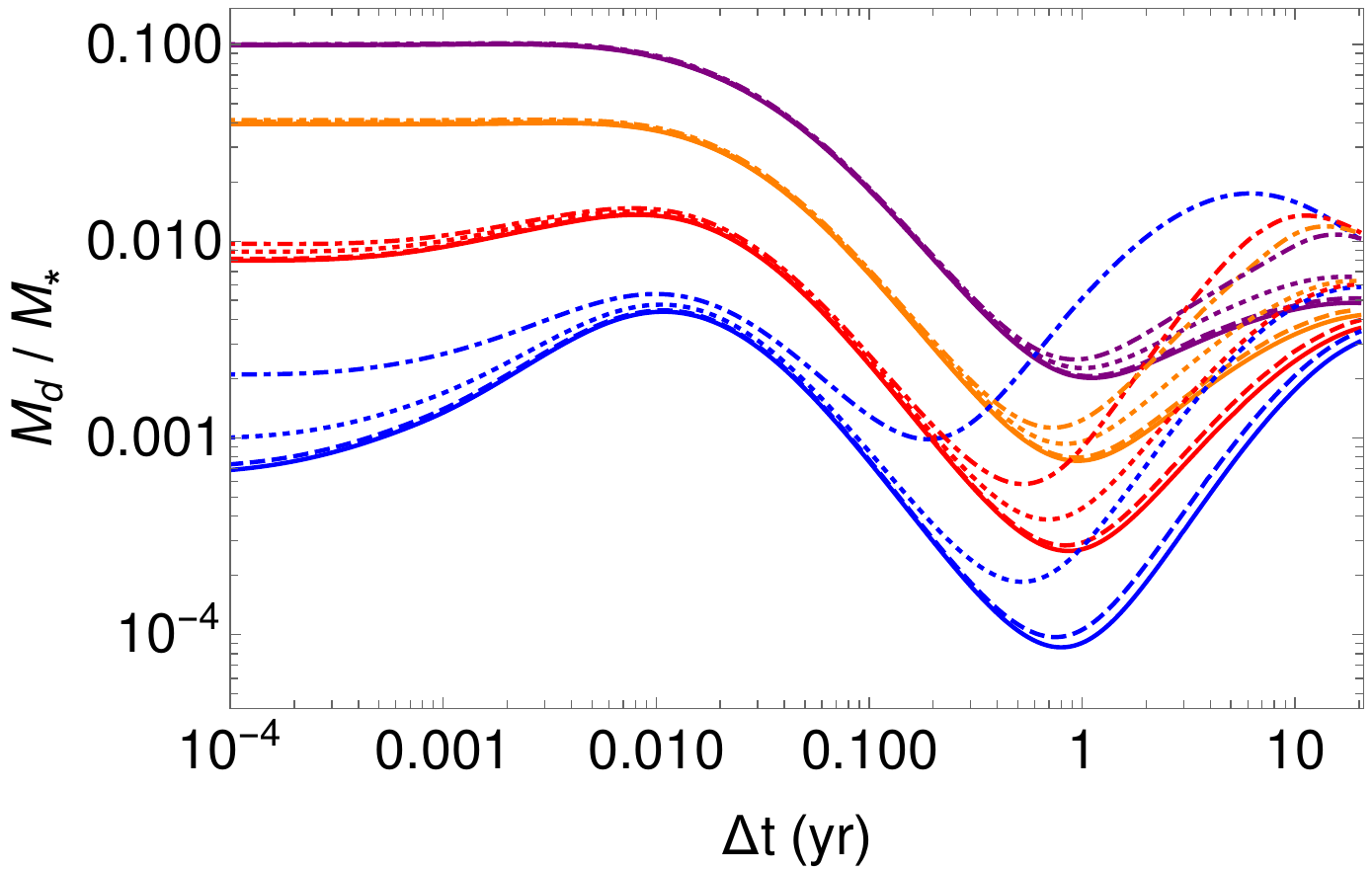}}
\end{center}
\caption{(a) The time evolution of mass accretion rate at inner radius $\dot{M}_a$ normalized to Eddington rate for case \Romannum{1} .(b) The time evolution of $n$ such that $\dot{M}_a \propto (t-t_c)^{n}$. See Table \ref{tctm} for $t_c$. The black dashed lines correspond to $n = -5/3$. (c) The ratio of mass accretion rate to mass fallback rate as a function of time. The mass accretion rate is close to the mass fallback rate but shows deviation with $b_2$ at late times. (d) The time evolution of disc mass $M_d$. The disc mass evolves with time that decreases when $\dot{M}_a > \dot{M}_{\rm fb}$ and increases for $\dot{M}_a < \dot{M}_{\rm fb}$. See section \ref{result} for more discussion.}
\label{macc}
\end{figure*}

\begin{figure}
\begin{center}
\subfigure[]{\includegraphics[scale=0.55]{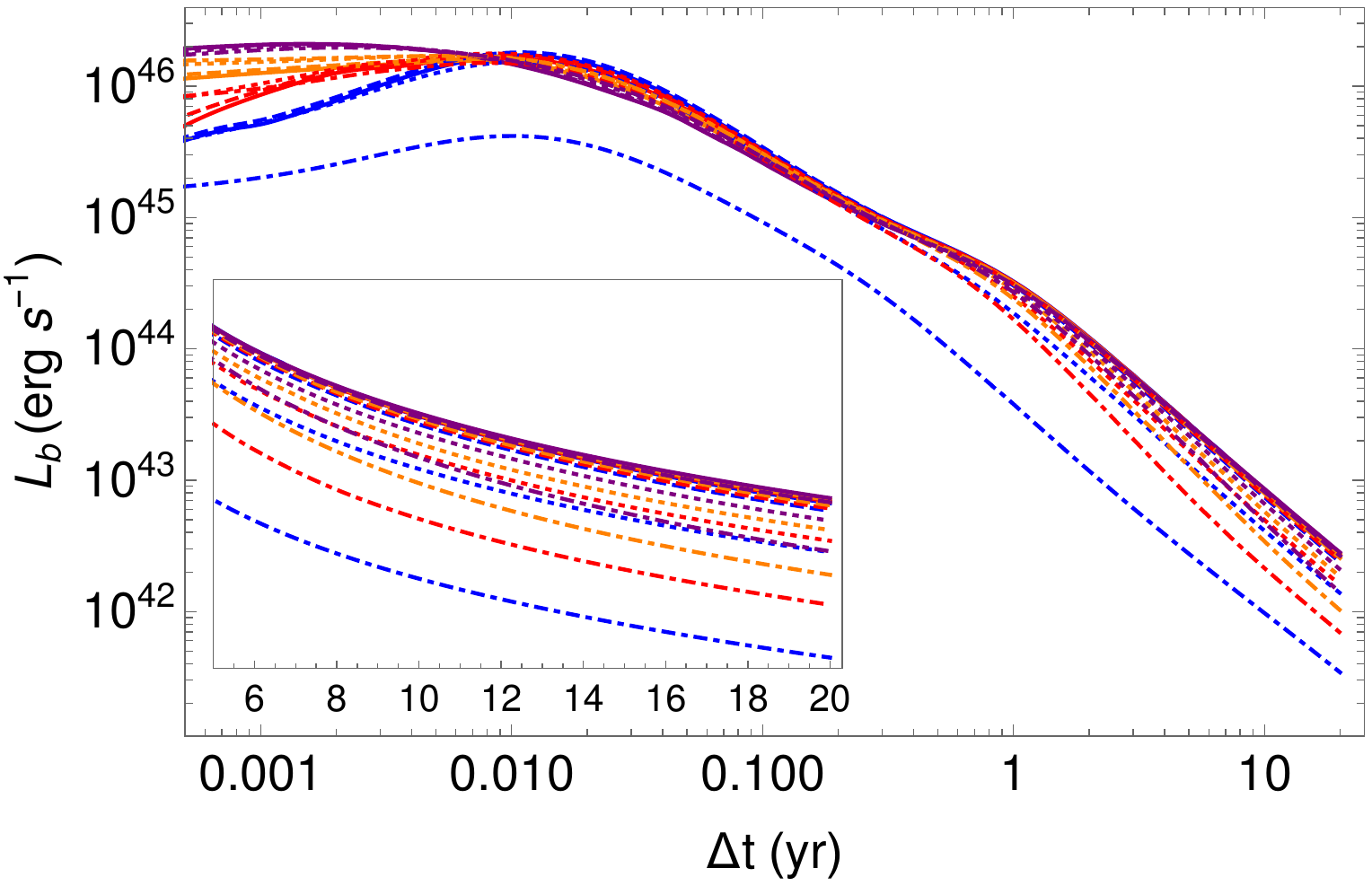}}
\subfigure[]{\includegraphics[scale=0.53]{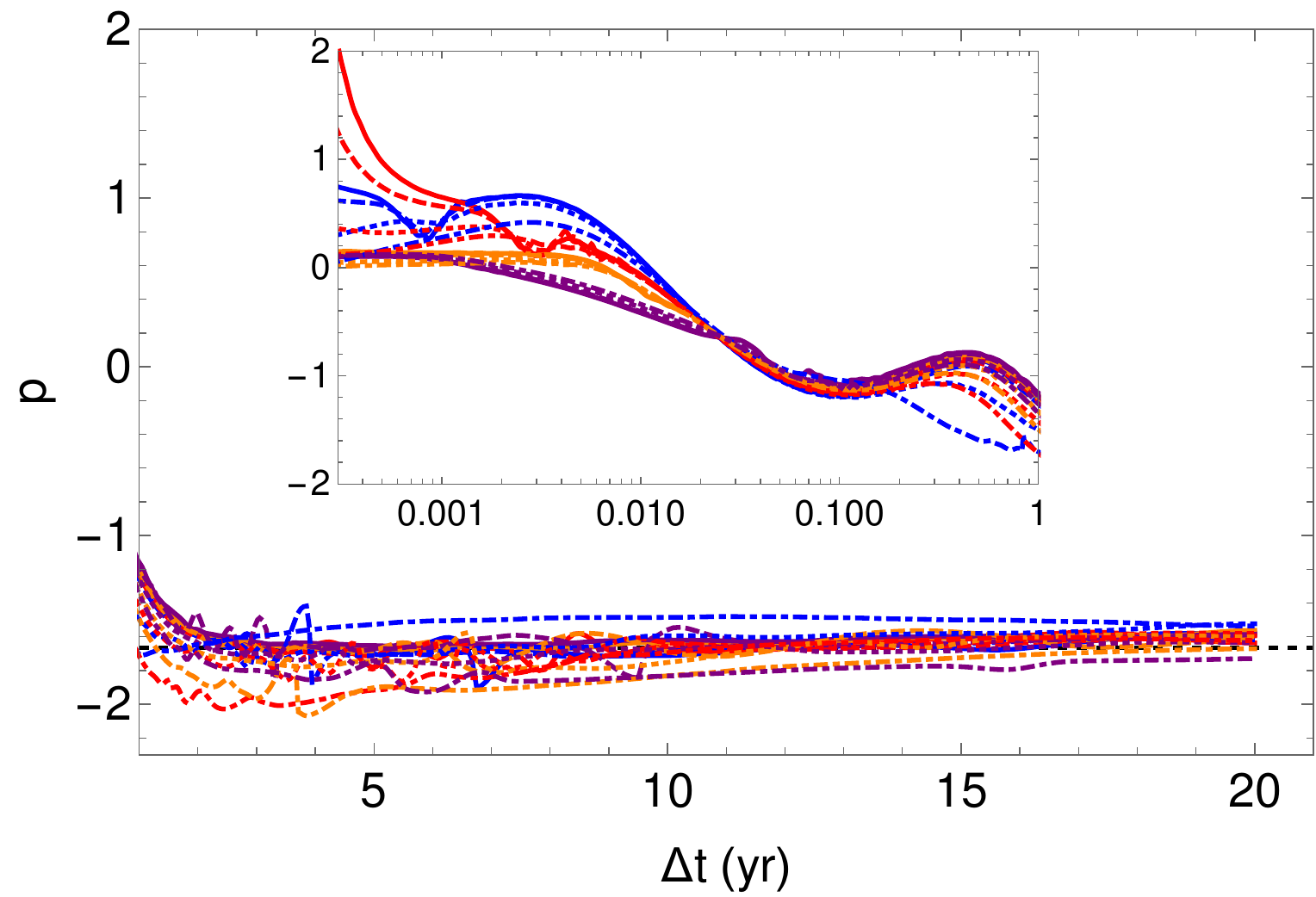}}
\subfigure[]{\includegraphics[scale=0.58]{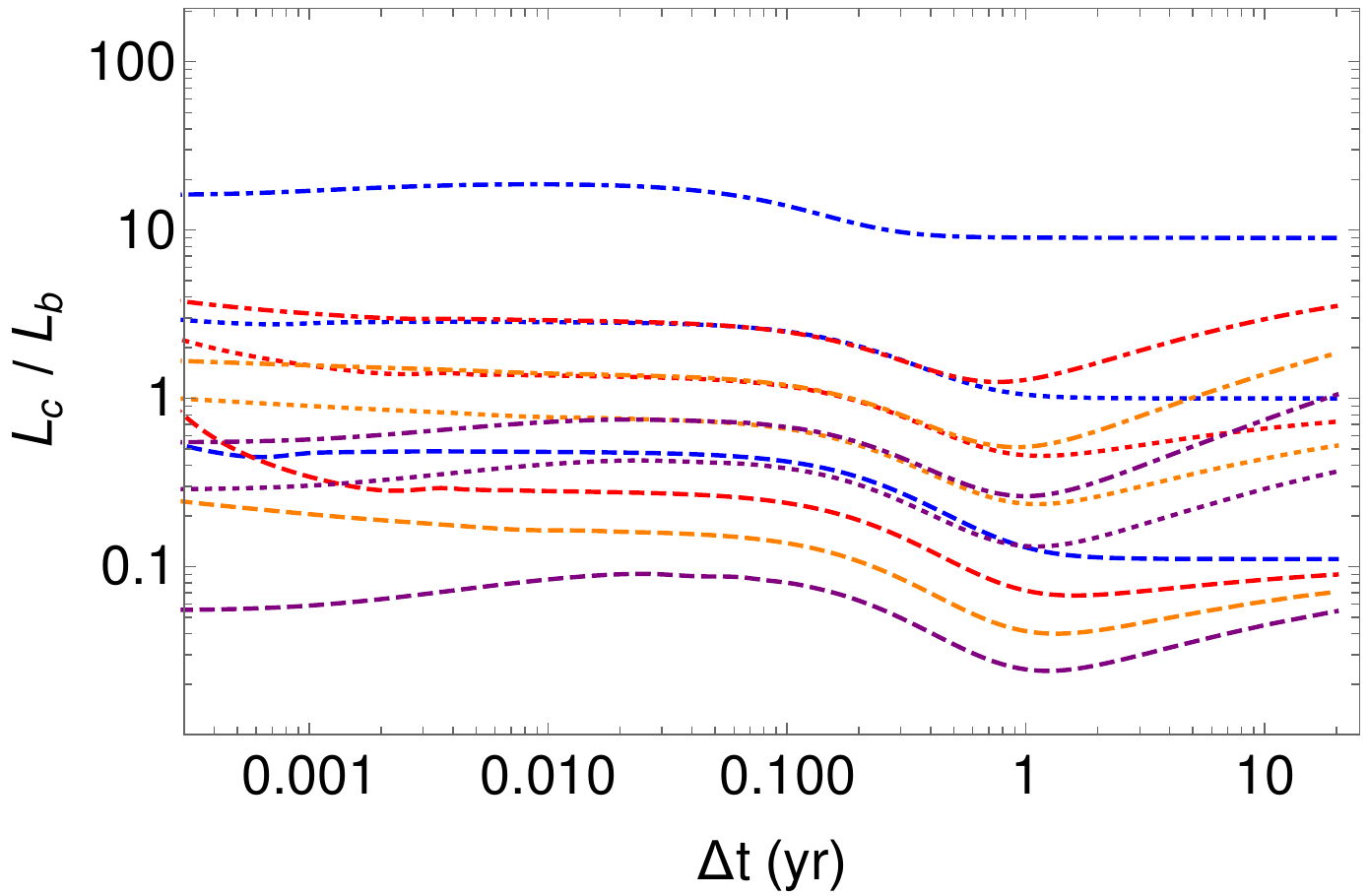}}
\end{center}
\caption{(a) The time evolution of disc luminosity for case \Romannum{1} and the plot's legend is same as given in Fig. \ref{macc}a. The disc luminosity decreases with an increase in energy transport to the corona. The luminosity shows a power-law decline at the late time and we approximate $L_b \propto (t-t_c)^{p}$. See Table \ref{tctm} for $t_c$. (b) The evolution of $p= \diff \log(L_b) / \diff \log(t)$. The black dashed line correspond to $p = -5/3$. The late time evolution of the disc luminosity is close to $t^{-5/3}$ evolution. (c) The total X-ray luminosity from the corona ($L_c$) to the disc bolometric luminosity $L_b$ ratio is shown. The ratio decreases with an increase in $\mu$ implying that the energy transport to corona decreases with an increase in gas pressure contribution to the viscous stress. For $\mu =1 $, at late time $L_c / L_b \simeq b_2 / (1- b_2) = 1$. See section \ref{result} for more discussion.}
\label{lbmu}
\end{figure}

The bolometric luminosity from the disc corresponding to case \Romannum{1} with various values of $b_2$ (cases B1, B2, B3, B4) and $\mu$ (cases A1, A2, A3, A4) is shown in Fig. \ref{lbmu}. The luminosity at the initial time increases with a decrease in $\mu$ due to an increase in mass accretion rate and thus the viscous emission. As time progress, the radiation pressure decreases, and the viscous stress tends toward gas pressure dominated at which the effect of $\mu$ on viscous stress is weak. Thus, the luminosity at late times is similar for various $\mu$ with no corona ($b_2 = 0$). However, the presence of corona affects the disc luminosity and with an increase in $b_2$, the disc luminosity decreases. With a decrease in $\mu$, the disc luminosity shows weak changes which implies that an increase in the gas pressure contribution to the viscous stress reduces the energy transport to the corona. The disc luminosity decreases with an increase in $b_2$ due to an increase in corona flux loss. The luminosity shows a transition from super-Eddington to sub-Eddington phase and the disc luminosity decreases with an increase in energy transport to the corona. The luminosity follows a power-law decline at late time and we approximate $L_b \propto (t-t_c)^{p}$. The time evolution of $p$ shows that at late times, the luminosity evolution is close to $t^{-5/3}$.

\begin{figure*}
\begin{center}
\subfigure[]{\includegraphics[scale=0.6]{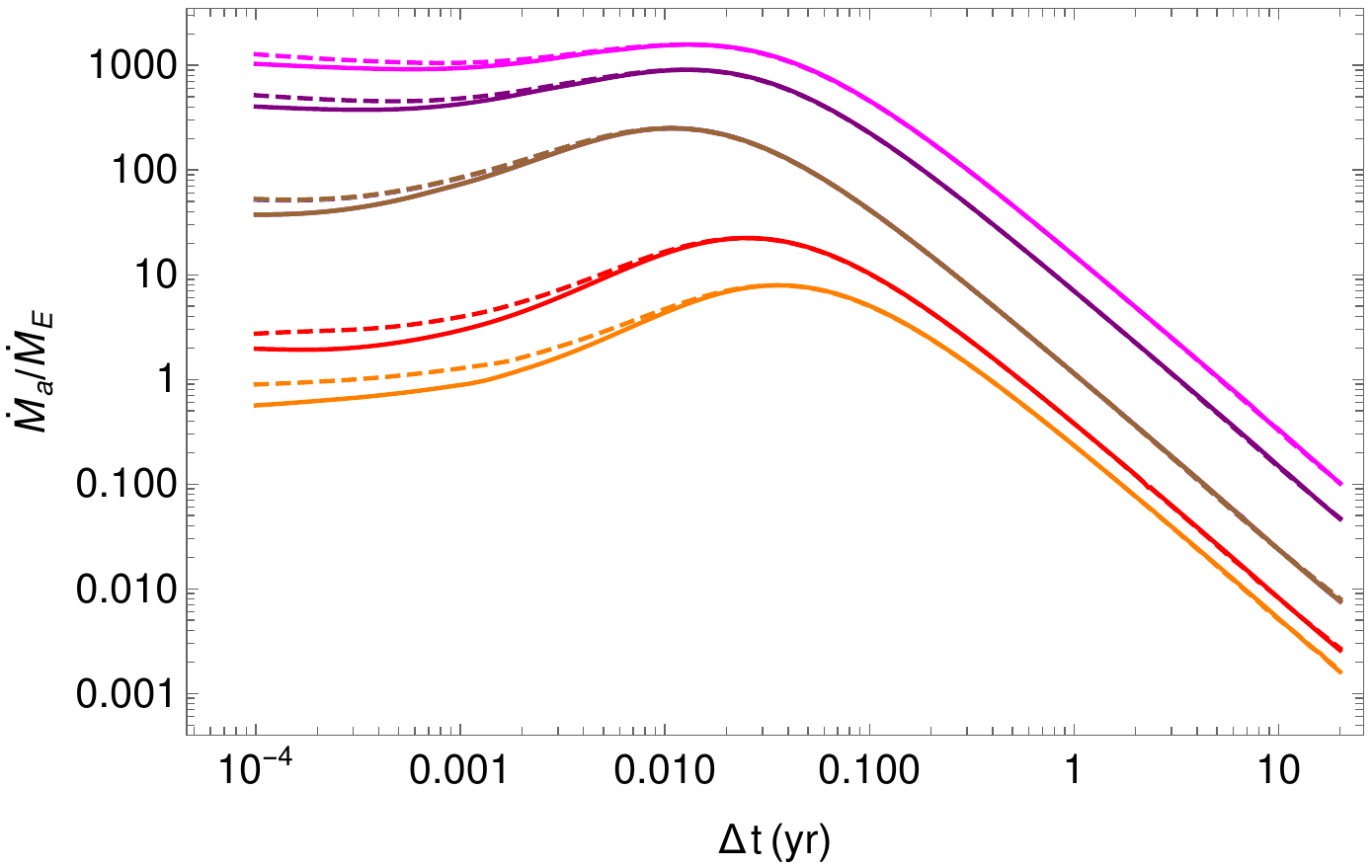}}~~
\subfigure[]{\includegraphics[scale=0.58]{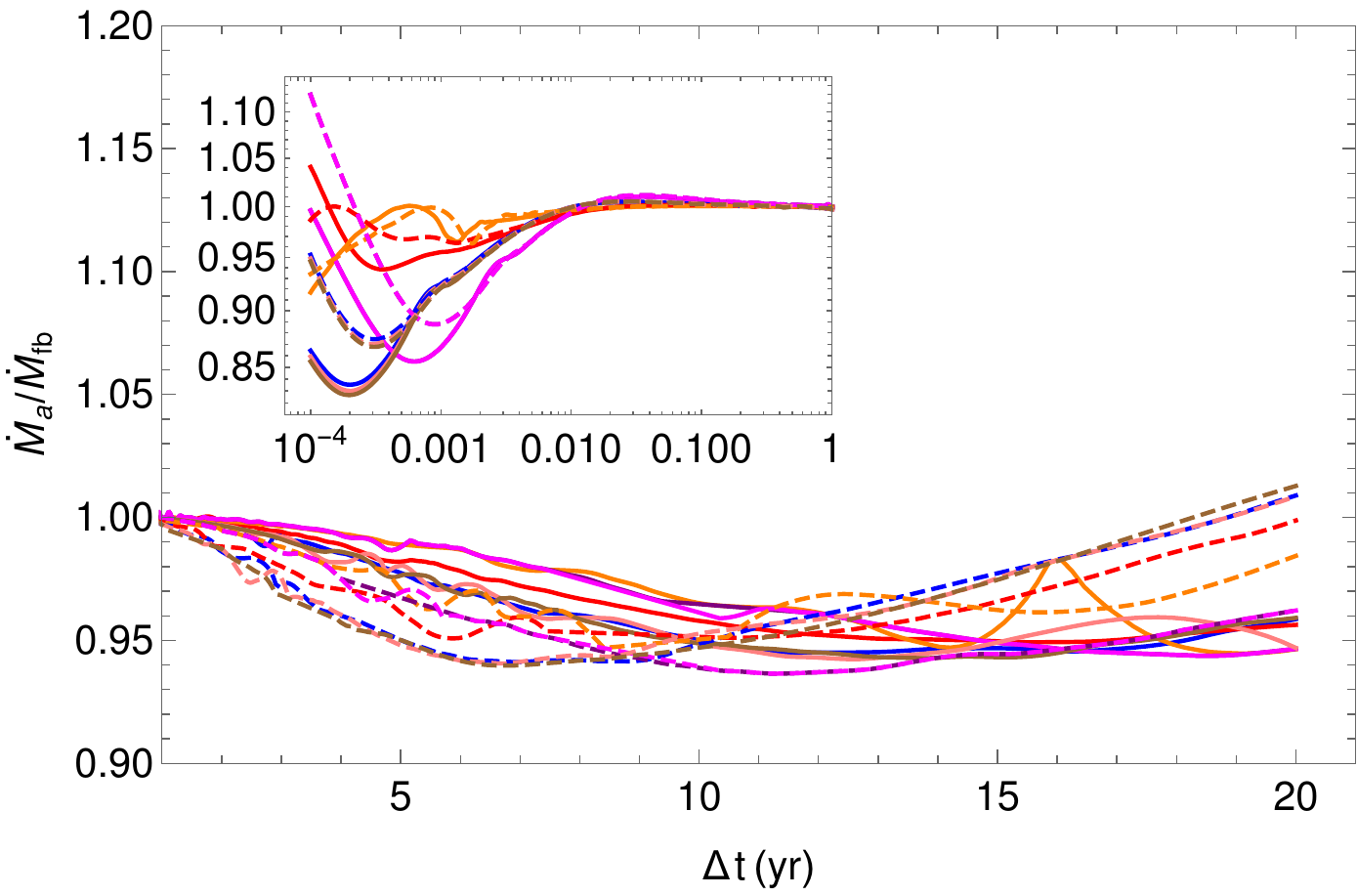}}
\subfigure[]{\includegraphics[scale=0.62]{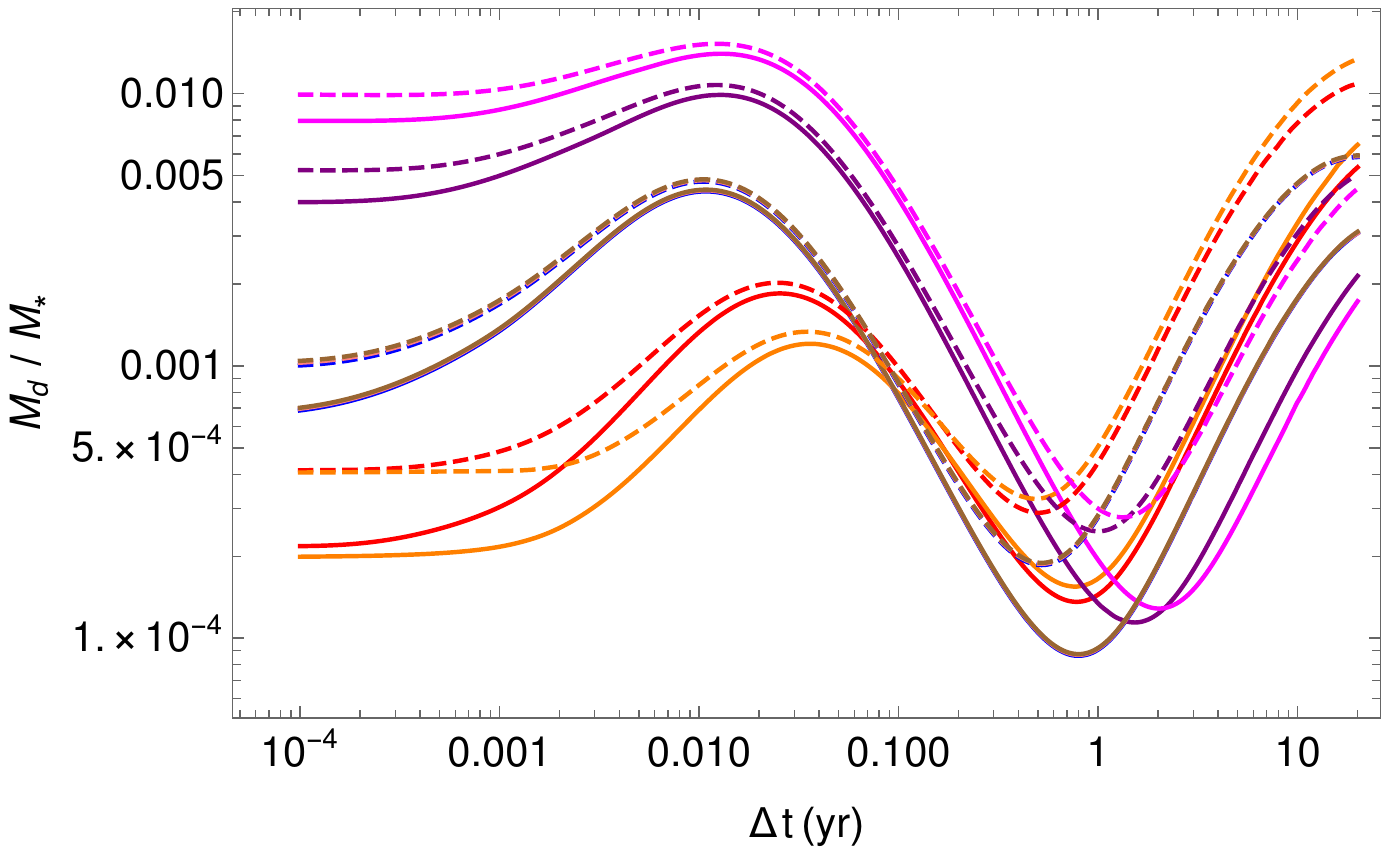}}~~~~~~~~~~~
\subfigure[]{\includegraphics[scale=0.43]{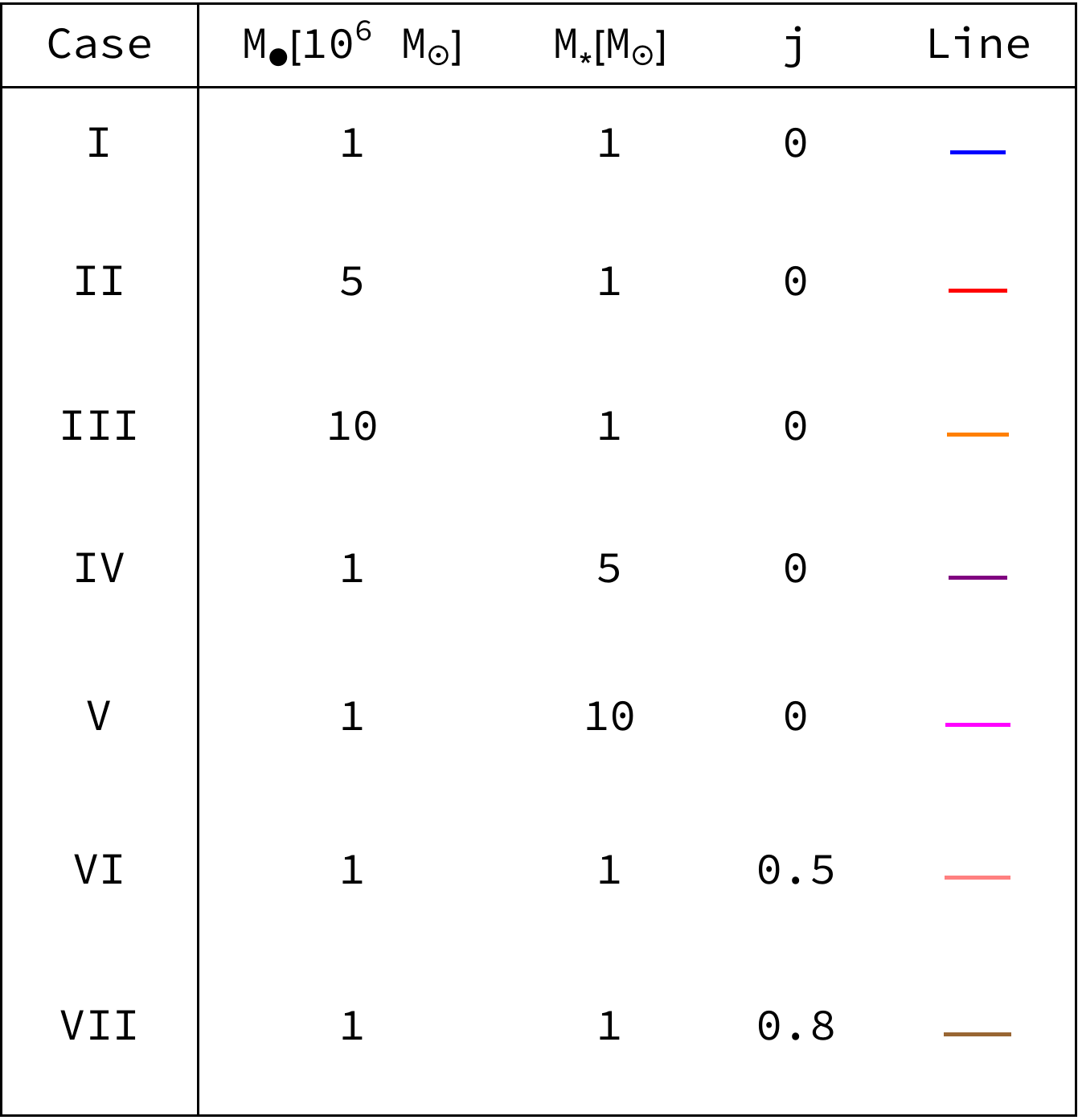}}
\end{center}
\caption{(a) The time evolution of mass accretion rate at inner radius $\dot{M}_a$ normalized to Eddington rate for no corona [$b_2 =0$, B1, solid lines] and with corona [$b_2 =0.5$, B3, dashed lines]. The $\mu = 1 $ correspond to case A1. The blue, pink, and brown lines are nearly overlapped implying that the black hole spin has an insignificant effect on $\dot{M}_a$. The normalized mass accretion rate increases with stellar mass but decreases with black hole mass implying that the disc is sub-Eddington for higher mass black holes. (b) The ratio of mass accretion rate to the mass fallback rate is shown and at late times, the ratio is dominated by a disc with corona. (c) The time evolution of disc mass and it evolves with time that decreases when $\dot{M}_a > \dot{M}_{\rm fb}$ and increases for $\dot{M}_a < \dot{M}_{\rm fb}$. (d) Shows the legend of various coloured curves.
}
\label{macct}
\end{figure*}

The time evolution of mass accretion rate for various values of black hole mass and spin and stellar mass is shown in Fig. \ref{macct}. The mass accretion rate decreases with time and tends to sub-Eddington phase at late times. The mass accretion rate normalized to the Eddington rate increases with stellar mass but decreases with black hole mass resulting in a sub-Eddington disc for higher mass black holes. The mass accretion rate is close to the mass fallback rate and the ratio increases at late times with the presence of corona. The time evolution of power index given by $n = \diff \log(\dot{M}_a) / \diff \log(t)$ is close to $n = -5/3$ corresponding to late time mass fallback rate. The disc mass decreases for $\dot{M}_a > \dot{M}_{\rm fb}$ and increases for $\dot{M}_a < \dot{M}_{\rm fb}$. The disc mass at an initial time in presence of corona is higher than in absence of corona because of an increase in the beginning time of accretion resulting in more debris mass for the initial disc (see Table \ref{tctm}).  Since the disc is formed at an early time, the initial disc mass in terms of the stellar mass is small. Since the mass accretion rate is comparable to the mass fallback rate, the amount of mass accreted by the black hole is of the order of mass added by the infalling debris. Thus, the disc mass remains small compared to the total stellar mass. 

\begin{figure}
\begin{center}
\subfigure[]{\includegraphics[scale=0.6]{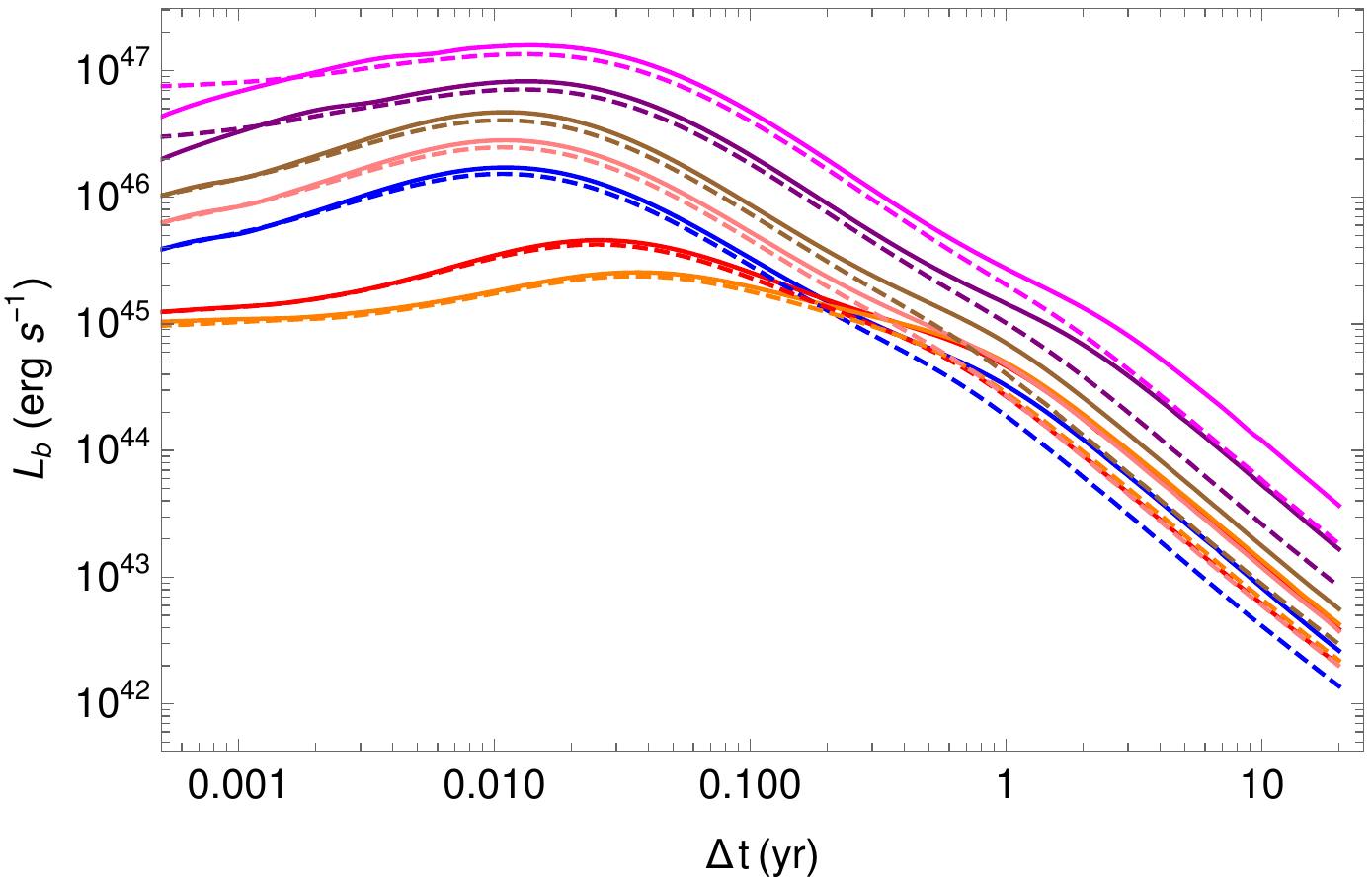}}
\subfigure[]{\includegraphics[scale=0.6]{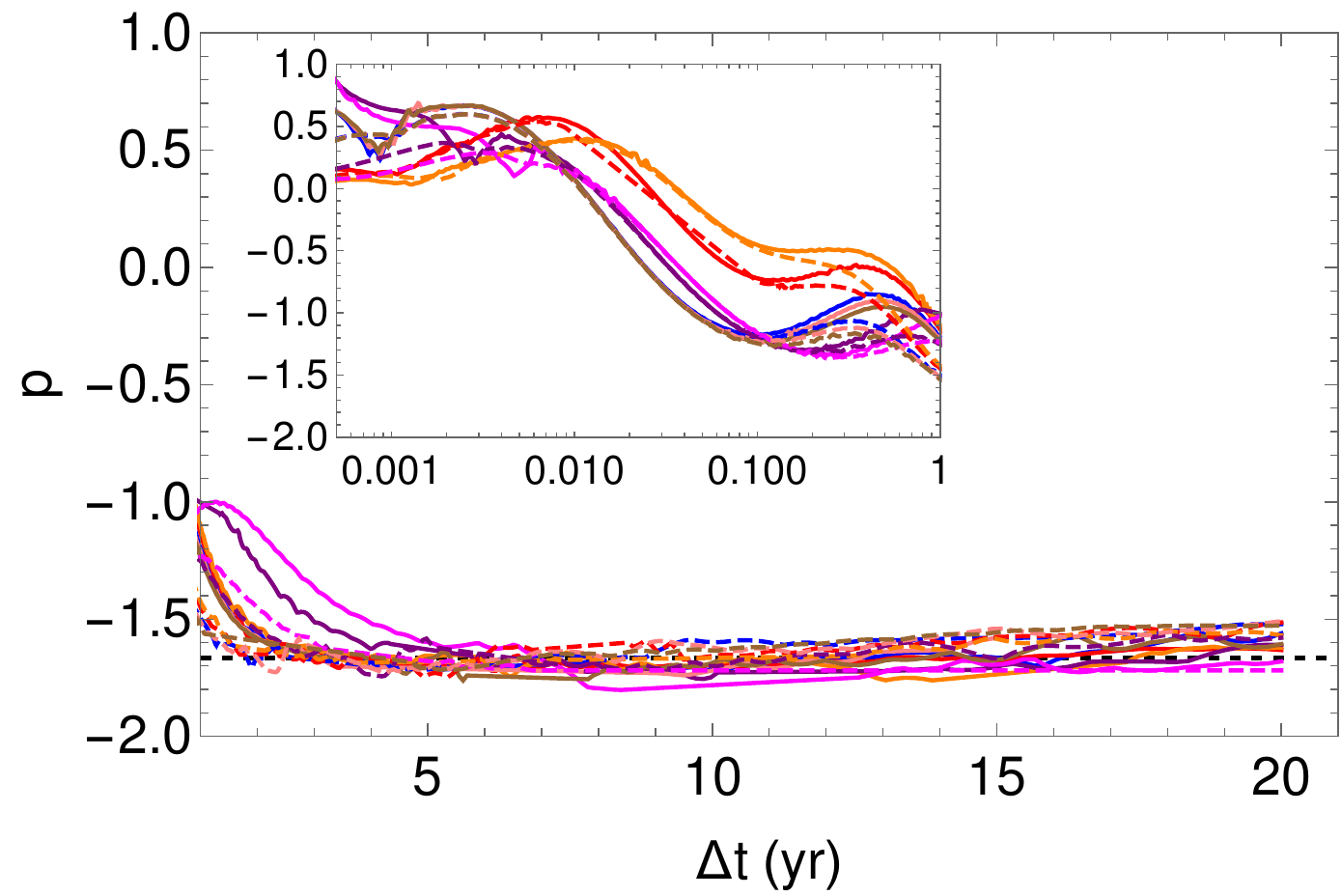}}
\subfigure[]{\includegraphics[scale=0.6]{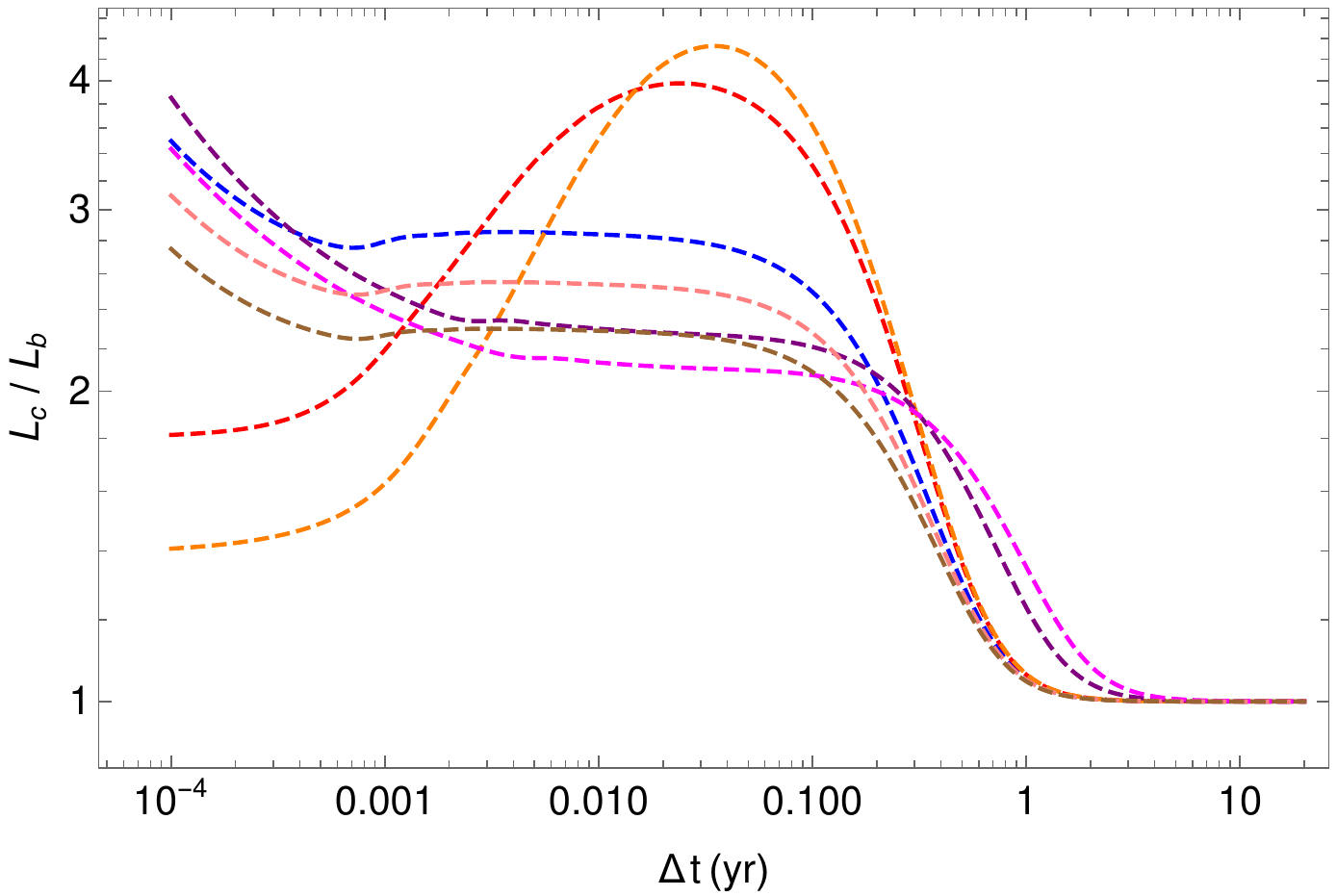}}
\end{center}
\caption{(a) The time evolution of disc bolometric luminosity $L_b$ for no corona [$b_2 =0$, B1, solid lines] and with corona [$b_2 =0.5$, B3, dashed lines]. The $\mu = 1 $ corresponding to case A1. The plot's legend is same as given in Fig. \ref{macct}d. The luminosity follows a power-law decline at late time and we approximate $L_b \propto (t-t_c)^{p}$. See Table \ref{tctm} for $t_c$. (b) The evolution of $p= \diff \log(L_b) / \diff \log(t)$. The black dashed line correspond to $p = -5/3$. The late time evolution of luminosity is close to $t^{-5/3}$ evolution. (c) The total X-ray luminosity from the corona ($L_c$) to the disc bolometric luminosity $L_b$ ratio for $b_2 =0.5$. See section \ref{result} for more discussion.}
\label{lbps}
\end{figure}

The time evolution of disc bolometric luminosity for various values of black hole mass and spin, and stellar mass is shown in Fig. \ref{lbps}. The disc luminosity decreases in presence of the corona due to energy loss to the corona. The bolometric luminosity increases with stellar mass and black hole spin but decreases with black hole mass at the initial time. At a late time, the luminosity increases with an increase in black hole mass. The duration of a disc in the super-Eddington phase increases with stellar mass and spin but reduces with black hole mass. Thus, the higher mass black holes will have a sub-Eddington disc. The luminosity follows a power-law decline at late times and is close to $t^{-5/3}$.

The ratio of total X-ray luminosity from the corona ($L_c$) to the disc bolometric luminosity $L_b$ is shown in Fig. \ref{lbmu}c. The ratio decreases with a decrease in $\mu$ implying that the energy transport to corona decreases with an increase in gas pressure contribution to the viscous stress. For $\mu = 1$, $f(\beta_g) = b_2$ and as the advection weakens with time, the radiative flux is given by $Q_{\rm rad} \approx (1 - f(\beta_g)) Q^{+}$ and thus the luminosity ratio at late time $L_c / L_b \simeq b_2 / (1- b_2)$. However, for $\mu \neq 1$, the $f(\beta_g)$ increases due to an increase in $\beta_g$ with time which results in an increase in $L_c / L_b$ as can be seen from the figure. The luminosity ratio for various stellar mass, black hole mass and spin is shown in Fig. \ref{lbps}c calculated for $b_2 = 0.5$ and $\mu = 1$ resulting in $f(\beta_g) = 0.5$. At the late time, $L_c / L_b \simeq 1$. 

In the previous calculations, we have taken $C_1 = 1$ such that $k_1 = 2/3$. We now consider various values of $C_1 = $ 0.17 \citep{2002ApJ...576..908J}, 0.5, 0.8 and 1 such that $k_1 = $ 0.113, 1/3, 8/15 and 2/3. For the case \Romannum{1} with $\mu = 1$, the beginning time of accretion $t_c ({\rm days}) = $ 3.447, 3.460, 3.468, 3.472 for $b_2 =0$ and 3.539, 3.56, 3.573 and 3.577 for $b_2 = 0.5$ calculated for $C_1 =$ 0.17, 0.5, 0.8 and 1.0 respectively. The onset time of accretion shows a weak increment with $C_1$. The mass accretion rate shows a small increment with $C_1$ at initial times as can be noticed from Fig. \ref{c1v}.  The beginning time $t_c$ for accretion increases with an increase in $C_1$ which suggest that the initial disc mass is higher for higher $C_1$. The disc mass decreases when $\dot{M}_a > \dot{M}_{\rm fb}$ and increases for $\dot{M}_a < \dot{M}_{\rm fb}$. The bolometric luminosity shows a weak decline with $C_1$ at the initial time but at a late time, the luminosity shows a negligible increment with $C_1$. The late time luminosity decline is close to $t^{-5/3}$ evolution. The total luminosity from the corona decreases with an increase in $C_1$ and implies that the energy transport to the corona is higher for lower $C_1$.

\begin{figure}
\begin{center}
\subfigure[]{\includegraphics[scale=0.6]{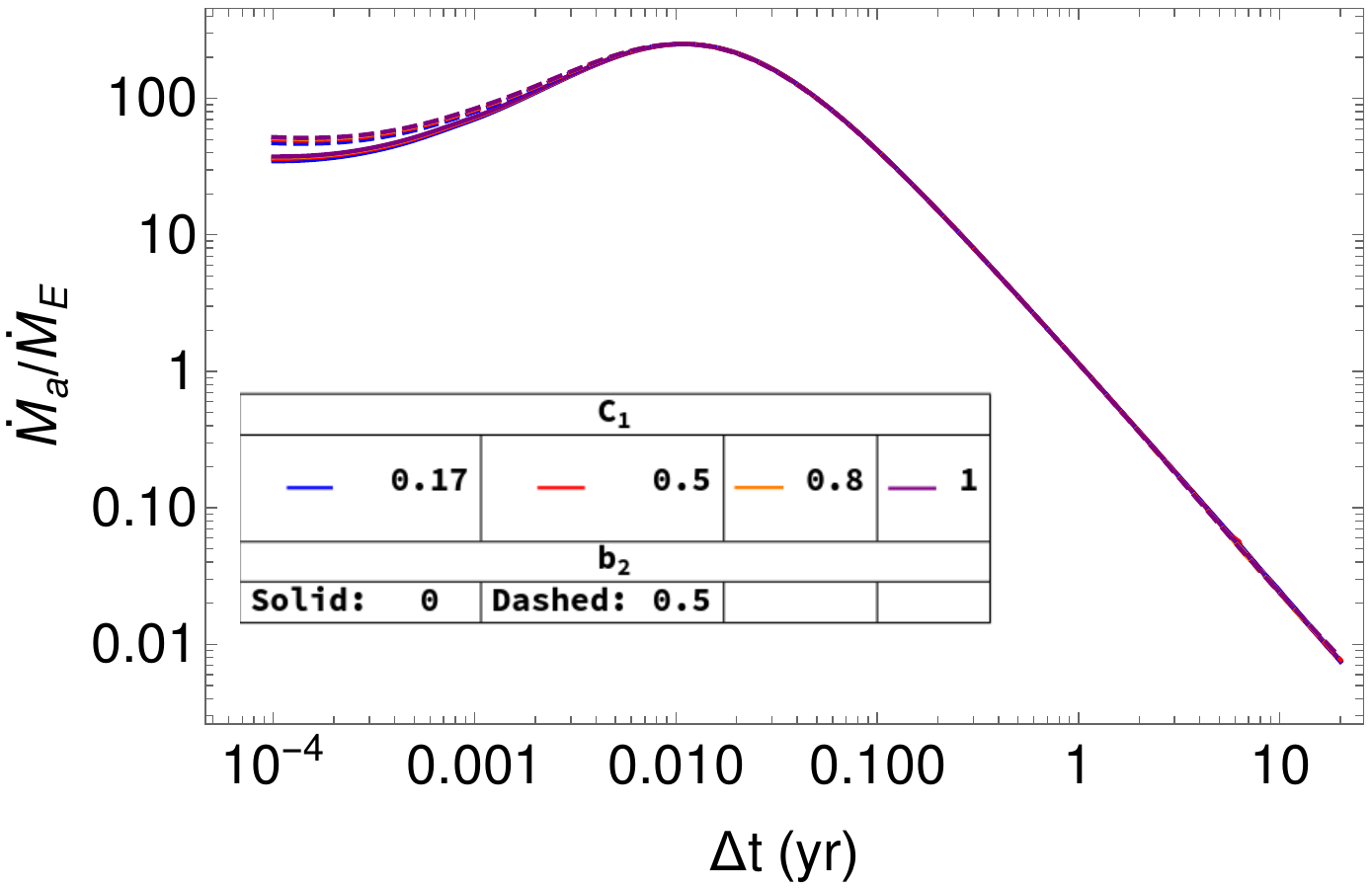}}
\subfigure[]{\includegraphics[scale=0.56]{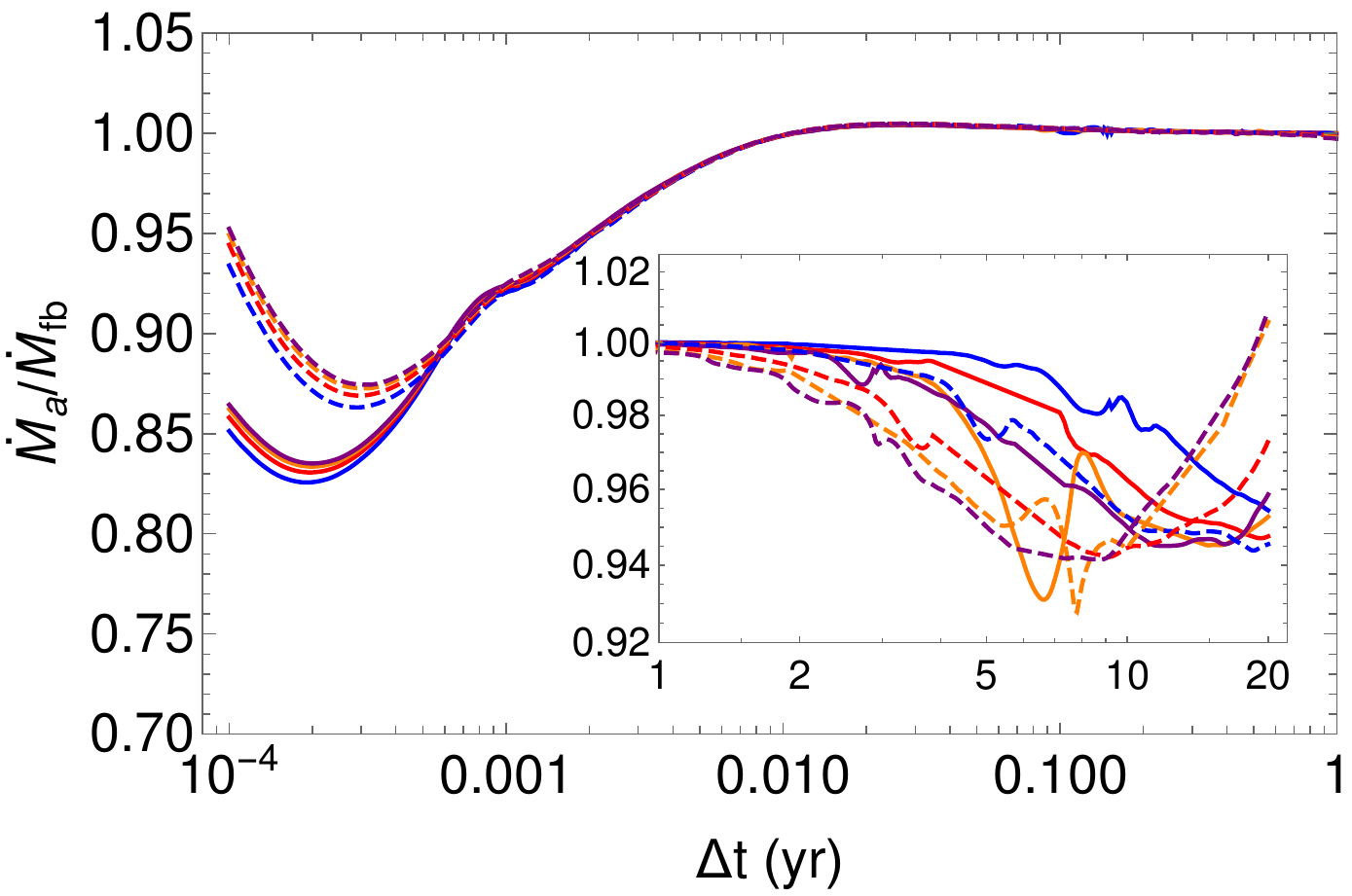}}
\subfigure[]{\includegraphics[scale=0.56]{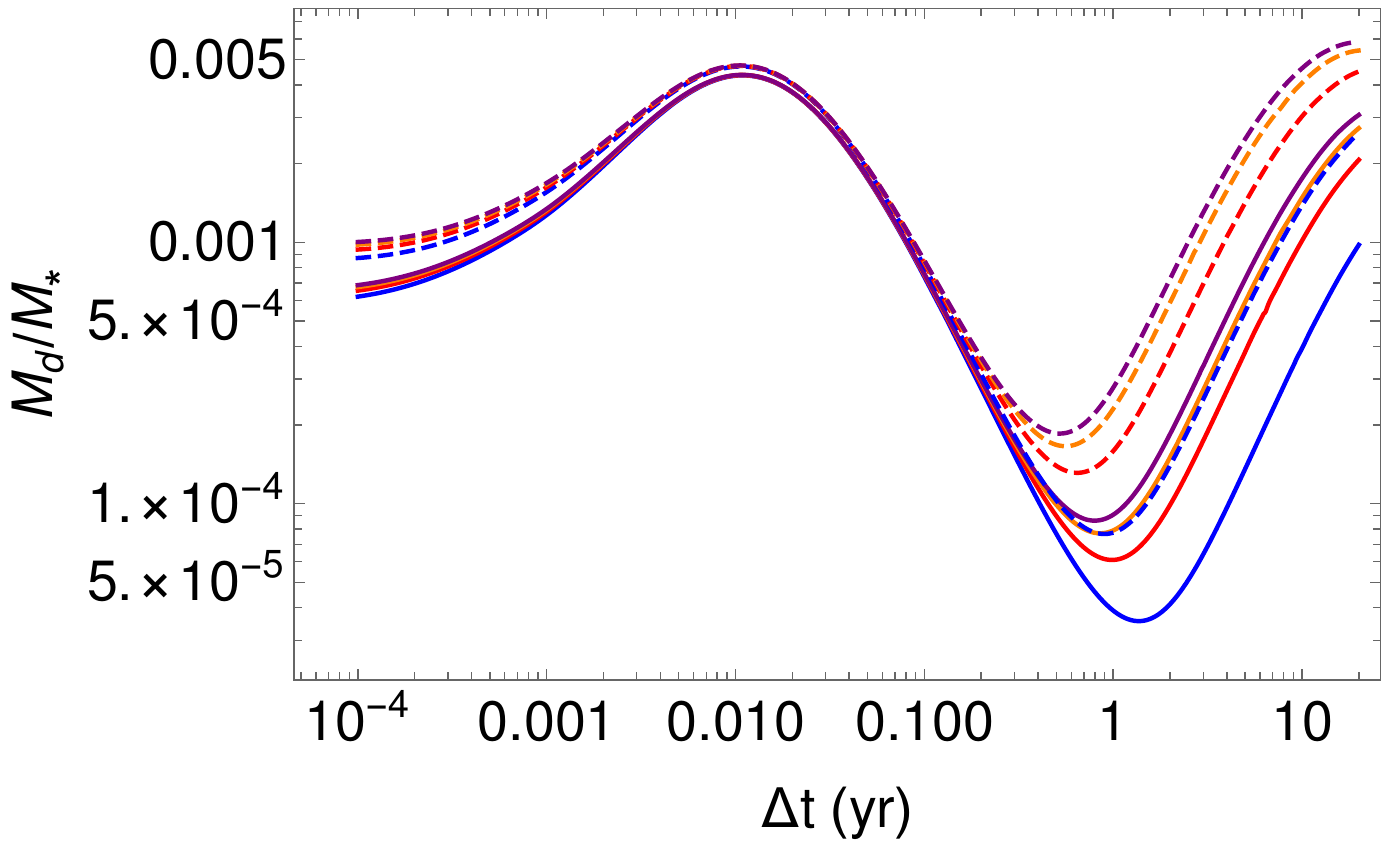}}
\end{center}
\caption{(a) The evolution of mass accretion rate $\dot{M}_a$ for the case \Romannum{1} with $\mu =1$ (A1). The solid lines correspond to no corona [$b_2 =0$] and dashed lines correspond to with corona [$b_2 =0.5$]. (b) The ratio of mass accretion rate to mass fallback rate. (c) The time evolution of disc mass. The disc mass evolves with time that decreases when $\dot{M}_a > \dot{M}_{\rm fb}$ and increases for $\dot{M}_a < \dot{M}_{\rm fb}$. The beginning time of accretion $t_c ({\rm days}) = $ 3.447, 3.460, 3.468, 3.472 for $b_2 =0$ and 3.539, 3.56, 3.573 and 3.577 for $b_2 = 0.5$.  See section \ref{result} for more discussion.}
\label{c1v}
\end{figure}

\begin{figure}
\begin{center}
\subfigure[]{\includegraphics[scale=0.58]{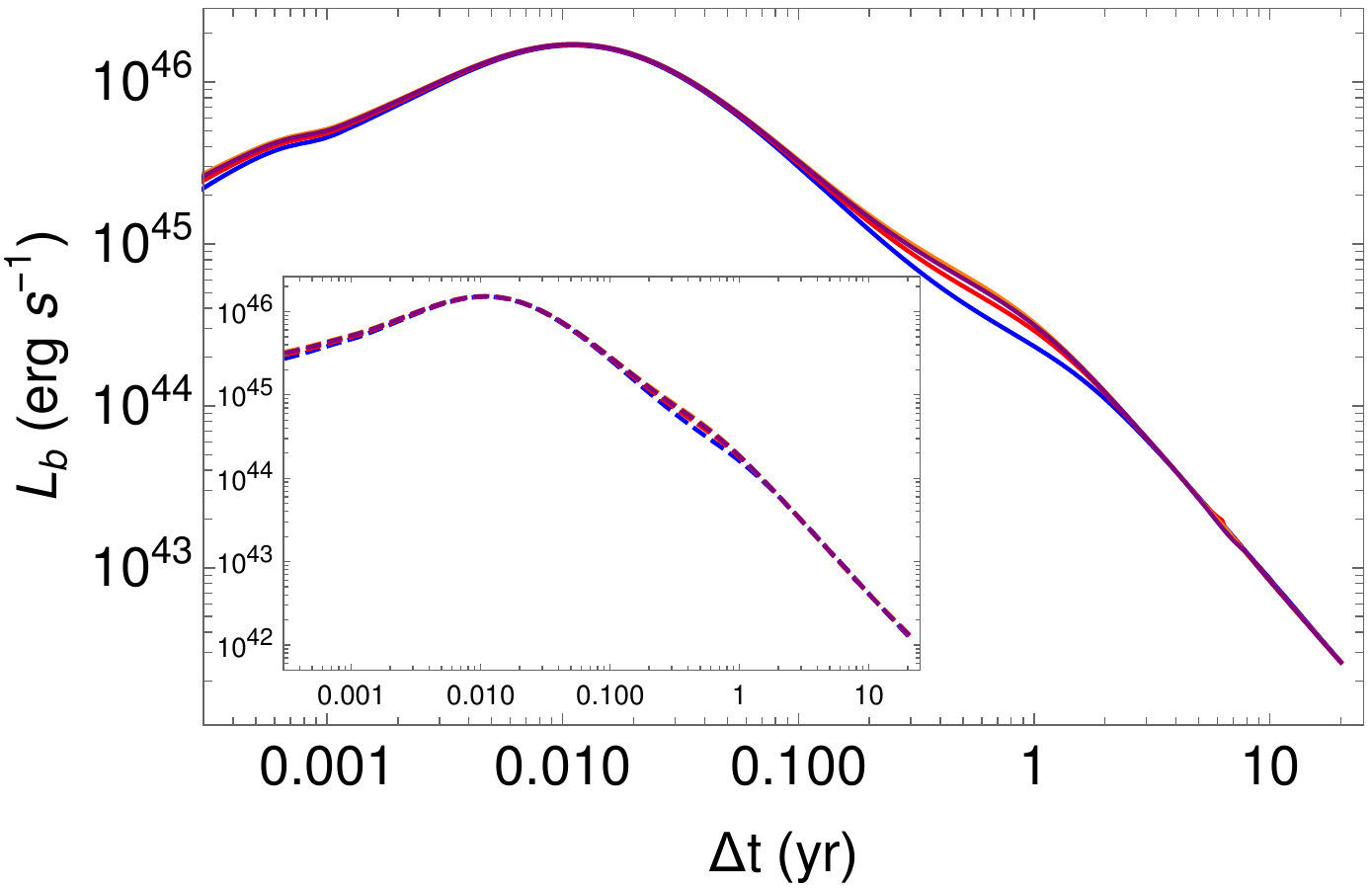}}
\subfigure[]{\includegraphics[scale=0.57]{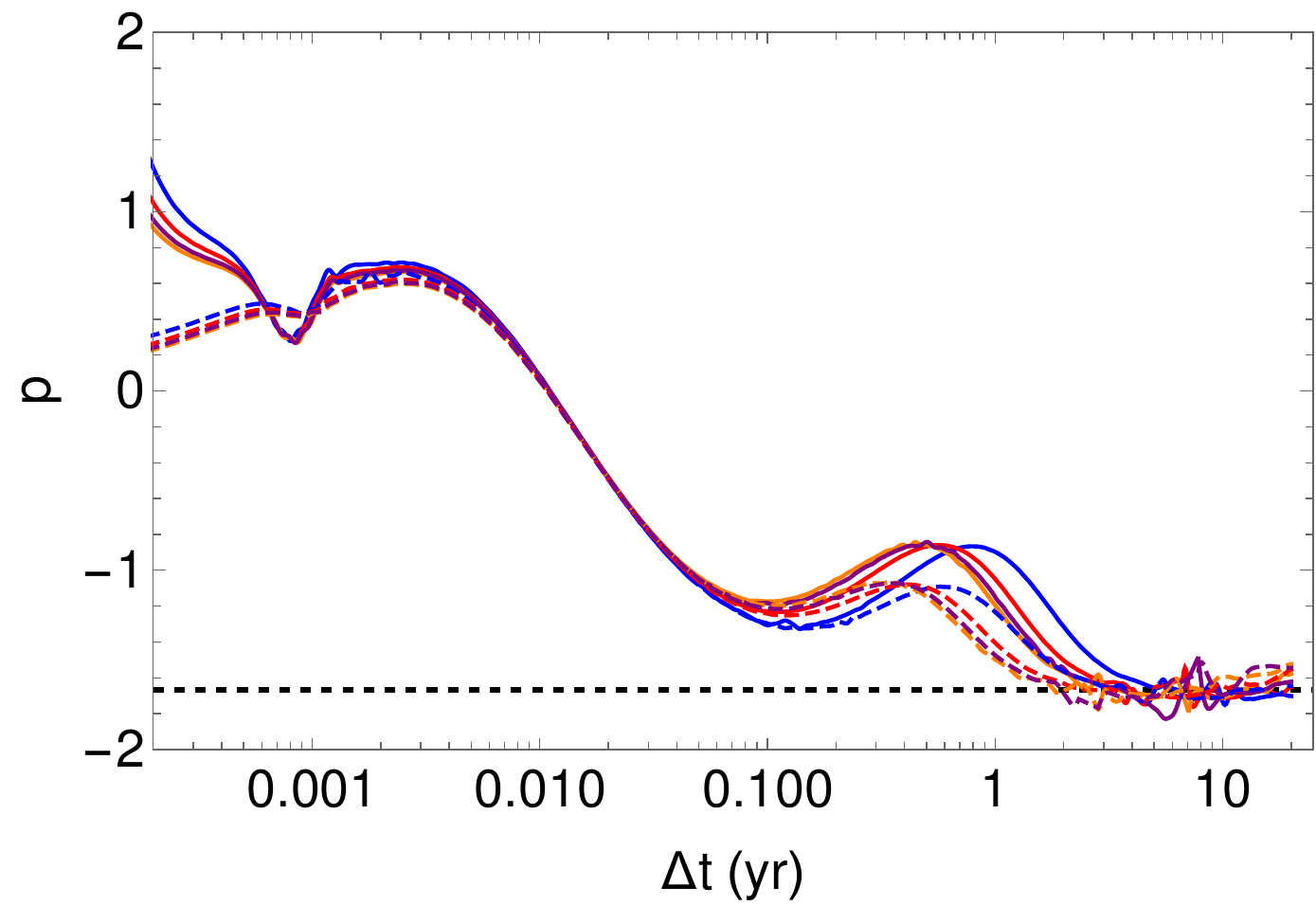}}
\subfigure[]{\includegraphics[scale=0.55]{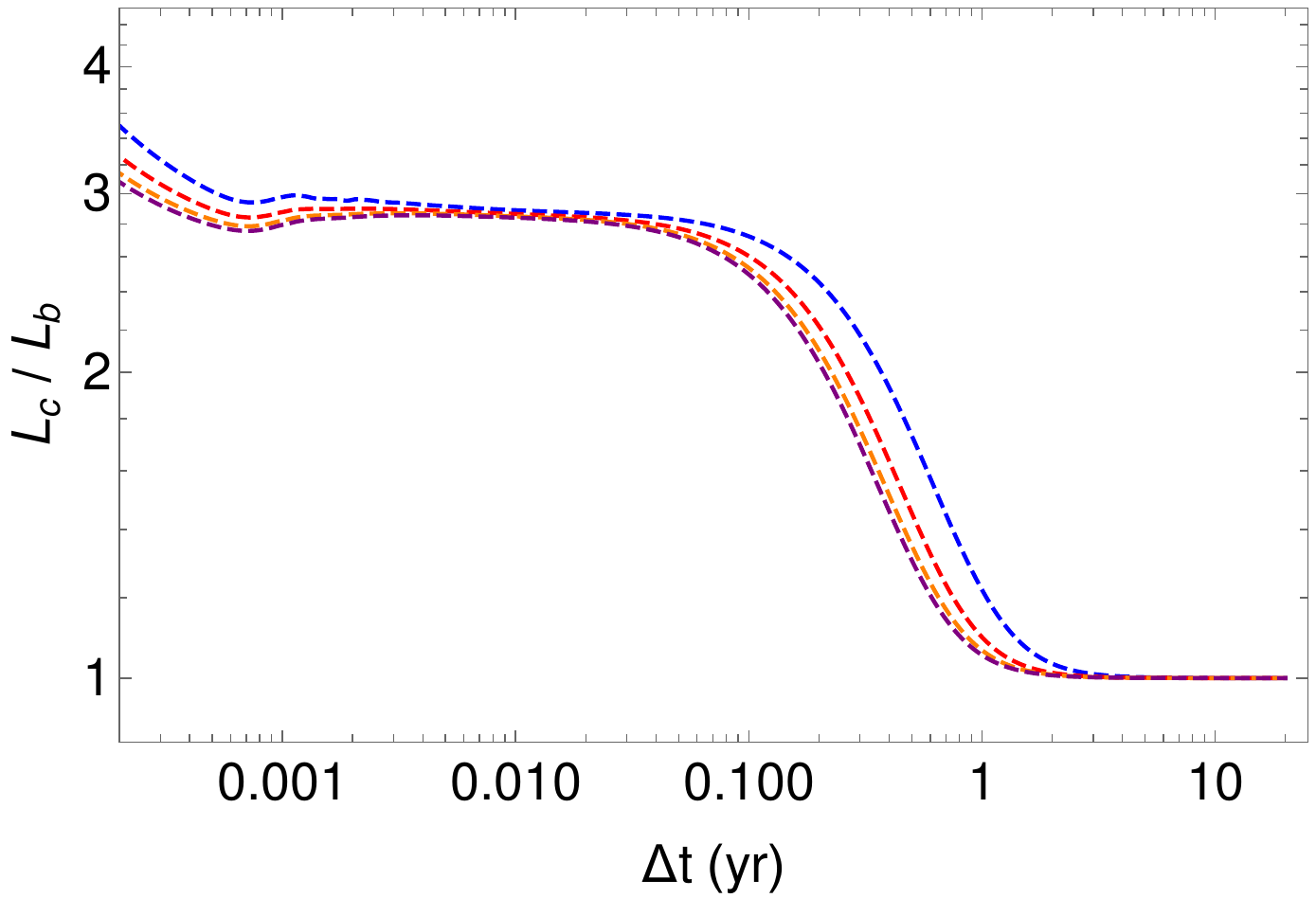}}
\end{center}
\caption{(a) The time evolution of disc bolometric luminosity for the case \Romannum{1} with $\mu =1$ (A1). The plot's legend is same as given in Fig. \ref{c1v}a. The luminosity follows a power-law decline at late time and we approximate $L_b \propto (t-t_c)^{p}$. See caption \ref{c1v} for $t_c$. (b) The evolution of $p= \diff \log(L_b) / \diff \log(t)$. The black dashed line correspond to $p = -5/3$. The late time evolution of luminosity is close to $t^{-5/3}$ evolution. (c) The total X-ray luminosity from the corona ($L_c$) to the disc bolometric luminosity $L_b$ ratio for $b_2 = 0.5$. The ratio decreases with $C_1$ which implies that the energy transport to corona is higher for lower $C_1$. See section \ref{result} for more discussion. }
\label{c1vl}
\end{figure}

\begin{figure*}
\begin{center}
\includegraphics[scale=0.7]{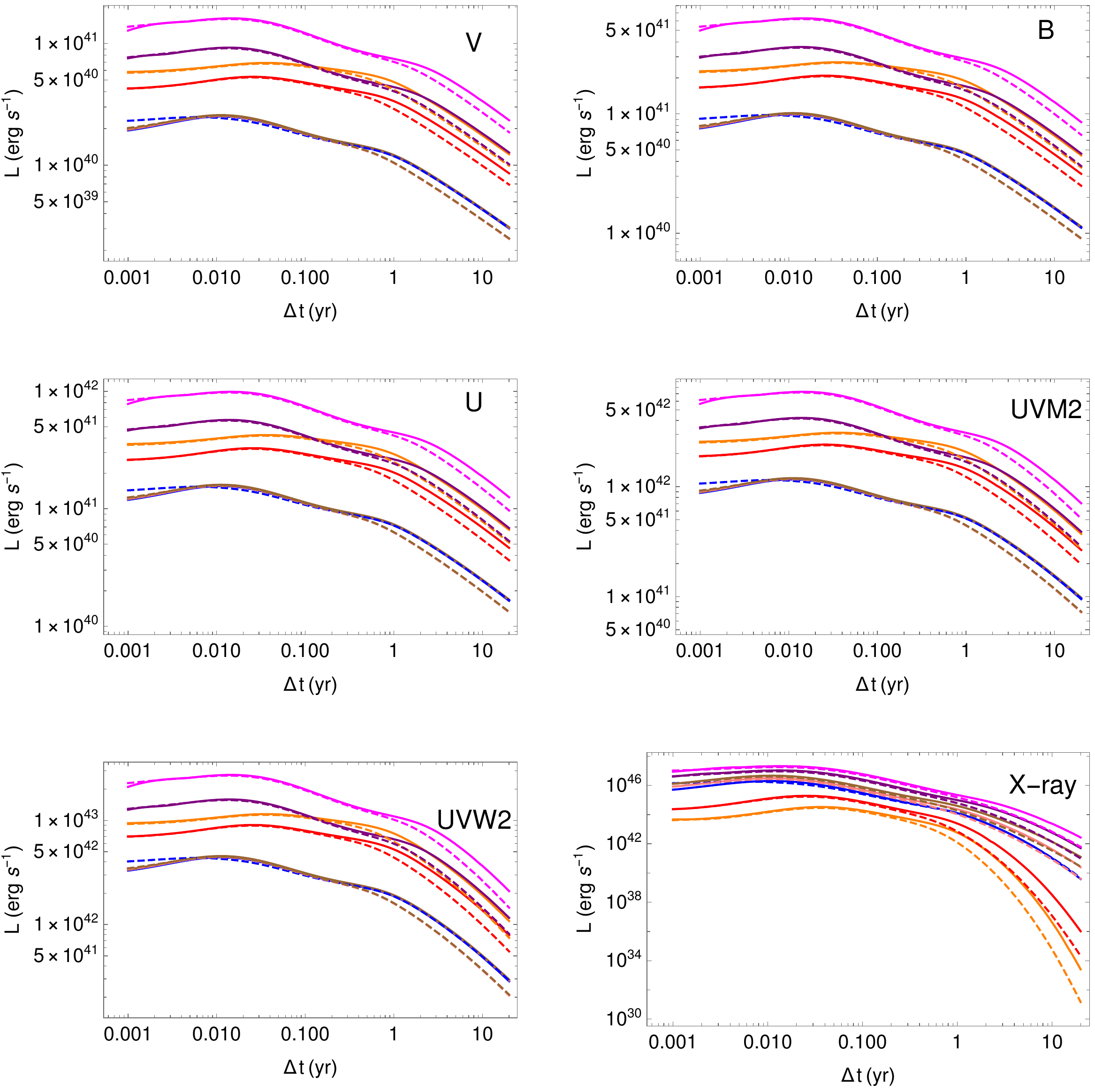}
\end{center}
\caption{The time evolution of disc luminosity in various spectral bands calculated using equation (\ref{fnuobs}) for $\mu =1$, $C_1 =1$ and $b_2 =$ 0 (solid) and 0.5 (dashed). The plot's legend is same as given in Fig. \ref{macct}d. The spectral bands are V (5000-6000 \AA), B (3600 - 5000 \AA), U (3000-4000 \AA), UVM2 (1660-2680 \AA), UVW2 (1120 - 2640 \AA) and X-ray (0.2 - 2 keV). The luminosity increases with black hole mass for optical/UV wavelengths but decreases in X-ray band. The presence of corona shows a considerable difference in spectral luminosity at late times. See section \ref{result} for more discussion.}
\label{spec}
\end{figure*}

The luminosity in various spectral bands calculated using equation (\ref{fnuobs}) is shown in Fig. \ref{spec}. The presence of corona has a severe effect on the spectral luminosity at late times by decreasing the spectral luminosity. The increase in stellar mass increases the mass accretion rate increasing the viscous stress and thus the luminosity. The luminosity increases with black hole mass for optical/UV wavelengths but decreases in the X-ray band. 

We now calculate the time-evolving properties of the corona using a two-temperature model from ions and electrons. The energy transport to the corona $Q_{\rm cor}$ heats the corona and the energy is transferred from ions to electrons through Coulomb coupling and the plasma cools via electron through various mechanisms such as bremsstrahlung, synchrotron, and Compton cooling. The bremsstrahlung and synchrotron cooling rates are obtained from \citet{1995ApJ...452..710N} and the Compton cooling is approximated via an amplification factor $\eta$ \citep{1991ApJ...369..410D}. Assuming equipartition between magnetic pressure and gas pressure in the corona and ion temperature equal to the 90\% of the virial temperature \citep{2009MNRAS.394..207C}, we calculate the electron temperature, the optical depth of electron scattering $\tau_{es}$ and Compton $y$ parameter in the corona. The details of the model are given in appendix \ref{corhc}.

\begin{figure}
\begin{center}
\subfigure[]{\includegraphics[scale=0.45]{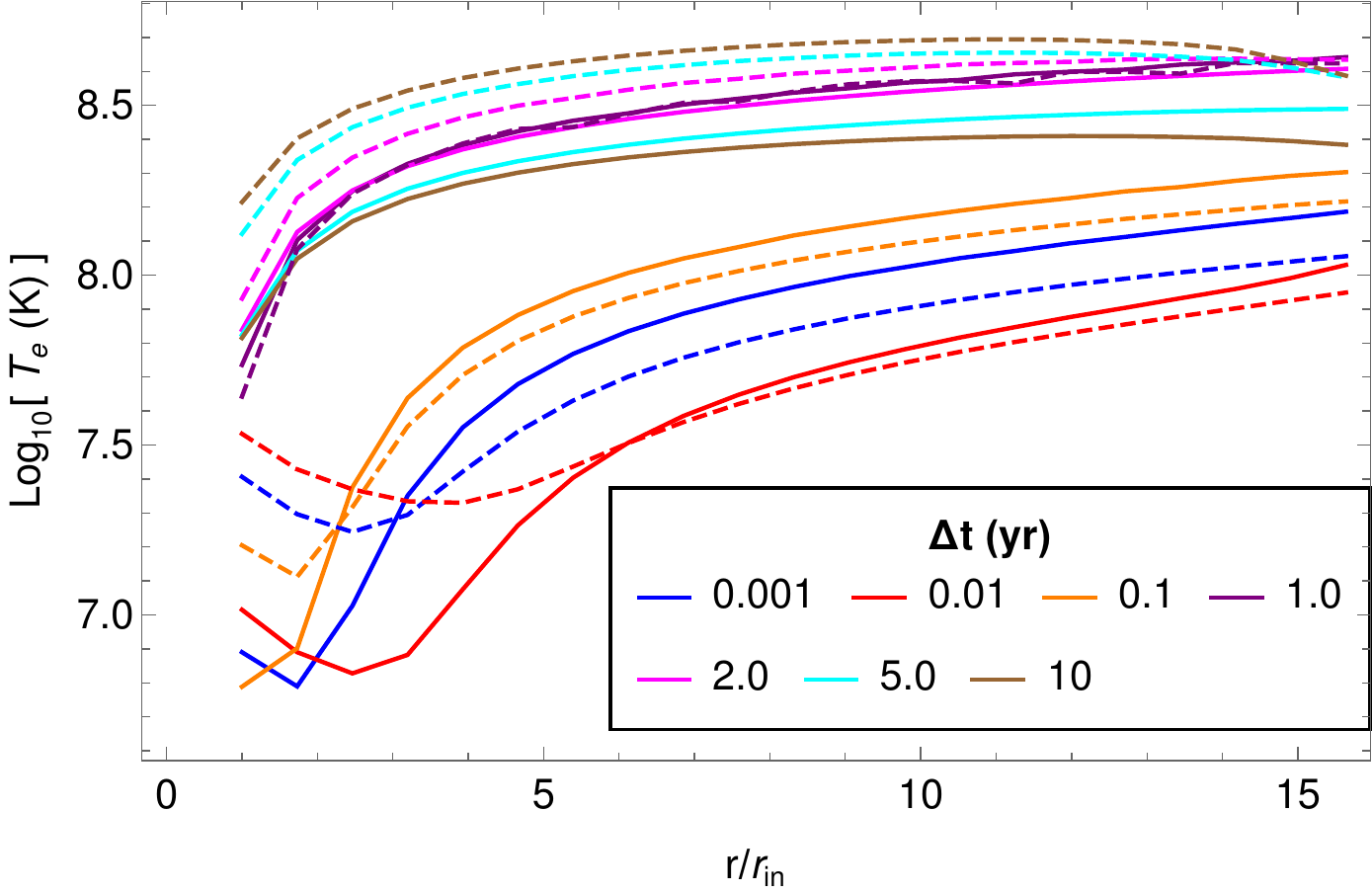}}
\subfigure[]{\includegraphics[scale=0.45]{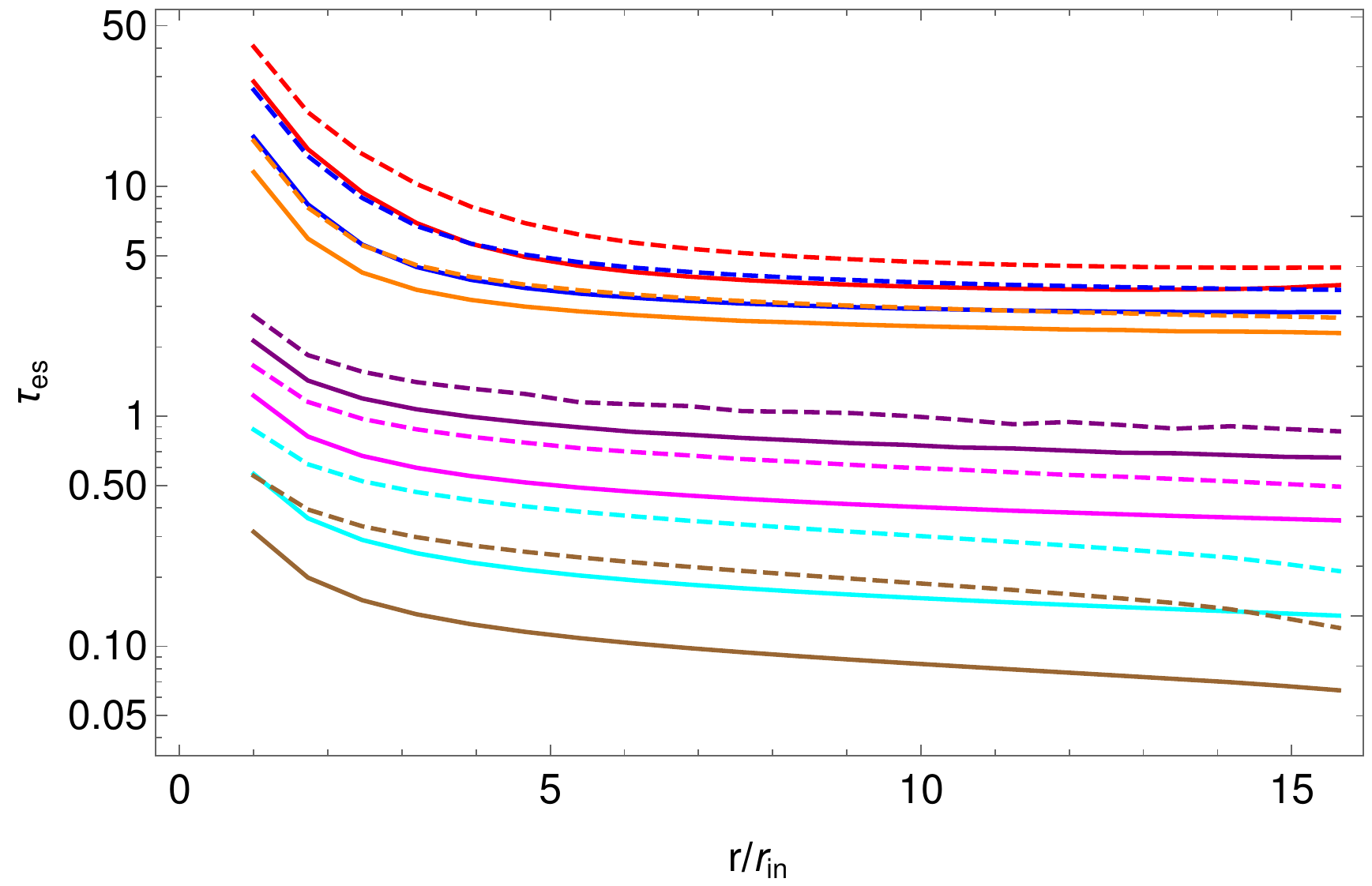}}
\subfigure[]{\includegraphics[scale=0.45]{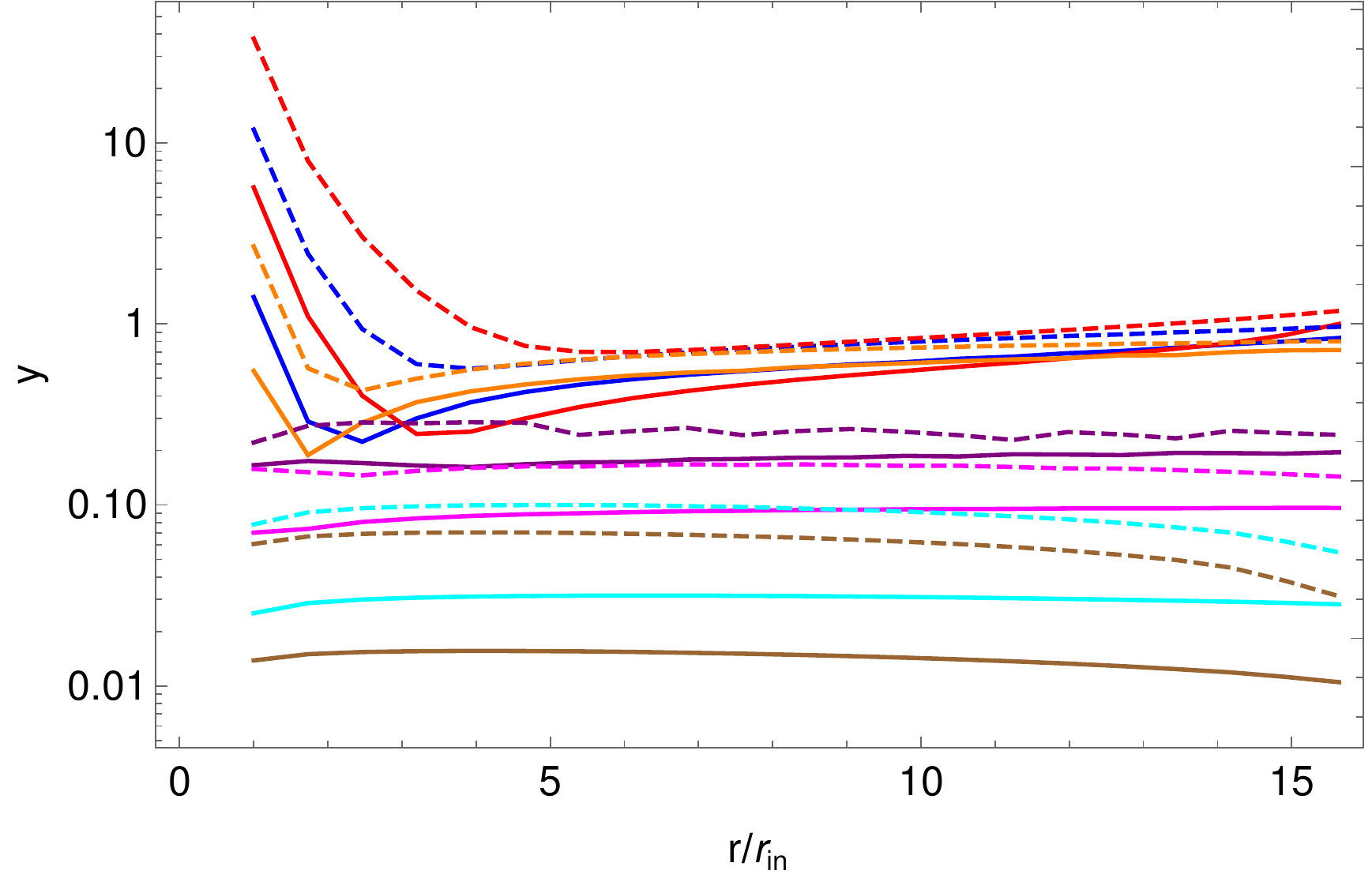}}
\end{center}
\caption{The electron temperature $T_e$, optical depth of electron scattering $\tau_{es}$ and Compton parameter $y$ are shown in (a), (b) and (c) respectively for the case \Romannum{1} with $\mu =1$ (A1) and $b_2 = $ 0.5 (B3: solid) and 0.9 (B4: dashed). The increase in $b_2$ increases the energy flux transport to the corona (see equation \ref{fcor}). The model is given in appendix \ref{corhc}. The corresponding bolometric disc luminosity is shown as blue dotted and dot-dashed lines in Fig. \ref{lbmu}. The decrease in optical depth with time indicates that the corona is transparent to the soft photons from the disc and the $y \ll 1$ suggests that there is no significant change in the photon energy and the comptonization will be negligible.  See section \ref{result} for more discussion.}
\label{cb2}
\end{figure}

Fig. \ref{cb2} shows the electron temperature, optical depth and Compton $y$ parameter for the case \Romannum{1} ($M_{\bullet,6} = 1$) with $\mu =1$ (A1) and with $b_2 = $ 0.5 (B3: solid) and 0.9 (B4: dashed). The electron temperature is comparable with that obtained by \citet{2009MNRAS.394..207C} for a steady accretion model without a fallback. The electron temperature increases whereas optical depth decrease as disc bolometric luminosity decreases. The decrease in optical depth with time implies that at late times, the corona is transparent to the soft photons from the disc. The Compton $y$ parameter that determines whether a photon will significantly change its energy in traversing the medium, decreases with time. For $y \ll 1$, we can conclude that there is no significant change in the photon energy. A similar result for case \Romannum{3} ($M_{\bullet,6} = 10$) with $\mu =1$ (A1) and $b_2 = 0.5$ (B3) is shown in Fig. \ref{clc}. 

\begin{figure}
\begin{center}
\subfigure[]{\includegraphics[scale=0.45]{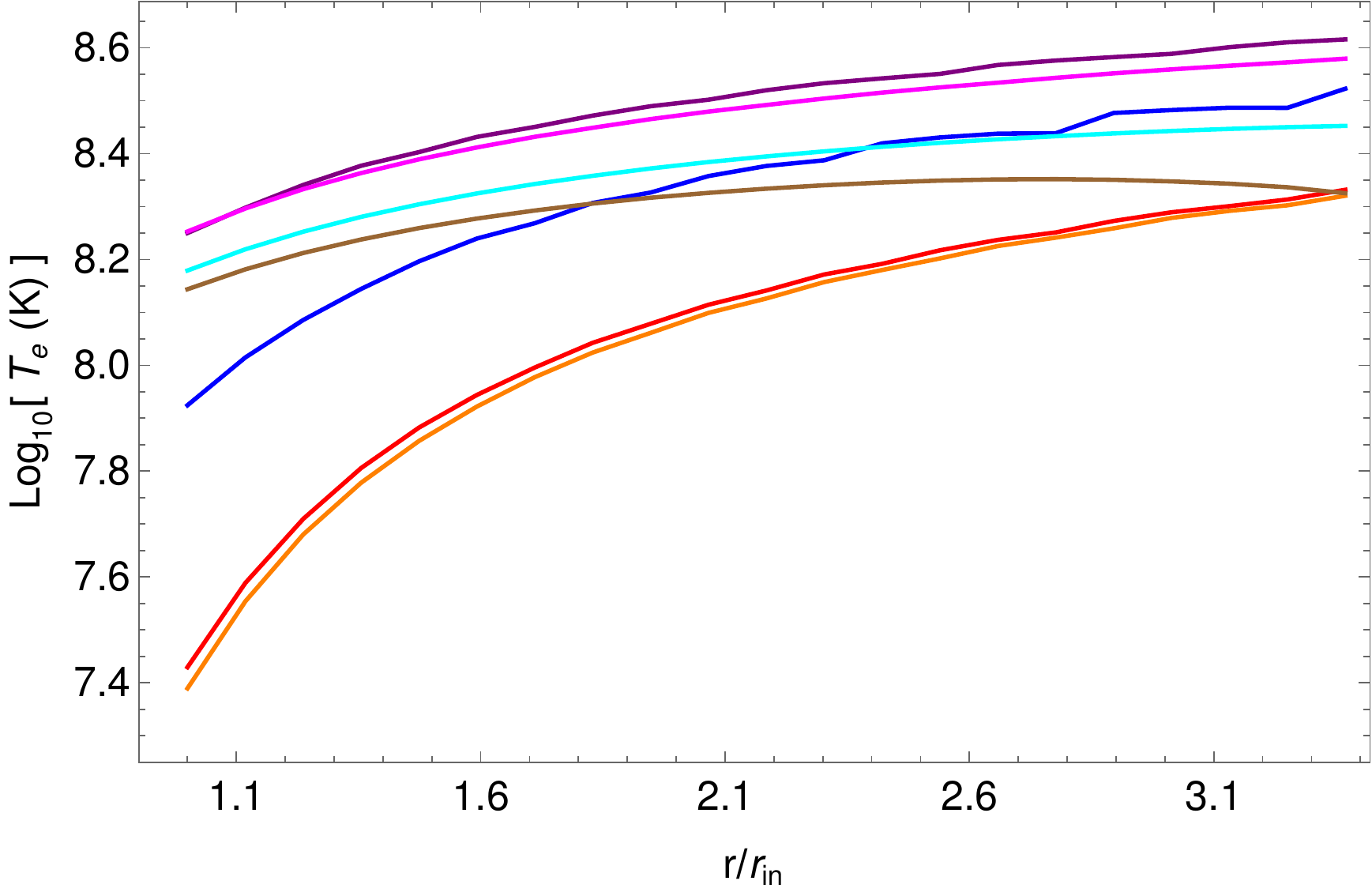}}
\subfigure[]{\includegraphics[scale=0.45]{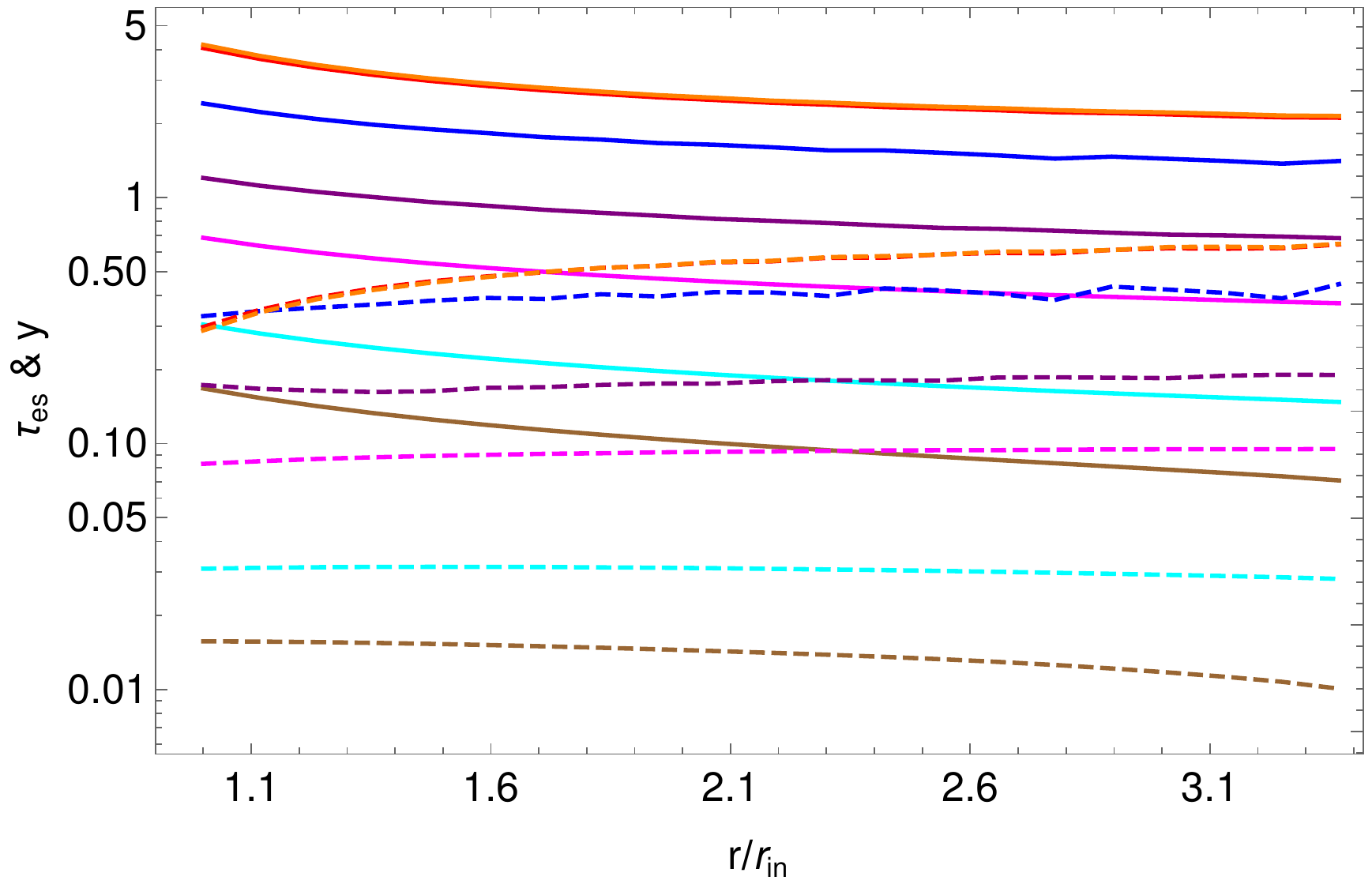}}
\end{center}
\caption{The electron temperature $T_e$ in (a), and optical depth of electron scattering $\tau_{es}$ (solid) and Compton parameter $y$ (dashed) in (b) for the case \Romannum{3} with $\mu =1$ (A1) and $b_2 = $ 0.5 (B3). The colour plots legend is the same as given in Fig. \ref{cb2}a. The model is given in appendix \ref{corhc}. The corresponding bolometric disc luminosity is shown as blue and red dashed lines in Fig. \ref{lbps}. See section \ref{result} for more discussion.}
\label{clc}
\end{figure}

\section{Discussion and conclusions}
\label{discuss}

TDEs are an essential phenomenon to perceive the accretion dynamics around supermassive black holes over a period of a few years. The numerical simulations on the circularization of the disrupted debris and the formation of an accretion disc, have been limited to a few parameter ranges of stellar mass and black hole mass and spin. The simulations have shown that the formed accretion disc can be a circular or an elliptical disc with or without still infalling debris \citep{2020A&A...642A.111C}. When the tidal radius lies above the ISCO, the stream interactions result in an inflow of matter to form a circular disc with an inner radius at ISCO. When the tidal radius lies below ISCO, some fraction of debris will plunge to the black hole during circularization and the amount of debris mass plunged to the black hole and the dynamics of circularization require detailed relativistic stream modelling. Due to uncertainty in the disc formation, here, we assume those cases where the tidal radius lies above ISCO. Thus, $r_t \geq r_g Z(j)$ which results in a condition given by $Z(j) \leq r_t/r_g$ and thus a minimum spin value given by $j_{m} = j_m(M_{\bullet,6},~m)$. We consider black holes with a prograde spin only ($j \geq 0$).

The infalling debris forms a seed accretion disc that evolves in the effect of viscous accretion and the mass fallback at the outer radius. We assume that the formed initial disc in time $t_c$ has the disc mass equal to the total debris mass that has infalled by the time $t_c$. There is no mass loss to black hole for $t < t_c$. Thus, $t_c$ is the time at which matter crosses the ISCO and the accretion to the black hole begins. During the formation of an accretion disc ($t < t_c$), the debris stream self-interaction will result in emissions, but we have not considered the dynamics of disc formation in this paper. The disc dynamics presented is after the matter crosses the disc inner radius. The obtained $t_c$ shows a weak increment with the presence of corona but increases with a decrease in $\mu$ which implies that an increase in gas pressure contribution to the disc delays the onset of accretion. An important aspect here is that we have used a semi-analytic formulation of the mass fallback rate calculated for a polytrope star using the impulse approximation. However, a star is tidally deformed before reaching the pericenter which influences the stellar density structure and the stellar density at the moment of disruption can be different from the original polytrope. The mass fallback rate obtained through simulation differs from the impulse approximation at initial times but the late time evolution is nearly the same \citep{2009MNRAS.392..332L,2019ApJ...872..163G}. This difference can affect the initial time $t_c$ and initial disc luminosity but the late time luminosity evolution will be similar to the luminosity evolution we have obtained using the semi-analytic model of the mass fallback rate. 

We include an energy loss to the corona in our model and the fraction $f$ of viscous heat energy transported to the corona is a function of $\beta_g$ and $\mu$. In the case of $\mu = 1$, the magnetic pressure and the viscous stress are a function of total pressure only and we found that the fraction is a constant, but with $\mu \neq 1$, the fraction $f \propto \beta_g^{(1-\mu)/2}$ and is small when the disc is dominated by radiation pressure ($\beta_g \ll 1$). The energy loss to the corona increases the accretion rate at the initial time but has an insignificant effect at the late times (see Fig. \ref{macc}), however, the bolometric and spectral luminosity declines in presence of the corona at late times. For $\mu \neq 1$, the increase in energy transport has a weak effect on disc luminosity at early times but shows a significant decline in late time luminosity. With an increase in gas pressure contribution to the magnetic pressure and viscous stress ($\mu$ decreases), $L_c / L_b$ decreases as can be seen from Fig. \ref{lbmu}c and thus implies a reduction in energy transport to the corona. The MRI growth rate depends on the ratio of radiation to gas pressure and with a decrease in $\mu$, the contribution of gas pressure increases compared to the radiation pressure which reduces the MRI and thus the magnetic stress \citep{2002ApJ...566..148T}. This indicates that the gas pressure acts as a support to the disc stability. 

\begin{figure}
\begin{center}
\includegraphics[scale=0.58]{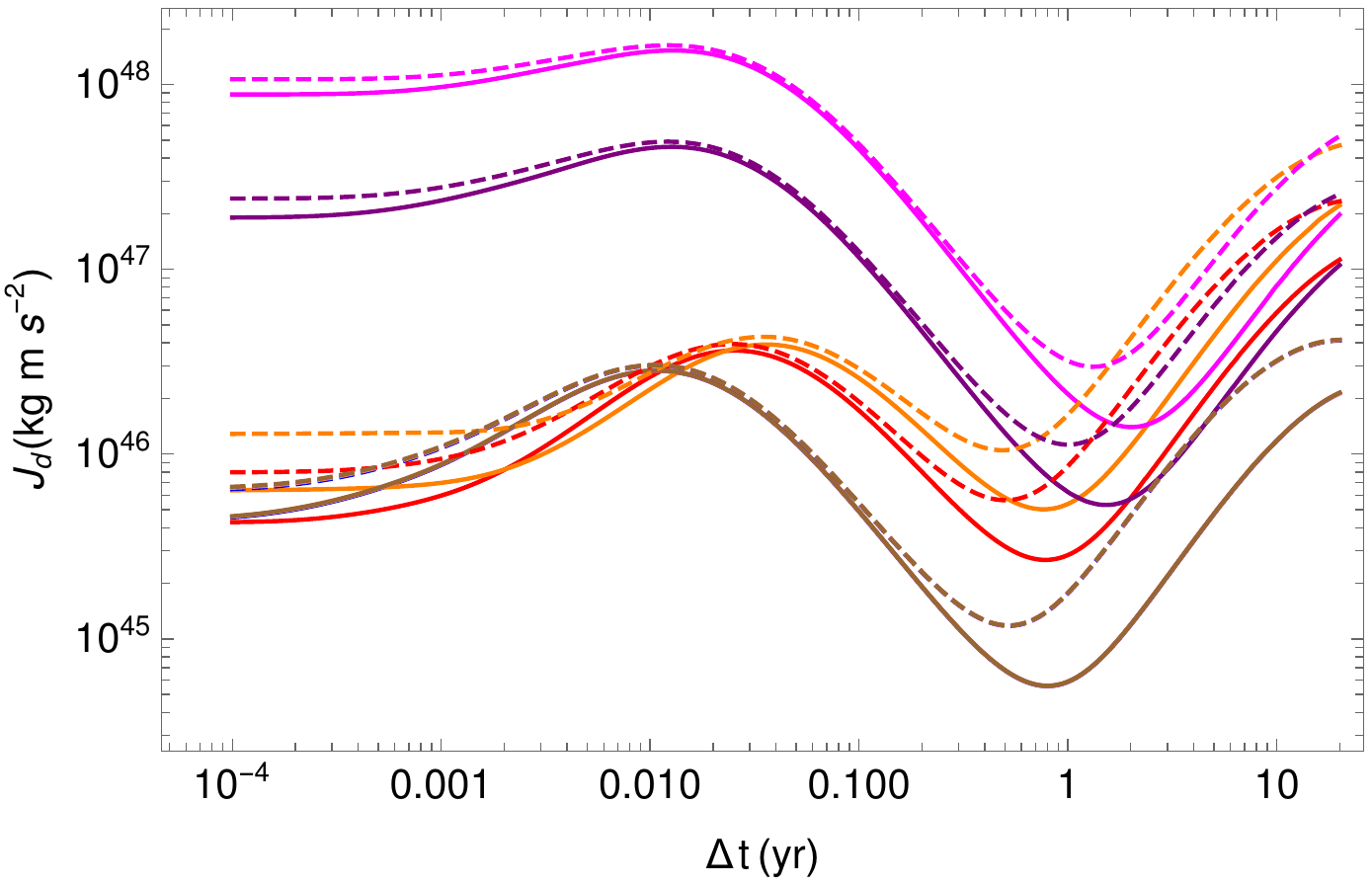}
\end{center}
\caption{The time evolution of total angular momentum of the disc for no corona [$b_2 =0$, B1, solid lines] and with corona [$b_2 =0.5$, B3, dashed lines]. The $\mu = 1 $ correspond to case A1. The color plots legend is the same as given in Fig. \ref{macct}d. The total angular momentum shows insignificant variation with black hole spin and thus pink and brown lines coincide with the blue line. See section \ref{discuss} for more discussion.}
\label{jdisk}
\end{figure}

We have taken the outer radius of the disc to be constant and the mass added at the outer radius is accreted inwardly through accretion. The total angular momentum ($J_d$) of the disc is not constant and evolves with time as shown in Fig. \ref{jdisk}. The total angular momentum shows insignificant variation with black hole spin. The $J_d$ is higher when there is an energy loss to corona ($b_2 \neq 0$).  The non-constant disc total angular momentum results in a late time luminosity evolution ($ L_b \propto t^{-5/3}$) that deviates from the self-similar solution of a sub-Eddington disc by \citet{1990ApJ...351...38C}, where the outer radius evolves with time and the luminosity $L_b \propto t^{-1.2}$.  

In the appendix \ref{sssol}, we present the self-similar solution of a non-advective accretion disc \citep{1990ApJ...351...38C} with energy loss to the corona included in the energy conservation equation. We compare the mass accretion rate and disc bolometric luminosity of the self-similar model with the advective disc-corona model in Figs. \ref{can} and \ref{can1}, and they show that the bolometric disc luminosity evolves linearly with the mass accretion rate for both advective disc and self-similar models in the sub-Eddington phase. The self-similar model is consistent when the evaluation is in the sub-Eddington phase, $\dot{M}/\dot{M}_E \ll 1$, where the pressure is dominated by gas pressure. Near the Eddington accretion, the pressure is dominated by radiation pressure and the advection is crucial to avoid the Lightman-Eardely instability. In the self-similar formulation, the outer radius evolves with time whereas we have taken it to be a constant. The advective model includes a fallback at the outer radius and the mass accretion rate follows the mass fallback rate whereas the fallback is not included in the self-similar model. 

The time-dependent advective accretion disc model with $\alpha-$viscosity developed by \citet{2002ApJ...576..908J} to study the radiation pressure instability driven variability assumed a steady solution at the initial time and truncated their model at a constant outer radius where the boundary condition is the stable steady solution with a constant mass accretion rate. This assumption implies
that some amount of the disc angular momentum is taken away at the outer radius of the computed disc region. \citet{2011ApJ...736..126M} developed a TDE advective model without corona and for $\beta-$viscosity with a mass fallback at the constant outer radius. In their numerical calculation, the mass accretion rate at the outer radius is equal to the mass fallback rate and they showed that the mass accretion rate and the disc bolometric luminosity at the late time evolves as $t^{-5/3}$. We have also considered a similar assumption at the outer radius. 

In our advective disc model, we consider a seed disc that is formed at an early time and we expect the formed disc to be in a non-steady state where the mass accretion rate varies with radius. The mass supply rate at the outer radius changes with time and the stable steady state solution may be inconsistent. We obtain a non-steady initial and boundary solution for the disc. We assume that the mass is added uniformly along the azimuthal direction and we truncate our advective disc model computation region to the circularization radius and the infalling matters at the outer radius lose their angular momentum following equation (\ref{mdot}) to the external infalling debris. This will induce an evaluation of the infalling debris but we do not consider its dynamics and assume that they maintain the mass fallback rate given by equation (\ref{mfbn}) at the outer radius. This assumption helps in maintaining both the angular momentum conservation equation and the mass accretion equal to the mass fallback at the outer radius. We understand that the outer radius can evolve with time depending on the mass accretion and mass fallback rates, but the considered assumptions for a simple disc-corona model with fallback provide a reasonable solution.  

Here, we calculate the total X-ray luminosity from the corona by integrating the energy transported to the corona from the disc. The photons emitted from the disc are scattered by the hot electrons in the corona and some fraction of the scattered photons travel downward to the disc known as the downward component. Some fraction of these downward-moving photons are reflected from the disc surface as disc albedo. The downward component and disc albedo evolve with time and can be a function of the hard component X-ray photon index. This analysis will require a detailed radiative transfer model of scattering in the corona which includes the synchrotron and bremsstrahlung emissivities that depend on the electron distribution in the corona. We have shown the effect of energy loss to corona on the disc evolution and this energy transported to the corona can be used to study the scattering dynamics in the corona and the evolution of X-ray hard components with time.    

In the appendix \ref{corhc}, we present an efficient yet approximate cooling mechanism of hot electrons to estimate the time-evolution of coronal properties and the Compton $y$ parameter. For the case of unsaturated repeated scattering by non-relativistic thermal electrons, up-scattered photons have a power-law energy distribution with photon index $\Gamma = -1/2 + \sqrt{9/4 + 4/y}$ \citep{2009PASP..121.1279S}. For the Compton $y$ parameter given in Fig. \ref{cb2} for $b_2 = 0.5$, the mean photon index at times $\Delta t{\rm (yr)} = $ 0.001, 0.01, 0.1, 1, 2, 5 and 10 are $\Gamma =$ 2.63, 2.64, 2.68, 4.448, 6.33, 11.12 and 16.50 respectively. The spectral index at initial times agrees with the spectral index from the observations \citep{2020MNRAS.497L...1W}. At late times, the spectral index is higher than the observed values. This is because of the low $y$ implying negligible Comptonization. The Comptonization approximation using an amplification factor given by equation (\ref{etacomp}) is derived for an optically thin medium with $\tau_{es} \ll 1$ \citep{1991ApJ...369..410D}. We can see from Figs. \ref{cb2} and \ref{clc}, that the optical depth shows significant evolution with time and thus we need to develop a more detailed Comptonization scattering model for the Compton cooling from an optically thick medium. We will construct such a detailed Comptonization process and the Comptonized spectra of the disc-corona model in the future.   

\begin{figure*}
\begin{center}
\includegraphics[scale=0.6]{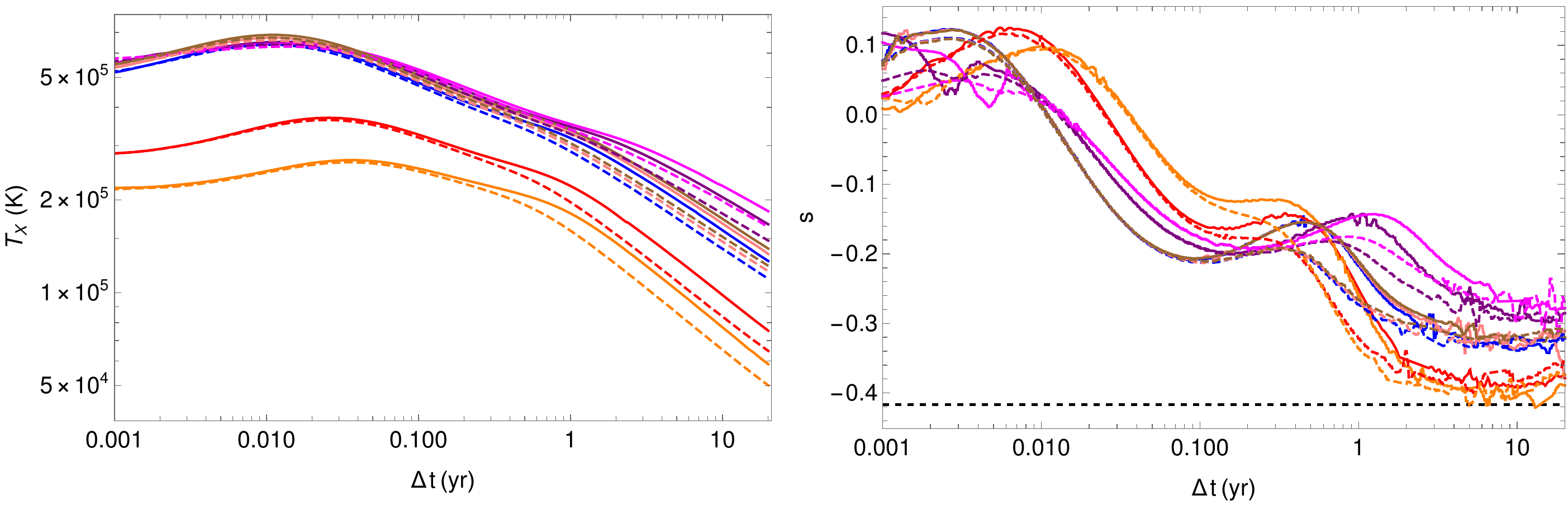}
\end{center}
\caption{Left: The X-ray temperature ($T_X$) obtained using procedure discussed below equation (\ref{xspecc}) is shown for no corona [$b_2 =0$, B1, solid lines] and with corona [$b_2 =0.5$, B3, dashed lines]. The $\mu = 1 $ corresponding to case A1. The color plots legend is the same as given in Fig. \ref{macct}d. The bolometric luminosity for each line is given in Fig. \ref{lbps}. The $T_X$ agrees with the blackbody temperature from the X-ray observations. The temperature shows a power-law decline at late time and we approximate $T_X \propto (t-t_c)^{s}$. See Table \ref{tctm} for $t_c$. (b) The evolution of $s= \diff \log(T_X) / \diff \log(t)$ is shown. The black dashed line correspond to $s = -5/12$ which indicates $T \propto \dot{M}_{\rm fb}^{1/4} \propto (t-t_c) ^{-5/12}$. See section \ref{discuss} for more discussion. }
\label{tsp}
\end{figure*}

The single blackbody temperature obtained from a blackbody model fit to the X-ray spectrum of the source such as XMMSL1 J061927.1-655311 is $\sim 1.4 \times 10^6~{\rm K}$ \citep{2014A&A...572A...1S}, ASAS-SN 14li is $\sim 10^{5}~{\rm K}$ \citep{2016MNRAS.455.2918H}, Abell-1795 is $\sim 1.2 \times 10^{6}~{\rm K}$ \citep{2013MNRAS.435.1904M} and NGC-3599 is $\sim 1.1 \times 10^6~{\rm K}$ \citep{2008A&A...489..543E}. Using equation (\ref{fnuobs}), the observed spectrum in the frequency range $\{\nu_l,~\nu_h\} = \{0.3,~10\}~{\rm keV}$, is given by 

\begin{equation}
F_{\nu_{\rm obs}} = \frac{\cos\theta_{\rm obs}}{D_L^2} \nu_{\rm obs} \int_{r_{\rm in}}^{r_{\rm out}} g^3 I_{\nu}\left(\frac{\nu_{\rm obs}}{g}\right) \, \diff A.
\label{xspecc}
\end{equation} 

\noindent Then, the luminosity is given by $L_{\nu_{\rm obs}} = 4 \pi D_L^2 F_{\nu_{\rm obs}}$. We fit the obtained spectrum with a luminosity model given by $L = 4 \pi \nu B(T,~\nu) A$, where $B(T,~\nu)$ is the blackbody function, and $T$ and $A$ are the single blackbody temperature and area of the disc. By taking the disc area ($A$) to be the $A = \pi (r_{\rm out}^2-r_{\rm in}^2)$, we calculate the X-ray temperature of the disc ($T_X$) and is shown in Fig. \ref{tsp} for various black hole mass and spin and stellar mass whose bolometric luminosity is given in Fig. \ref{lbps}. The obtained X-ray temperature agrees with the blackbody temperature from the X-ray observations. At late time, $T_X$ declines as a power-law given by $T_X \propto (t-t_c)^{s}$. The effective temperature for a steady thin disc accretion model is given by $T \propto \dot{M}_{\rm fb}^{1/4}$ \citep{2011MNRAS.410..359L} and decreases as $T \propto t^{-5/12}$. The late-time evolution of the disc X-ray temperature is slower than the temperature evolution in a steady thin disc.  

\begin{figure*}
\begin{center}
\subfigure[]{\includegraphics[scale=0.55]{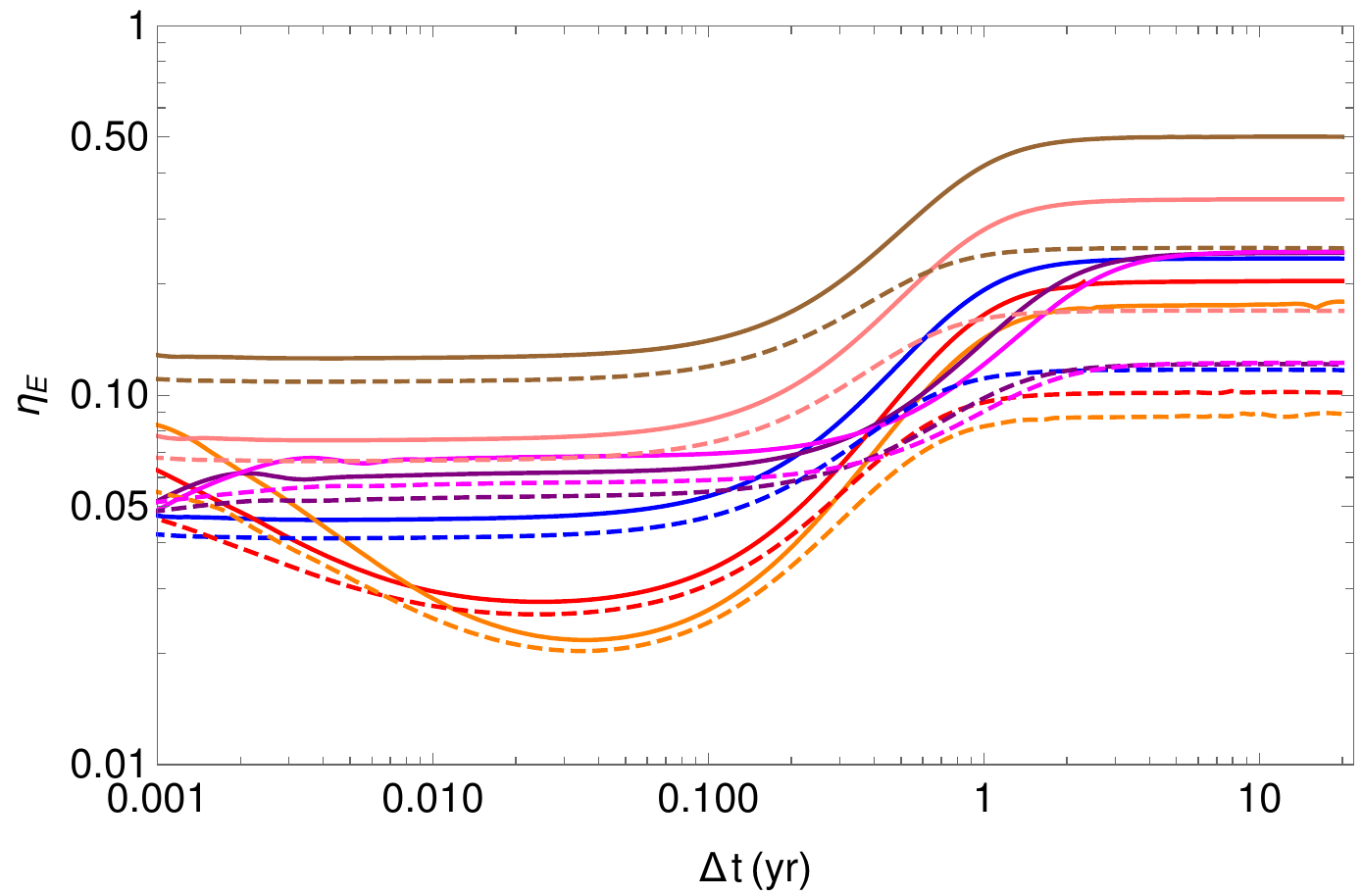}}~~
\subfigure[]{\includegraphics[scale=0.55]{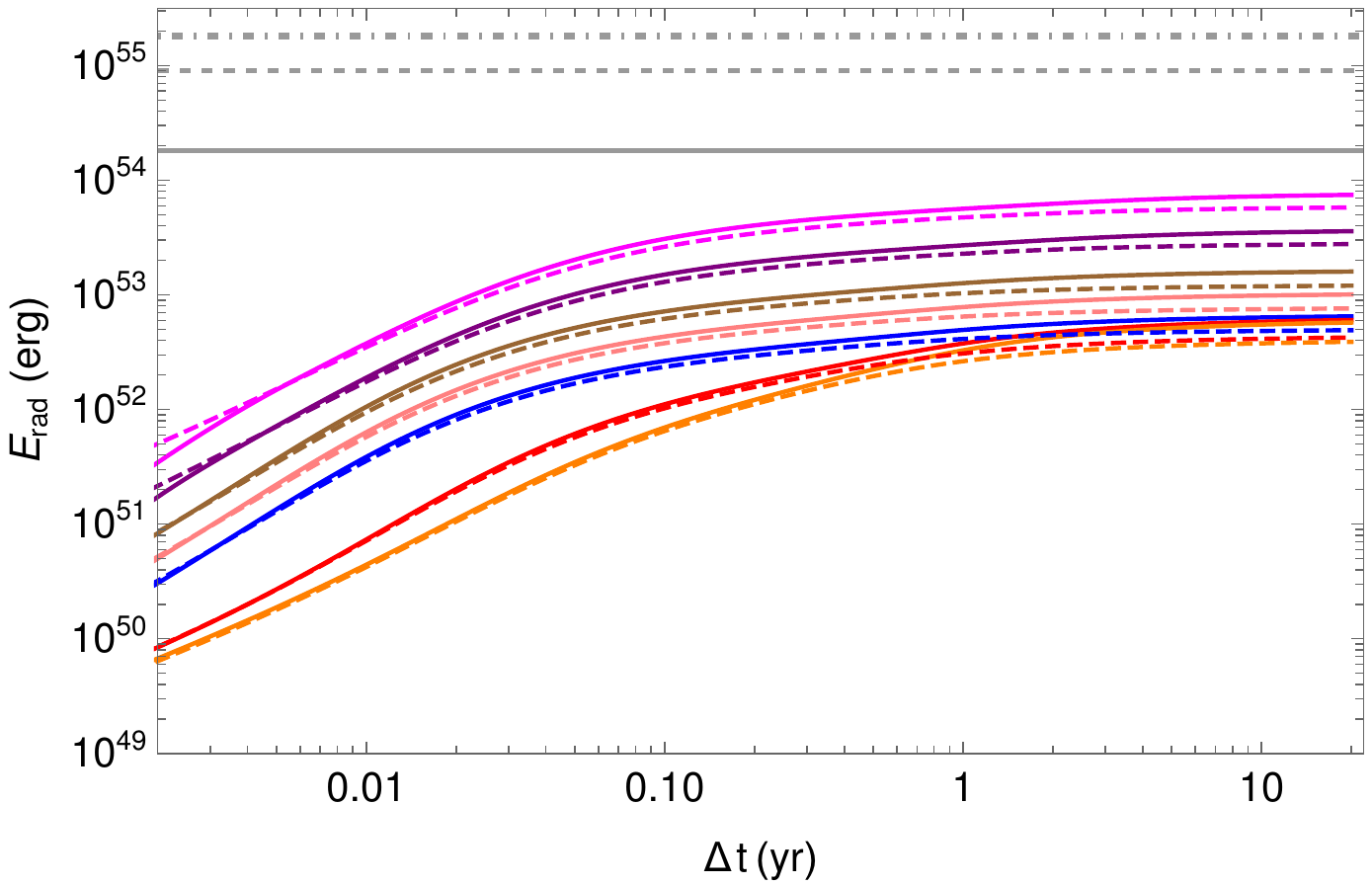}}
\end{center}
\caption{(a) The time evolution of radiative efficiency in the disc for no corona [$b_2 =0$, B1, solid lines] and with corona [$b_2 =0.5$, B3, dashed lines]. The $\mu = 1 $ correspond to case A1. (b) The integrated radiated energy given by equation (\ref{enint}) is shown. The gray lines corresponds to stellar mass energy ($m M_{\odot} c^2$) with solid, dashed and dot-dashed  represents $m =1,~5,~{\rm and}~10$ respectively. The color plots legend is the same as given in Fig. \ref{macct}d. See section \ref{discuss} for more discussion.}
\label{raden}
\end{figure*}

The bolometric luminosity $L_b$ of the disc shown in Figs. \ref{lbmu}a, \ref{lbps}a and \ref{c1vl}a are high even though the disc mass is small compared to the stellar mass. The radiative efficiency given by $\eta_E = L_b/(\dot{M} c^2)$ where $\dot{M}$ is the mass accretion rate at the inner radius, is shown in Fig \ref{raden}. The radiative efficiency at the initial time is small as is expected for an advective disc in near Eddington to super-Eddington phase. With time, the disc mass accretion rate and the bolometric luminosity decrease, and the disc tends to a sub-Eddington phase. Thus, the radiative efficiency increases with time. At late times, the disc is sub-Eddington and the radiative efficiency attains a saturation value. The integrated radiated energy $E_{\rm rad}$ is given by

\begin{equation}
E_{\rm rad}(\Delta t) = \int_{0}^{\Delta t} L_b(\Delta t') \, \diff \Delta t',
\label{enint}
\end{equation}

\noindent where the bolometric luminosity is given by equation (\ref{lbol}). The integrated energy at various times is shown in Fig. \ref{raden}b and is smaller than the total stellar mass energy reservoir of $M_{\star} c^2$. 
 
The temperature obtained using a blackbody model fit to UV and optical observations is ten times smaller than the X-ray temperature. This is explained by considering the optical/UV emissions either from the outflowing winds \citep{2009MNRAS.400.2070S}, the reprocessing of X-ray emission \citep{2014ApJ...783...23G}, or the emission at a higher radius from an elliptical disc \citep{2021ApJ...908..179L}. We consider a time-dependent circular disc, and the reprocessing of the disc luminosity is the only plausible explanation for the observed temperature difference. The reprocessing layer can be either the outflowing wind \citep{2020ApJ...894....2P} or a static atmosphere \citep{2016ApJ...827....3R}. From Fig \ref{spec}, we can see that the X-ray luminosity from the disc is higher than the UV/optical luminosity. However, the observations have shown that the UV/optical luminosity dominates over the X-ray luminosity with many TDE showing no significant X-ray observations. We present a simple reprocessing model in the appendix \ref{repr} and it shows the dominance of optical/UV flux over the X-ray flux. Thus, the reprocessing of the disc luminosity is crucial to explaining the observations and the reprocessing analysis requires a detailed radiative transfer of disc luminosity through the medium. The disc-corona model presented in this paper shows that the spectral luminosity dominates in X-rays and thus TDEs with presented disc evolution is more suitable for X-ray TDEs.

The mass accretion rate $\dot{M}$ exceeds the Eddington rate $\dot{M}_E$ for low mass supermassive black holes and with an increase in black hole mass, the mass accretion rate to Eddington rate decreases, and the disc is sub-Eddington for higher mass black holes.  \citet{2019ApJ...885...93F} showed through global steady disc solution that the outﬂows are driven from the disc surface if the mass accretion rate is $\dot{M} / \dot{M}_E \gtrsim \dot{m}_{\rm crit} = 1.78-1.91$, where $\dot{m}_{\rm crit}$ is the critical mass accretion rate normalized to the Eddington rate. When the mass accretion rate at the outer radius is less than the critical mass accretion rate, there is no outflow, and the disc is similar to a slim disc. They showed that the mid-plane temperature of these slim discs is substantially lower than that of the discs with outflows, but their effective temperature does not deviate much from that of the discs with outflows. The radiation in the wind-driving region is automatically regulated by the wind, which sets an upper limit on the effective temperature. Thus, the disc bolometric luminosity increases slowly with the total mass loss rate when there is wind in the disc. The presence of an outflow results in the disc losing its mass, angular momentum, and energy resulting in a cooling of the accretion flow. The luminosity that a disc without outflow emits will decrease when the outflow is included because the outflowing wind will carry some of the energy generated via viscous heating. The increase in the mass outflow will reduce the mass accretion rate. The radial velocity is comparable to the azimuthal velocity in a super-Eddington disc, and the disc surface density at the outer radius decreases with an increase in the radial velocity if the mass accretion rate is equal to the mass fallback rate. The time evolution of disc surface density is a function of the mass accretion and outflow rates. For such a disc, it is necessary to include the outflow in conservation equations, and the resulting disc will be the advection-dominated inflow-outflow (ADIO) disc \citep{1999MNRAS.303L...1B}.

The reprocessing models for TDEs have shown that the outflow is crucial as it forms the atmosphere for the reprocessing of the disc emissions. However, the presence of a corona surrounding the disc that has an outflow is uncertain, as the strong wind may destroy the low-density corona. If the strong wind sweeps the low-density corona, then there will be no coronal emission, and the spectrum will be dominated by the disc. As the disc evolves with time and the radiation force decreases, the mass outflow rate decreases. As the wind strength decreases, the corona will tend toward a stable structure and its emission will increase. In the sub-Eddington phase, the coronal emission dominates the X-ray spectrum as a non-thermal emission. Our model is limited to an accretion disc without outflow and with a stable corona and is applicable to the near to sub-Eddington accretion phase. The time-dependent accretion model with an outflowing wind and the dynamics of the corona surrounding such a disc are complex. The global disc solution of a TDE disc from super-Eddington with outflow to sub-Eddington requires a detailed numerical simulation. Our present accretion model that begins at a very early time after disruption (see Table \ref{tctm} for initial time $t_c$) shows the disc evolution from Eddington to sub-Eddington phases. Our accretion model explores the evolution of the accretion disc in the presence of infalling debris and energy loss to the corona. Our advective disc-corona model fit to the observations will return the physical parameters such as star mass, black hole mass and spin, and the circularization time. 

\section{Summary}
\label{summary}

We have constructed a non-relativistic and time-dependent advective accretion disc model with corona and fallback at the constant outer radius. The stress tensor is assumed to be dominated by Maxwell stresses ($\tau_{r\phi} \propto P_{\rm mag}$) with magnetic pressure given by $P_{\rm mag} \propto P_g^{1-\mu}P_t^{\mu}$. We have obtained initial and boundary solutions. 

The increase in the contribution of gas pressure to the viscous stress (decreasing $\mu$) increases $t_c$ (see Table \ref{tctm}) and thus delays the beginning of disc accretion. The increase in the black hole and stellar mass also increases $t_c$ whereas the black hole spin does not affect much. 

The gas to total pressure $\beta_g$ increases and disc height decreases with time implying that the disc evolves from Eddington to sub-Eddington phase and thus our model presents the disc transition through different phases (see Fig. \ref{sigt}). 

The mass accretion rate is of the order of the mass fallback rate and the time evolution of mass accretion rate depends on the viscous dynamics. The mass accretion rate deviates from the mass fallback rate at the initial time, then follows it for a certain time and shows a small variation at the later times. The mass accretion rate at late times is close to the $t^{-5/3}$ evolution. The presence of corona increases the mass accretion rate at the initial time followed by a small variation at late times. The variation increases with an increase in energy transport to the corona and can be reduced by decreasing $\mu$.

The beginning time $t_c$ for accretion increases with a decrease in $\mu$ which results in a higher initial disc mass. The disc mass decreases when $\dot{M}_a > \dot{M}_{\rm fb}$ and increases for $\dot{M}_a < \dot{M}_{\rm fb}$. The disc mass at late times shows a significant variation in evolution with an increase in $b_2$ (increase in energy transport to corona) as can be seen from Figs. \ref{macc} and \ref{macct}. 

The total angular momentum of the disc is not conserved and evolves with time. This is due to the considered assumption of constant outer radius. The angular momentum is higher for a disc with energy loss to corona as can be seen from Fig. \ref{jdisk}. 

The bolometric luminosity increases with stellar mass and black hole spin but decreases with black hole mass at the initial time. The duration of the disc in the super-Eddington phase increases with stellar mass and spin but reduces with an increase in black hole mass. The luminosity at late times is close to $t^{-5/3}$ evolution. 

The radiative efficiency ($\eta_E$) increases with time and attains a saturation value at late times in the sub-Eddington phase (Fig. \ref{raden}). The $\eta_E$ shows a significant increment with the black hole spin and decreases when energy loss to the corona is included. The integrated radiated energy from the disc is smaller than the stellar mass energy ($M_{\star} c^2$). 

The ratio of total X-ray luminosity from corona to bolometric disc luminosity increases with $\mu$ implying that the energy transport to corona decreases with an increase in gas pressure contribution to the viscous stress. The disc bolometric luminosity decreases with an increase in energy transport to the corona. The ratio is constant at late times for $\mu =1 $ but increases for $\mu \neq 1$.  

The presence of corona impacts the spectral luminosity at late times by decreasing the spectral luminosity. The luminosity increases with black hole mass for optical/UV wavelengths but decreases in X-ray band as can be seen from Fig. \ref{spec}. The increase in stellar mass increases the bolometric and spectral luminosity. The X-ray luminosity declines faster with an increase in the black hole mass.

The electron temperature $T_e$ in the corona is $10^7 -10^{8.5}$ depending on the disc mass accretion rate and the energy transport to the corona. The electron temperature increases with a decrease in the mass accretion rate. The Compton $y$ parameter decreases with mass accretion rate. The decrease in optical depth with time indicates that the corona is transparent to the soft photons from the disc and the $y \ll 1$ suggests that the comptonization is negligible. The spectral index $\Gamma$ calculated at earlier times agrees with the observations. The evaluation of the optical depth and $y$ with time suggest a more detailed Comptonization scattering model for the Compton cooling.

The X-ray temperature $T_X$ of the disc is in agreement with the blackbody temperature of $\sim ~{\rm few}~\times~10^{5}~{\rm K}$ from the X-ray observations. The late time evaluation of $T_X$ (see Fig. \ref{tsp}) is slower than the temperature evolution in a steady thin disc ($\propto t^{-5/12}$). The spectral luminosity dominates in X-ray and we have shown that the reprocessing is crucial to explain the observed dominance of optical/UV emission over the disc emission. Without reprocessing, the disc-corona model presented is suitable for X-ray TDEs.

We have neglected the mass loss due to an outflow in our model. The outflow is crucial when the disc is in super-Eddington accretion where the outflow will carry the mass, angular momentum, and energy from the disc. The outflow is also important as it forms the atmosphere for the reprocessing of the disc emission. However, the possibility of a corona around a disc with an outflow is uncertain as the strong outflow may destroy the low-density corona. \citet{2020MNRAS.497L...1W} has shown that the X-ray spectrum evolves from a disc dominated to a non-thermal power-law dominated as the luminosity decreases. We have shown in Figs. \ref{lbmu}, \ref{lbps} and \ref{spec} that the disc bolometric and spectral luminosity shows weak variation in presence of corona when the luminosity is high but declines significantly at low luminosity. The presence of an outflow will impact the disc evolution and the luminosity. A global disc solution of a disc with outflow and including corona is highly complex. Within the framework of our disc-corona model, we have shown the disc evolution and the corresponding time-evolution of disc mass, angular momentum and luminosity in the presence and absence of corona. A caveat is the possibility of an accretion disc with an evolving outer radius and mass fallback; such an accretion disc-corona system is computationally tedious and will be taken up in the future.

\section*{Acknowledgements} 

We thank the anonymous referee for the constructive suggestions that have improved the paper. MT has been supported by the Basic Science Research Program through the National Research Foundation of Korea (NRF) funded by the Ministry of Education (2016R1A5A1013277 and 2020R1A2C1007219).

\section*{Data availability} 

No new data were generated or analysed in support of this research. Any other relevant data will be available on request.

%%%%%%%%%%%%%%%%%%%%%%%%%%%%%%%%%%%%%%%%%%%%%%%%%%

%%%%%%%%%%%%%%%%%%%% REFERENCES %%%%%%%%%%%%%%%%%%

% The best way to enter references is to use BibTeX:

\bibliographystyle{mnras}
\bibliography{reference} % if your bibtex file is called example.bib

\begin{thebibliography}{}
\makeatletter
\relax
\def\mn@urlcharsother{\let\do\@makeother \do\$\do\&\do\#\do\^\do\_\do\%\do\~}
\def\mn@doi{\begingroup\mn@urlcharsother \@ifnextchar [ {\mn@doi@}
  {\mn@doi@[]}}
\def\mn@doi@[#1]#2{\def\@tempa{#1}\ifx\@tempa\@empty \href
  {http://dx.doi.org/#2} {doi:#2}\else \href {http://dx.doi.org/#2} {#1}\fi
  \endgroup}
\def\mn@eprint#1#2{\mn@eprint@#1:#2::\@nil}
\def\mn@eprint@arXiv#1{\href {http://arxiv.org/abs/#1} {{\tt arXiv:#1}}}
\def\mn@eprint@dblp#1{\href {http://dblp.uni-trier.de/rec/bibtex/#1.xml}
  {dblp:#1}}
\def\mn@eprint@#1:#2:#3:#4\@nil{\def\@tempa {#1}\def\@tempb {#2}\def\@tempc
  {#3}\ifx \@tempc \@empty \let \@tempc \@tempb \let \@tempb \@tempa \fi \ifx
  \@tempb \@empty \def\@tempb {arXiv}\fi \@ifundefined
  {mn@eprint@\@tempb}{\@tempb:\@tempc}{\expandafter \expandafter \csname
  mn@eprint@\@tempb\endcsname \expandafter{\@tempc}}}

\bibitem[\protect\citeauthoryear{{Abramowicz}, {Chen}, {Kato}, {Lasota}  \&
  {Regev}}{{Abramowicz} et~al.}{1995}]{1995ApJ...438L..37A}
{Abramowicz} M.~A.,  {Chen} X.,  {Kato} S.,  {Lasota} J.-P.,   {Regev} O.,
  1995, \mn@doi [\apjl] {10.1086/187709}, \href
  {https://ui.adsabs.harvard.edu/abs/1995ApJ...438L..37A} {438, L37}

\bibitem[\protect\citeauthoryear{{Alexander} \& {Kumar}}{{Alexander} \&
  {Kumar}}{2001}]{2001ApJ...549..948A}
{Alexander} T.,  {Kumar} P.,  2001, \mn@doi [\apj] {10.1086/319436}, \href
  {http://adsabs.harvard.edu/abs/2001ApJ...549..948A} {549, 948}

\bibitem[\protect\citeauthoryear{{Arcodia}, {Merloni}, {Nandra}  \&
  {Ponti}}{{Arcodia} et~al.}{2019}]{2019A&A...628A.135A}
{Arcodia} R.,  {Merloni} A.,  {Nandra} K.,   {Ponti} G.,  2019, \mn@doi [\aap]
  {10.1051/0004-6361/201935874}, \href
  {https://ui.adsabs.harvard.edu/abs/2019A&A...628A.135A} {628, A135}

\bibitem[\protect\citeauthoryear{{Auchettl}, {Guillochon}  \&
  {Ramirez-Ruiz}}{{Auchettl} et~al.}{2017}]{2017ApJ...838..149A}
{Auchettl} K.,  {Guillochon} J.,   {Ramirez-Ruiz} E.,  2017, \mn@doi [\apj]
  {10.3847/1538-4357/aa633b}, \href
  {https://ui.adsabs.harvard.edu/abs/2017ApJ...838..149A} {838, 149}

\bibitem[\protect\citeauthoryear{{Blaes} \& {Socrates}}{{Blaes} \&
  {Socrates}}{2001}]{2001ApJ...553..987B}
{Blaes} O.,  {Socrates} A.,  2001, \mn@doi [\apj] {10.1086/320968}, \href
  {https://ui.adsabs.harvard.edu/abs/2001ApJ...553..987B} {553, 987}

\bibitem[\protect\citeauthoryear{{Blandford} \& {Begelman}}{{Blandford} \&
  {Begelman}}{1999}]{1999MNRAS.303L...1B}
{Blandford} R.~D.,  {Begelman} M.~C.,  1999, \mn@doi [\mnras]
  {10.1046/j.1365-8711.1999.02358.x}, \href
  {https://ui.adsabs.harvard.edu/abs/1999MNRAS.303L...1B} {303, L1}

\bibitem[\protect\citeauthoryear{{Cannizzo}, {Lee}  \& {Goodman}}{{Cannizzo}
  et~al.}{1990}]{1990ApJ...351...38C}
{Cannizzo} J.~K.,  {Lee} H.~M.,   {Goodman} J.,  1990, \mn@doi [\apj]
  {10.1086/168442}, \href {http://adsabs.harvard.edu/abs/1990ApJ...351...38C}
  {351, 38}

\bibitem[\protect\citeauthoryear{{Cao}}{{Cao}}{2009}]{2009MNRAS.394..207C}
{Cao} X.,  2009, \mn@doi [\mnras] {10.1111/j.1365-2966.2008.14347.x}, \href
  {https://ui.adsabs.harvard.edu/abs/2009MNRAS.394..207C} {394, 207}

\bibitem[\protect\citeauthoryear{{Chandrasekhar}}{{Chandrasekhar}}{1943}]{1943ApJ....97..255C}
{Chandrasekhar} S.,  1943, \mn@doi [\apj] {10.1086/144517}, \href
  {http://adsabs.harvard.edu/abs/1943ApJ....97..255C} {97, 255}

\bibitem[\protect\citeauthoryear{{Chornock} et~al.,}{{Chornock}
  et~al.}{2014}]{2014ApJ...780...44C}
{Chornock} R.,  et~al., 2014, \mn@doi [\apj] {10.1088/0004-637X/780/1/44},
  \href {http://adsabs.harvard.edu/abs/2014ApJ...780...44C} {780, 44}

\bibitem[\protect\citeauthoryear{{Clerici} \& {Gomboc}}{{Clerici} \&
  {Gomboc}}{2020}]{2020A&A...642A.111C}
{Clerici} A.,  {Gomboc} A.,  2020, \mn@doi [\aap]
  {10.1051/0004-6361/202037641}, \href
  {https://ui.adsabs.harvard.edu/abs/2020A&A...642A.111C} {642, A111}

\bibitem[\protect\citeauthoryear{{Dermer}, {Liang}  \& {Canfield}}{{Dermer}
  et~al.}{1991}]{1991ApJ...369..410D}
{Dermer} C.~D.,  {Liang} E.~P.,   {Canfield} E.,  1991, \mn@doi [\apj]
  {10.1086/169770}, \href
  {https://ui.adsabs.harvard.edu/abs/1991ApJ...369..410D} {369, 410}

\bibitem[\protect\citeauthoryear{{Done} \& {Davis}}{{Done} \&
  {Davis}}{2008}]{2008ApJ...683..389D}
{Done} C.,  {Davis} S.~W.,  2008, \mn@doi [\apj] {10.1086/589647}, \href
  {https://ui.adsabs.harvard.edu/abs/2008ApJ...683..389D} {683, 389}

\bibitem[\protect\citeauthoryear{{Esquej} et~al.,}{{Esquej}
  et~al.}{2008}]{2008A&A...489..543E}
{Esquej} P.,  et~al., 2008, \mn@doi [\aap] {10.1051/0004-6361:200810110}, \href
  {http://adsabs.harvard.edu/abs/2008A%26A...489..543E} {489, 543}

\bibitem[\protect\citeauthoryear{{Feng}, {Cao}, {Gu}  \& {Ma}}{{Feng}
  et~al.}{2019}]{2019ApJ...885...93F}
{Feng} J.,  {Cao} X.,  {Gu} W.-M.,   {Ma} R.-Y.,  2019, \mn@doi [\apj]
  {10.3847/1538-4357/ab4592}, \href
  {https://ui.adsabs.harvard.edu/abs/2019ApJ...885...93F} {885, 93}

\bibitem[\protect\citeauthoryear{{Frank} \& {Rees}}{{Frank} \&
  {Rees}}{1976}]{1976MNRAS.176..633F}
{Frank} J.,  {Rees} M.~J.,  1976, \mnras, \href
  {http://adsabs.harvard.edu/abs/1976MNRAS.176..633F} {176, 633}

\bibitem[\protect\citeauthoryear{{Frank}, {King}  \& {Raine}}{{Frank}
  et~al.}{2002}]{2002apa..book.....F}
{Frank} J.,  {King} A.,   {Raine} D.~J.,  2002, {Accretion Power in
  Astrophysics: Third Edition}.
Cambridge University Press, Cambridge, UK

\bibitem[\protect\citeauthoryear{{Gezari} et~al.,}{{Gezari}
  et~al.}{2012}]{2012Natur.485..217G}
{Gezari} S.,  et~al., 2012, \mn@doi [\nat] {10.1038/nature10990}, \href
  {http://adsabs.harvard.edu/abs/2012Natur.485..217G} {485, 217}

\bibitem[\protect\citeauthoryear{{Gezari}, {Cenko}  \& {Arcavi}}{{Gezari}
  et~al.}{2017}]{2017ApJ...851L..47G}
{Gezari} S.,  {Cenko} S.~B.,   {Arcavi} I.,  2017, \mn@doi [\apjl]
  {10.3847/2041-8213/aaa0c2}, \href
  {https://ui.adsabs.harvard.edu/abs/2017ApJ...851L..47G} {851, L47}

\bibitem[\protect\citeauthoryear{{Golightly}, {Coughlin}  \&
  {Nixon}}{{Golightly} et~al.}{2019}]{2019ApJ...872..163G}
{Golightly} E. C.~A.,  {Coughlin} E.~R.,   {Nixon} C.~J.,  2019, \mn@doi [\apj]
  {10.3847/1538-4357/aafd2f}, \href
  {https://ui.adsabs.harvard.edu/abs/2019ApJ...872..163G} {872, 163}

\bibitem[\protect\citeauthoryear{{Guillochon}, {Manukian}  \&
  {Ramirez-Ruiz}}{{Guillochon} et~al.}{2014}]{2014ApJ...783...23G}
{Guillochon} J.,  {Manukian} H.,   {Ramirez-Ruiz} E.,  2014, \mn@doi [\apj]
  {10.1088/0004-637X/783/1/23}, \href
  {https://ui.adsabs.harvard.edu/abs/2014ApJ...783...23G} {783, 23}

\bibitem[\protect\citeauthoryear{{Hayasaki}, {Stone}  \& {Loeb}}{{Hayasaki}
  et~al.}{2016}]{2016MNRAS.461.3760H}
{Hayasaki} K.,  {Stone} N.,   {Loeb} A.,  2016, \mn@doi [\mnras]
  {10.1093/mnras/stw1387}, \href
  {https://ui.adsabs.harvard.edu/abs/2016MNRAS.461.3760H} {461, 3760}

\bibitem[\protect\citeauthoryear{{Holoien} et~al.,}{{Holoien}
  et~al.}{2016a}]{2016MNRAS.455.2918H}
{Holoien} T.~W.-S.,  et~al., 2016a, \mn@doi [\mnras] {10.1093/mnras/stv2486},
  \href {http://adsabs.harvard.edu/abs/2016MNRAS.455.2918H} {455, 2918}

\bibitem[\protect\citeauthoryear{{Holoien} et~al.,}{{Holoien}
  et~al.}{2016b}]{2016MNRAS.463.3813H}
{Holoien} T.~W.~S.,  et~al., 2016b, \mn@doi [\mnras] {10.1093/mnras/stw2272},
  \href {https://ui.adsabs.harvard.edu/abs/2016MNRAS.463.3813H} {463, 3813}

\bibitem[\protect\citeauthoryear{{Janiuk}, {Czerny}  \&
  {Siemiginowska}}{{Janiuk} et~al.}{2002}]{2002ApJ...576..908J}
{Janiuk} A.,  {Czerny} B.,   {Siemiginowska} A.,  2002, \mn@doi [\apj]
  {10.1086/341804}, \href
  {https://ui.adsabs.harvard.edu/abs/2002ApJ...576..908J} {576, 908}

\bibitem[\protect\citeauthoryear{{Kajava}, {Giustini}, {Saxton}  \&
  {Miniutti}}{{Kajava} et~al.}{2020}]{2020A&A...639A.100K}
{Kajava} J. J.~E.,  {Giustini} M.,  {Saxton} R.~D.,   {Miniutti} G.,  2020,
  \mn@doi [\aap] {10.1051/0004-6361/202038165}, \href
  {https://ui.adsabs.harvard.edu/abs/2020A&A...639A.100K} {639, A100}

\bibitem[\protect\citeauthoryear{{Kippenhahn} \& {Weigert}}{{Kippenhahn} \&
  {Weigert}}{1994}]{1994sse..book.....K}
{Kippenhahn} R.,  {Weigert} A.,  1994, {Stellar Structure and Evolution}.
Springer-Verlag press, Berlin Heidelberg New York

\bibitem[\protect\citeauthoryear{{Kochanek}}{{Kochanek}}{1992}]{1992ApJ...385..604K}
{Kochanek} C.~S.,  1992, \mn@doi [\apj] {10.1086/170966}, \href
  {https://ui.adsabs.harvard.edu/abs/1992ApJ...385..604K} {385, 604}

\bibitem[\protect\citeauthoryear{{Kochanek}}{{Kochanek}}{1994}]{1994ApJ...422..508K}
{Kochanek} C.~S.,  1994, \mn@doi [\apj] {10.1086/173745}, \href
  {https://ui.adsabs.harvard.edu/abs/1994ApJ...422..508K} {422, 508}

\bibitem[\protect\citeauthoryear{{Li}, {Narayan}  \& {Menou}}{{Li}
  et~al.}{2002}]{2002ApJ...576..753L}
{Li} L.-X.,  {Narayan} R.,   {Menou} K.,  2002, \mn@doi [\apj]
  {10.1086/341890}, \href
  {https://ui.adsabs.harvard.edu/abs/2002ApJ...576..753L} {576, 753}

\bibitem[\protect\citeauthoryear{{Liu}, {Cao}, {Abramowicz}, {Wielgus}, {Cao}
  \& {Zhou}}{{Liu} et~al.}{2021}]{2021ApJ...908..179L}
{Liu} F.~K.,  {Cao} C.~Y.,  {Abramowicz} M.~A.,  {Wielgus} M.,  {Cao} R.,
  {Zhou} Z.~Q.,  2021, \mn@doi [\apj] {10.3847/1538-4357/abd2b6}, \href
  {https://ui.adsabs.harvard.edu/abs/2021ApJ...908..179L} {908, 179}

\bibitem[\protect\citeauthoryear{{Lodato} \& {Rossi}}{{Lodato} \&
  {Rossi}}{2011}]{2011MNRAS.410..359L}
{Lodato} G.,  {Rossi} E.~M.,  2011, \mn@doi [\mnras]
  {10.1111/j.1365-2966.2010.17448.x}, \href
  {http://adsabs.harvard.edu/abs/2011MNRAS.410..359L} {410, 359}

\bibitem[\protect\citeauthoryear{{Lodato}, {King}  \& {Pringle}}{{Lodato}
  et~al.}{2009}]{2009MNRAS.392..332L}
{Lodato} G.,  {King} A.~R.,   {Pringle} J.~E.,  2009, \mn@doi [\mnras]
  {10.1111/j.1365-2966.2008.14049.x}, \href
  {http://adsabs.harvard.edu/abs/2009MNRAS.392..332L} {392, 332}

\bibitem[\protect\citeauthoryear{{Mageshwaran} \&
  {Bhattacharyya}}{{Mageshwaran} \&
  {Bhattacharyya}}{2020}]{2020MNRAS.496.1784M}
{Mageshwaran} T.,  {Bhattacharyya} S.,  2020, \mn@doi [\mnras]
  {10.1093/mnras/staa1604}, \href
  {https://ui.adsabs.harvard.edu/abs/2020MNRAS.496.1784M} {496, 1784}

\bibitem[\protect\citeauthoryear{{Mageshwaran} \& {Mangalam}}{{Mageshwaran} \&
  {Mangalam}}{2021}]{2021NewA...8301491M}
{Mageshwaran} T.,  {Mangalam} A.,  2021, \mn@doi [\na]
  {10.1016/j.newast.2020.101491}, \href
  {https://ui.adsabs.harvard.edu/abs/2021NewA...8301491M} {83, 101491}

\bibitem[\protect\citeauthoryear{{Maksym}, {Ulmer}, {Eracleous}, {Guennou}  \&
  {Ho}}{{Maksym} et~al.}{2013}]{2013MNRAS.435.1904M}
{Maksym} W.~P.,  {Ulmer} M.~P.,  {Eracleous} M.~C.,  {Guennou} L.,   {Ho}
  L.~C.,  2013, \mn@doi [\mnras] {10.1093/mnras/stt1379}, \href
  {http://adsabs.harvard.edu/abs/2013MNRAS.435.1904M} {435, 1904}

\bibitem[\protect\citeauthoryear{{Merloni}}{{Merloni}}{2003}]{2003MNRAS.341.1051M}
{Merloni} A.,  2003, \mn@doi [\mnras] {10.1046/j.1365-8711.2003.06496.x}, \href
  {https://ui.adsabs.harvard.edu/abs/2003MNRAS.341.1051M} {341, 1051}

\bibitem[\protect\citeauthoryear{{Merloni} \& {Fabian}}{{Merloni} \&
  {Fabian}}{2002}]{2002MNRAS.332..165M}
{Merloni} A.,  {Fabian} A.~C.,  2002, \mn@doi [\mnras]
  {10.1046/j.1365-8711.2002.05288.x}, \href
  {https://ui.adsabs.harvard.edu/abs/2002MNRAS.332..165M} {332, 165}

\bibitem[\protect\citeauthoryear{{Minoshima}, {Hirose}  \& {Sano}}{{Minoshima}
  et~al.}{2015}]{2015ApJ...808...54M}
{Minoshima} T.,  {Hirose} S.,   {Sano} T.,  2015, \mn@doi [\apj]
  {10.1088/0004-637X/808/1/54}, \href
  {https://ui.adsabs.harvard.edu/abs/2015ApJ...808...54M} {808, 54}

\bibitem[\protect\citeauthoryear{{Montesinos Armijo} \& {de Freitas
  Pacheco}}{{Montesinos Armijo} \& {de Freitas
  Pacheco}}{2011}]{2011ApJ...736..126M}
{Montesinos Armijo} M.,  {de Freitas Pacheco} J.~A.,  2011, \mn@doi [\apj]
  {10.1088/0004-637X/736/2/126}, \href
  {http://adsabs.harvard.edu/abs/2011ApJ...736..126M} {736, 126}

\bibitem[\protect\citeauthoryear{{Mummery} \& {Balbus}}{{Mummery} \&
  {Balbus}}{2019}]{2019MNRAS.489..132M}
{Mummery} A.,  {Balbus} S.~A.,  2019, \mn@doi [\mnras] {10.1093/mnras/stz2141},
  \href {https://ui.adsabs.harvard.edu/abs/2019MNRAS.489..132M} {489, 132}

\bibitem[\protect\citeauthoryear{{Narayan} \& {Yi}}{{Narayan} \&
  {Yi}}{1995}]{1995ApJ...452..710N}
{Narayan} R.,  {Yi} I.,  1995, \mn@doi [\apj] {10.1086/176343}, \href
  {https://ui.adsabs.harvard.edu/abs/1995ApJ...452..710N} {452, 710}

\bibitem[\protect\citeauthoryear{{Piro} \& {Lu}}{{Piro} \&
  {Lu}}{2020}]{2020ApJ...894....2P}
{Piro} A.~L.,  {Lu} W.,  2020, \mn@doi [\apj] {10.3847/1538-4357/ab83f6}, \href
  {https://ui.adsabs.harvard.edu/abs/2020ApJ...894....2P} {894, 2}

\bibitem[\protect\citeauthoryear{{Rees}}{{Rees}}{1988}]{1988Natur.333..523R}
{Rees} M.~J.,  1988, \mn@doi [\nat] {10.1038/333523a0}, \href
  {http://adsabs.harvard.edu/abs/1988Natur.333..523R} {333, 523}

\bibitem[\protect\citeauthoryear{{Roth}, {Kasen}, {Guillochon}  \&
  {Ramirez-Ruiz}}{{Roth} et~al.}{2016}]{2016ApJ...827....3R}
{Roth} N.,  {Kasen} D.,  {Guillochon} J.,   {Ramirez-Ruiz} E.,  2016, \mn@doi
  [\apj] {10.3847/0004-637X/827/1/3}, \href
  {https://ui.adsabs.harvard.edu/abs/2016ApJ...827....3R} {827, 3}

\bibitem[\protect\citeauthoryear{{Ruan}, {Anderson}, {Eracleous}, {Green},
  {Haggard}, {MacLeod}, {Runnoe}  \& {Sobolewska}}{{Ruan}
  et~al.}{2019}]{2019ApJ...883...76R}
{Ruan} J.~J.,  {Anderson} S.~F.,  {Eracleous} M.,  {Green} P.~J.,  {Haggard}
  D.,  {MacLeod} C.~L.,  {Runnoe} J.~C.,   {Sobolewska} M.~A.,  2019, \mn@doi
  [\apj] {10.3847/1538-4357/ab3c1a}, \href
  {https://ui.adsabs.harvard.edu/abs/2019ApJ...883...76R} {883, 76}

\bibitem[\protect\citeauthoryear{{Rybicki} \& {Lightman}}{{Rybicki} \&
  {Lightman}}{1979}]{1979rpa..book.....R}
{Rybicki} G.~B.,  {Lightman} A.~P.,  1979, {Radiative processes in
  astrophysics}

\bibitem[\protect\citeauthoryear{{Saxton} et~al.,}{{Saxton}
  et~al.}{2014}]{2014A&A...572A...1S}
{Saxton} R.~D.,  et~al., 2014, \mn@doi [\aap] {10.1051/0004-6361/201424347},
  \href {http://adsabs.harvard.edu/abs/2014A%26A...572A...1S} {572, A1}

\bibitem[\protect\citeauthoryear{{Saxton}, {Read}, {Komossa}, {Lira},
  {Alexander}  \& {Wieringa}}{{Saxton} et~al.}{2017}]{2017A&A...598A..29S}
{Saxton} R.~D.,  {Read} A.~M.,  {Komossa} S.,  {Lira} P.,  {Alexander} K.~D.,
  {Wieringa} M.~H.,  2017, \mn@doi [\aap] {10.1051/0004-6361/201629015}, \href
  {https://ui.adsabs.harvard.edu/abs/2017A&A...598A..29S} {598, A29}

\bibitem[\protect\citeauthoryear{{Shen} \& {Matzner}}{{Shen} \&
  {Matzner}}{2014}]{2014ApJ...784...87S}
{Shen} R.-F.,  {Matzner} C.~D.,  2014, \mn@doi [\apj]
  {10.1088/0004-637X/784/2/87}, \href
  {http://adsabs.harvard.edu/abs/2014ApJ...784...87S} {784, 87}

\bibitem[\protect\citeauthoryear{{Steiner}, {Narayan}, {McClintock}  \&
  {Ebisawa}}{{Steiner} et~al.}{2009}]{2009PASP..121.1279S}
{Steiner} J.~F.,  {Narayan} R.,  {McClintock} J.~E.,   {Ebisawa} K.,  2009,
  \mn@doi [\pasp] {10.1086/648535}, \href
  {https://ui.adsabs.harvard.edu/abs/2009PASP..121.1279S} {121, 1279}

\bibitem[\protect\citeauthoryear{{Strubbe} \& {Quataert}}{{Strubbe} \&
  {Quataert}}{2009}]{2009MNRAS.400.2070S}
{Strubbe} L.~E.,  {Quataert} E.,  2009, \mn@doi [\mnras]
  {10.1111/j.1365-2966.2009.15599.x}, \href
  {http://adsabs.harvard.edu/abs/2009MNRAS.400.2070S} {400, 2070}

\bibitem[\protect\citeauthoryear{{Svensson} \& {Zdziarski}}{{Svensson} \&
  {Zdziarski}}{1994}]{1994ApJ...436..599S}
{Svensson} R.,  {Zdziarski} A.~A.,  1994, \mn@doi [\apj] {10.1086/174934},
  \href {https://ui.adsabs.harvard.edu/abs/1994ApJ...436..599S} {436, 599}

\bibitem[\protect\citeauthoryear{{Turner}, {Stone}  \& {Sano}}{{Turner}
  et~al.}{2002}]{2002ApJ...566..148T}
{Turner} N.~J.,  {Stone} J.~M.,   {Sano} T.,  2002, \mn@doi [\apj]
  {10.1086/338081}, \href
  {https://ui.adsabs.harvard.edu/abs/2002ApJ...566..148T} {566, 148}

\bibitem[\protect\citeauthoryear{{Ulmer}}{{Ulmer}}{1999}]{1999ApJ...514..180U}
{Ulmer} A.,  1999, \mn@doi [\apj] {10.1086/306909}, \href
  {http://adsabs.harvard.edu/abs/1999ApJ...514..180U} {514, 180}

\bibitem[\protect\citeauthoryear{{Wevers}}{{Wevers}}{2020}]{2020MNRAS.497L...1W}
{Wevers} T.,  2020, \mn@doi [\mnras] {10.1093/mnrasl/slaa097}, \href
  {https://ui.adsabs.harvard.edu/abs/2020MNRAS.497L...1W} {497, L1}

\makeatother
\end{thebibliography}

% Alternatively you could enter them by hand, like this:
% This method is tedious and prone to error if you have lots of references
%\begin{thebibliography}{99}
%\bibitem[\protect\citeauthoryear{Author}{2012}]{Author2012}
%Author A.~N., 2013, Journal of Improbable Astronomy, 1, 1
%\bibitem[\protect\citeauthoryear{Others}{2013}]{Others2013}
%Others S., 2012, Journal of Interesting Stuff, 17, 198
%\end{thebibliography}

%%%%%%%%%%%%%%%%%%%%%%%%%%%%%%%%%%%%%%%%%%%%%%%%%%

%%%%%%%%%%%%%%%%% APPENDICES %%%%%%%%%%%%%%%%%%%%%

\appendix

\section{Heating and cooling of a two temperature corona}
\label{corhc}

We determine the properties of the corona using two temperature model where we assume that the energy is transferred from ions to electrons through Coulomb coupling and the plasma cools via electrons through various mechanisms. 

The energy equation describing the two-temperature corona is given by

\begin{equation}
Q_{\rm cor} = Q_{\rm cor}^{ie} + \delta Q_{\rm cor} = F_{\rm cor}^{-},
\label{encor}
\end{equation}
 
\noindent where $F_{\rm cor}^{-}$ is the cooling rate in per unit surface area of the corona and $Q_{\rm cor}^{ie}$ is the energy transfer rate from the ions to the electrons via Coulomb interactions. The fraction $\delta$ indicates the amount of energy directly heats the electrons and can be as high as $\sim 0.5$ \citep{2009MNRAS.394..207C}. We consider the plasma to be mostly of hydrogen element such that the Coulomb interaction between ions and electrons is given by 

\begin{multline}
Q_{\rm cor}^{ie} = 1.5 \frac{m_e}{m_p} \frac{\mu_e}{\mu_H} n_e^2 c \sigma_T H_{\rm cor} \frac{k_B(T_i - T_e)}{K_2(\theta_e^{-1})K_2(\theta_H^{-1})} \ln\Lambda~ \times \\ \left[\frac{2(\theta_e+\theta_H)^2+1}{\theta_e+\theta_H}K_1\left(\frac{\theta_e+\theta_H}{\theta_e\theta_H}\right)+2K_0\left(\frac{\theta_e+\theta_H}{\theta_e\theta_H}\right)\right],
\label{qcorie}
\end{multline} 

\noindent where $m_e$ and $m_p$ are the electron and proton mass, electron and ions mean molecular weight are $\mu_e = 2/(1+X)$ and $\mu_H = 4/(1+3 X)$ with H mass fraction $X = 0.7$, $n_e$ is the electron density in the corona, $H_{\rm cor}$ is the corona height, $T_i$ and $T_e$ are the ion and electron temperatures, $\ln\Lambda = 20$ and $K_0$, $K_1$ and $K_2$ are the modified Bessel function of second kind. The $\theta_e = k_B T_e / m_e c^2$ and $\theta_H = k_B T_i / m_p c^2$. Here, we adopt the ion temperature $T_i = 0.9 T_{\rm vir}$, where the virial temperature $T_{\rm vir} = G M_{\bullet} m_p/(3 k_B r)$ \citep{2002MNRAS.332..165M,2009MNRAS.394..207C}.

The total pressure in the corona is given by 

\begin{equation}
p_{\rm cor} = \frac{\rho_{\rm cor} k_B T_i}{\mu_H m_p} + \frac{\rho_{\rm cor} k_B T_e}{\mu_e m_p} + p_{\rm cor,m},
\end{equation}

\noindent where $\rho_{\rm cor}$ is the corona density and $p_{\rm cor,m}$ is the magnetic pressure. Here, we assume that the magnetic pressure is in equipartition with the gas pressure in the corona. Following a vertical hydrodynamical equilibrium, the corona height is $H_{\rm cor} = 1/\Omega_K \sqrt{p_{\rm cor}/\rho_{\rm cor}}$.

The electrons are cooled by various processes that include bremsstrahlung, synchrotron, and Compton cooling. In case of Bremsstrahlung cooling, the electron cool via electron-ion and electron-electron bremsstrahlung such that the cooling rate per unit volume is $q_{\rm br} = q_{\rm ei} + q_{\rm ee}$. Following \citet{1995ApJ...452..710N} [hereafter NY95], these rates are given by 

\begin{multline}
q_{\rm ei} = 1.25 n_e^2 \sigma_T c \alpha_f m_e c^2 \times \\ \left\{\begin{array}{ll}
4 \left(\frac{2 \theta_e}{\pi^3}\right)^{1/2} \left[1+1.781 \theta_e^{1.34}\right]& \theta_e <1 \\
&\\
 \frac{9 \theta_e}{2\pi} \left[\log(1.123\theta_e + 0.481)+1.5\right]& \theta_e >1,
\end{array}
\right.
\end{multline}

\noindent and

\begin{equation}
q_{\rm ee} = n_e^2 c r_e^2 \alpha_f m_e c^2 \left\{\begin{array}{ll}
\frac{20}{9\sqrt{\pi}}(44 - 3 \pi^2)\theta_e^{3/2} \times & \\
\left[1+ 1.1 \theta_e + \theta_e^2 - 1.25 \theta_e^{5/2}\right]& \theta_e <1 \\
&\\
 24 \theta_e \left[\log(2.246\theta_e) + 1.28\right]& \theta_e >1,
\end{array}
\right.
\end{equation}

\noindent where $r_e$ is the classical electron radius and $\alpha_f$ is the fine structure constant. For Synchrotron cooling, we again follow NY95 and is given by

\begin{equation}
q_{\rm syn} = \frac{2 \pi}{3 c^2} k_B T_e \frac{\nu_c^3}{r},
\end{equation}

\noindent where the critical frequency below which emission is self-absorbed is $\nu_c = 1.5 \nu_0 \theta_e^2 x_M$, with $\nu_0 = 2.8 \times 10^6~B~{\rm Hz}$ and magnetic field $B$ is in unit of Gauss. To calculate the magnetic field, we use $P_{\rm mag} = B^2/(8 \pi) = \alpha_0 P_g^{1-\mu} P_t^{\mu}$ discussed below equation (\ref{feq}). The $x_M$ is given by

\begin{multline}
\exp\left(1.8899 x_M^{1/3}\right) = 2.49 \times 10^{-10} \frac{4 \pi n_e r}{B} \frac{1}{\theta_e^3 K_2(\theta_e^{-1})} \\ \left[\frac{1}{x_M^{7/6}}+\frac{0.4}{x_M^{17/12}}+\frac{0.5316}{x_M^{5/3}}\right]
\end{multline}

The scattering of photons by hot thermal electrons in the corona is also an essential mechanism to cool the electrons. \citet{1991ApJ...369..410D} gave an efficient approximation method for the Comptonized energy enhancement factor $\eta$ which is the average change in energy of a photon between injection and escape from the scattering medium. It is given by 

\begin{equation}
\eta = 1 + \frac{P (A-1)}{1 - P A} \left[1 - \left(\frac{x}{3 \theta_e}\right)^{-1-\ln P /\ln A}\right],
\label{etacomp}
\end{equation}

\noindent where $x = h \nu /m_e c^2$, probability of photon escaping $P = 1 - \exp(-\tau_{es})$ and mean amplification factor in the energy of scattered photons $A = 1 + 4\theta_e + 16 \theta_e^2$. The electron scattering optical depth $\tau_{es} = \sigma_T n_e H_{\rm cor}$. The energy density of the injected photons per unit time from the disc is approximated to $\pi I_{\nu}/H$, where $I_{\nu}$ is the blackbody intensity and $H$ is the disc height. Then, the Comptonization of the flux is (following NY95)

\begin{equation}
q_{\rm C} = \int_{\nu_{\rm min}}^{\nu_{\rm max}} \frac{\pi I_{\nu}}{H} (\eta-1) \, \diff \nu,
\end{equation}

\noindent where $\nu_{\rm min}$ is taken to be $\nu_c$ and $\nu_{\rm max} = {\rm Max[}\nu_l, 3 \theta_e{\rm ]}$. The $\nu_l$ is a typical high value which we take to be $100~{\rm keV}$ as the blackbody emission above this is negligible. Then, the total cooling rate per unit surface area following a plane parallel corona is given by

\begin{equation}
F_{\rm cor}^{-} (t,~r) = H_{\rm cor}(q_{\rm br} + q_{\rm syn} + q_C). 
\label{fcort}
\end{equation} 

\noindent Using $Q_{\rm cor}$ obtained from equation (\ref{feq}), we use equations (\ref{encor}) and (\ref{qcorie}) to calculate the electron density $n_e$ in terms of $t$, $r$ and $T_e$. Then, we use equations (\ref{encor}) and (\ref{fcort}) to calculate the electron temperature $T_e$ in terms of $t$ and $r$. 

A Compton $y$ parameter determines whether a photon will significantly change its energy in traversing the medium. In general when $y \geq 1$, the total photon spectrum and energy will be significantly altered whereas for $y \ll 1$, there is no significant change in the total energy. In the non-relativistic limit, the Compton y parameter is given by \citep{1979rpa..book.....R}

\begin{equation}
y = \frac{4 k_B T_e}{m_e c^2} {\rm Max[}\tau_{es},\tau_{es}^2{\rm ]}.
\end{equation}

\section{Non-advective disc-corona self-similar solution}
\label{sssol}

We now formulate a solution for a disc with no advection and pressure is dominated by gas pressure, $p = p_g$. In this case, equation (\ref{nuvis}) results in a $\nu \Sigma = (4 \alpha /3) H /\Omega_K p_g$, and $Q^{+} = 3 \alpha H \Omega_K p_g$. Since, the ratio of gas to total pressure $\beta_g = 1$, equation (\ref{fcor}) results in the fraction of energy transported to corona, $f(\beta_g) = Q_{\rm cor}/Q^{+} = b_2$. Thus, the energy conservation equation is $Q_{\rm rad} = Q^{+} - Q_{\rm cor} = (1-b_2) Q^{+}$. Then, for $H = c_s / \Omega_K$ and $\rho = \Sigma/2H$, we get

\begin{equation}
\nu\Sigma = \chi \Sigma^{5/3} r;~~~~\chi = \frac{2 \alpha}{3} \frac{k_B}{\mu_m m_p} \left[\frac{3\alpha}{2} \frac{k_B}{\mu_m m_p} \frac{3\kappa}{4 \sigma} \frac{1}{G M_{\bullet}} \right]^{1/3} (1-b_2)^{1/3}.
\end{equation}  

\noindent The mass and angular momentum conservation equations (\ref{mcons}) and (\ref{mdot}) results in 

\begin{equation}
\frac{\partial \Sigma}{\partial t} = \frac{3 \chi}{r} \frac{\partial }{\partial r} \left[\sqrt{r} \frac{\partial }{\partial r} (r^{3/2} \Sigma^{5/3})\right],
\end{equation}

\noindent and the self-similar solution assuming total angular momentum of the disc remains constant is given by \citep{1990ApJ...351...38C}

\begin{equation}
\Sigma(r,~t) = \Sigma_0 \left(\frac{t}{t_0}\right)^{-15/16} u^{-3/5} (1-u^{7/5})^{3/2},
\end{equation}

\noindent where $u = r/r_0 (t/t_0)^{-3/8}$, and $\Sigma_0$, $r_0$ and $t_0$ are the self-similar constants that satisfy $t_0^{-1} = \chi \Sigma_0^{2/3} / r_0$. The $u=1$ corresponds to the outer radius such that the outer radius evolves as $r_{\rm out} = r_0 (t/t_0)^{3/8}$. The total angular momentum of the disc is given by

\begin{equation}
J_d = 28^{-3/2} \frac{10 \pi}{7} B[5/2,~19/14] r_0^{5/2} \Sigma_0 \sqrt{G M_{\bullet}},
\end{equation}

\noindent where $B$ is the usual Beta function and for a TDE disc, the total angular momentum of the bound debris is $J_d = \sqrt{2 G M_{\bullet}r_t}M_{\star}/2$. We assume that at a time $t = t_i$ after disruption, the mass of the disc is $M_{\star}/2$ and outer radius at initial time $r_{\rm out,i} = 2 r_t$. Thus, we get

\begin{eqnarray}
u &=& \frac{r}{r_0} \left(\frac{t}{t_0}\right)^{-3/8} = \frac{r}{r_{\rm out,i}} \left(\frac{t}{t_i}\right)^{-3/8}, \\
t_i &=& \frac{M_{\star}}{2} \frac{1}{2 \pi \chi} \left[\frac{J_d}{28^{-3/2} \frac{10 \pi}{7} B[5/2,~19/14] \sqrt{G M_{\bullet}}}\right]^{-5/3}r_{\rm out,i}^{19/6} \times \nonumber \\
&&\left[\int_{u_{\rm in,i}}^{1} u^{2/5} (1-u^{7/5})^{3/2} \, \diff u \right]^{-1}. \label{tiini}
\end{eqnarray}

\noindent Then, the mass accretion rate at the inner radius, $r_{\rm in} = r_{\rm ISCO}$, and the disc bolometric luminosity is given by

\begin{eqnarray}
\dot{M} &=& \frac{6 \pi}{28^{5/2}} \chi \left[\frac{J_d}{28^{-3/2} \frac{10 \pi}{7} B[5/2,~19/14] \sqrt{G M_{\bullet}}}\right]^{5/3}r_{\rm out,i}^{-19/6} \times \nonumber \\ 
&& \left(\frac{t}{t_i}\right)^{-19/16} \left[\sqrt{u} \frac{\partial }{\partial u} \left\{\sqrt{u} (1-u^{7/5})^{5/2}\right\}\right]_{r_{\rm in}}, \label{mdotcan}\\
L_b &=& \frac{9 \pi}{2} \frac{G M_{\bullet} \chi }{28^{5/2}} \left[\frac{J_d}{28^{-3/2} \frac{10 \pi}{7} B[5/2,~19/14] \sqrt{G M_{\bullet}}}\right]^{5/3}r_{\rm out,i}^{-19/6} \times \nonumber \\
&&  \left(\frac{t}{t_i}\right)^{-19/16}\int_{r_{\rm in}}^{r_{\rm out}} r^{-2} \left[1-\left\{\frac{r}{r_{\rm out,i}} \left(\frac{t}{t_i}\right)^{-3/8}\right\}^{7/5}\right]^{5/2} \, \diff r. \label{lumcan}
\end{eqnarray}

\begin{figure}
\begin{center}
\includegraphics[scale=0.56]{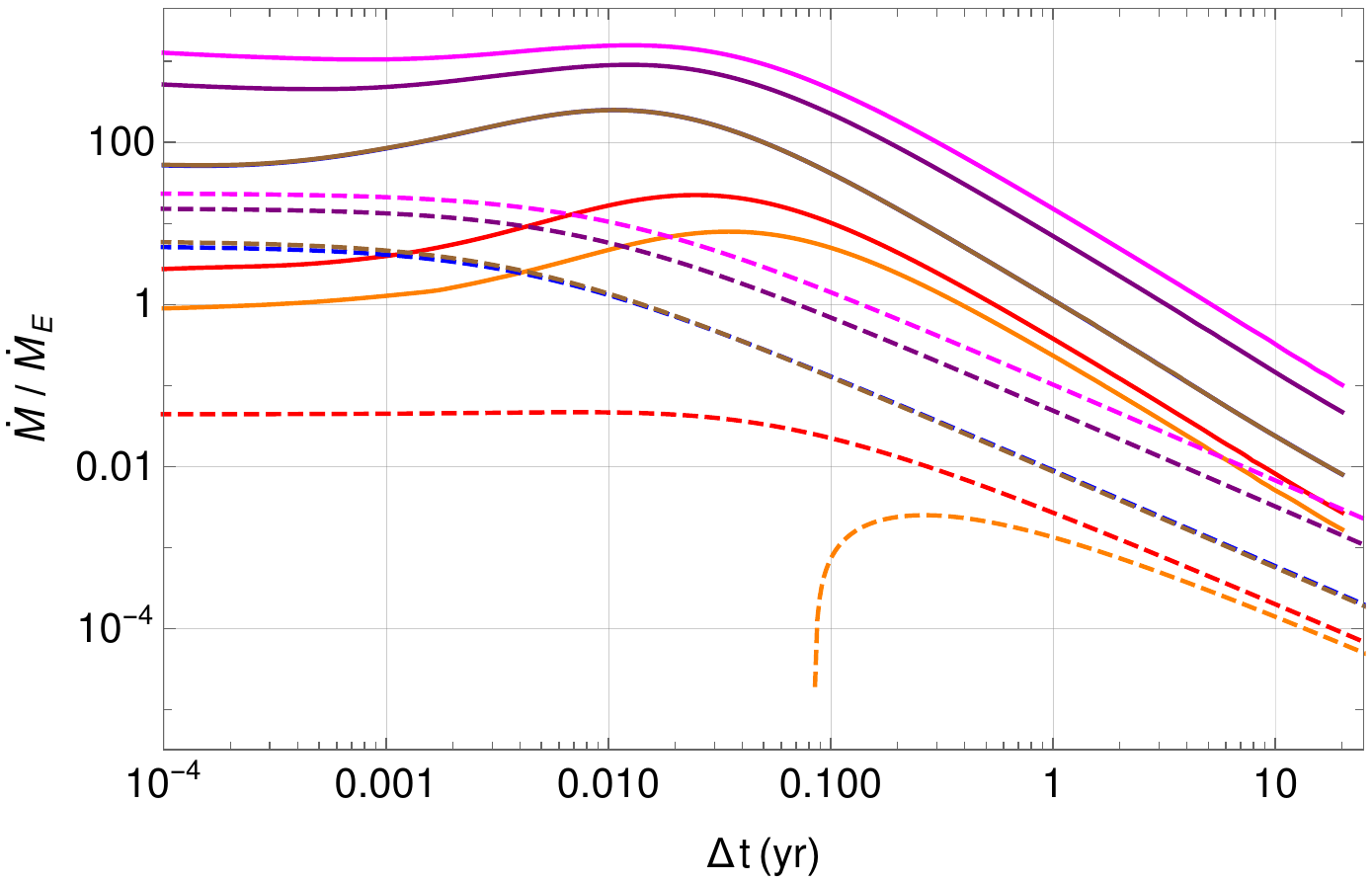}
\end{center}
\caption{The time evolution of disc mass accretion rate is shown for the sub-Eddington self-similar model (dashed lines) and the advective disc-corona model (solid lines). The blue, red, orange, purple, magenta and brown lines corresponds to $\{M_{\bullet}[10^6 M_{\odot}],~M_{\star}[M_{\odot}],~j\} = \{1,~1,~0\},~\{5,~1,~0\},~\{10,~1,~0\},~\{1,~5,~0\},~\{1,~10,~0\},~{\rm and}~\{1,~1,~0.8\}$ respectively. The fraction of energy transported to corona $b_2 = 0.5$. The initial time $t_i$ (see equation \ref{tiini}) for the disc is $t_i {\rm (days)}=$ 1.44 (blue), 10.51 (red), 29.23 (orange), 2.75 (purple), 3.65 (magenta) and 1.39 (brown).}
\label{can}
\end{figure}

\begin{figure*}
\begin{center}
\subfigure[]{\includegraphics[scale=0.56]{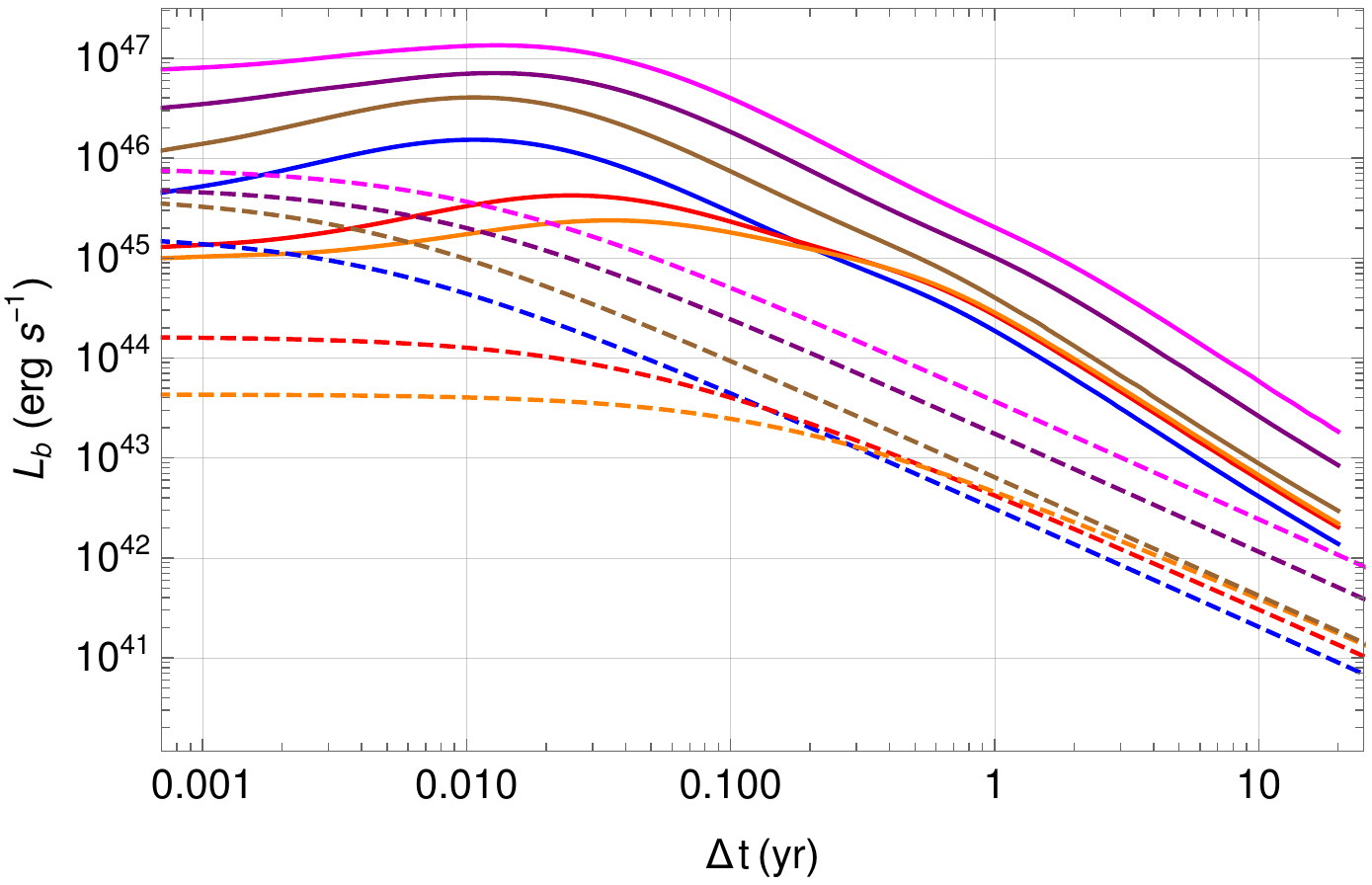}}
\subfigure[]{\includegraphics[scale=0.56]{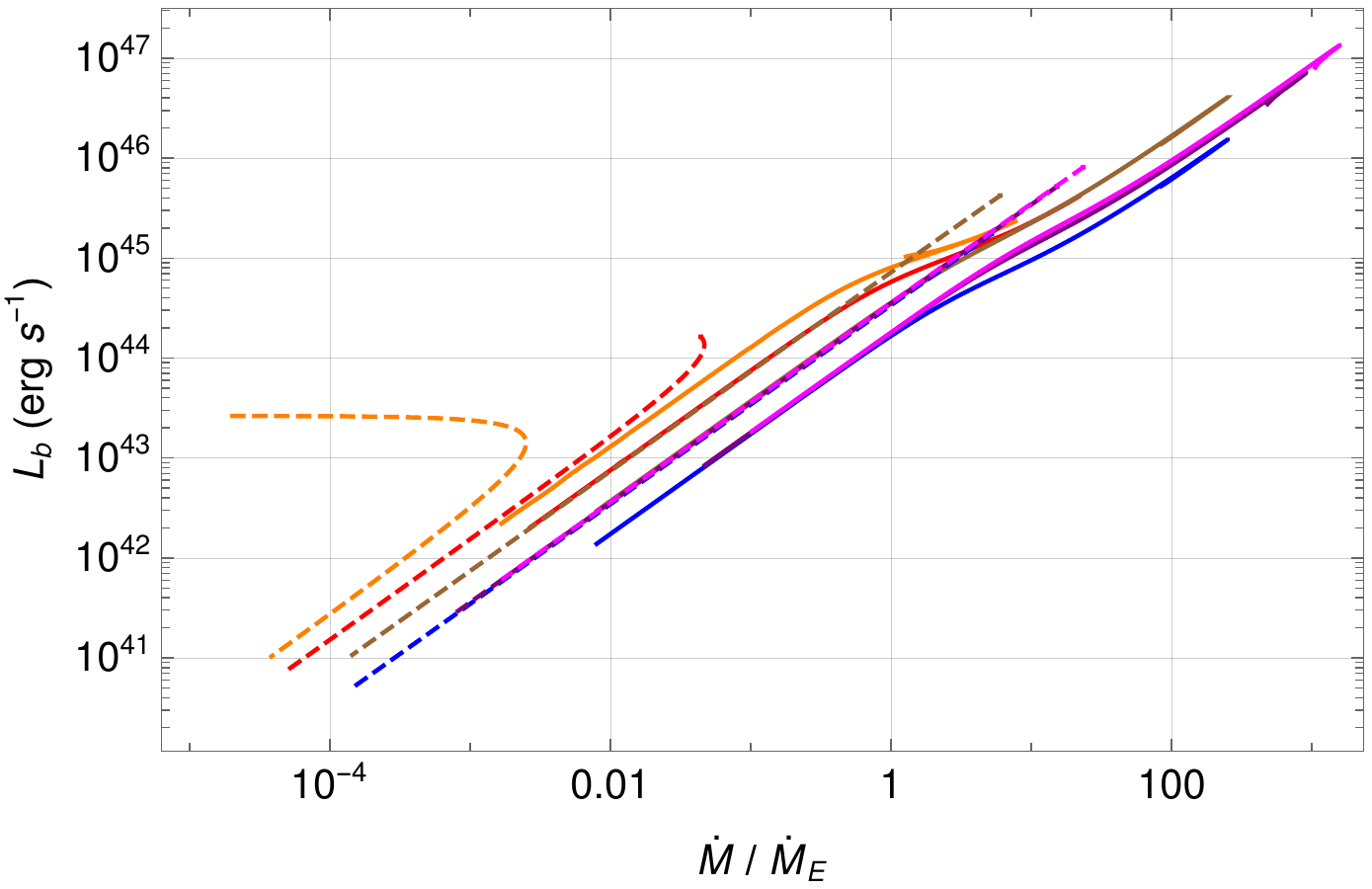}}
\end{center}
\caption{The time evolution of disc bolometric luminosity and the parametric plots of bolometric luminosity to mass accretion rate are shown for the sub-Eddington self-similar model (dashed lines) and the advective disc-corona model (solid lines). The blue, red, orange, purple, magenta and brown lines corresponds to $\{M_{\bullet}[10^6 M_{\odot}],~M_{\star}[M_{\odot}],~j\} = \{1,~1,~0\},~\{5,~1,~0\},~\{10,~1,~0\},~\{1,~5,~0\},~\{1,~10,~0\},~{\rm and}~\{1,~1,~0.8\}$ respectively. The fraction of energy transported to corona $b_2 = 0.5$. The initial time $t_i$ (see equation \ref{tiini}) for the disc is $t_i {\rm (days)}=$ 1.44 (blue), 10.51 (red), 29.23 (orange), 2.75 (purple), 3.65 (magenta) and 1.39 (brown). }
\label{can1}
\end{figure*}

\section{Reprocessing of disc emission}
\label{repr}

\begin{figure*}
\begin{center}
\includegraphics[scale=0.6]{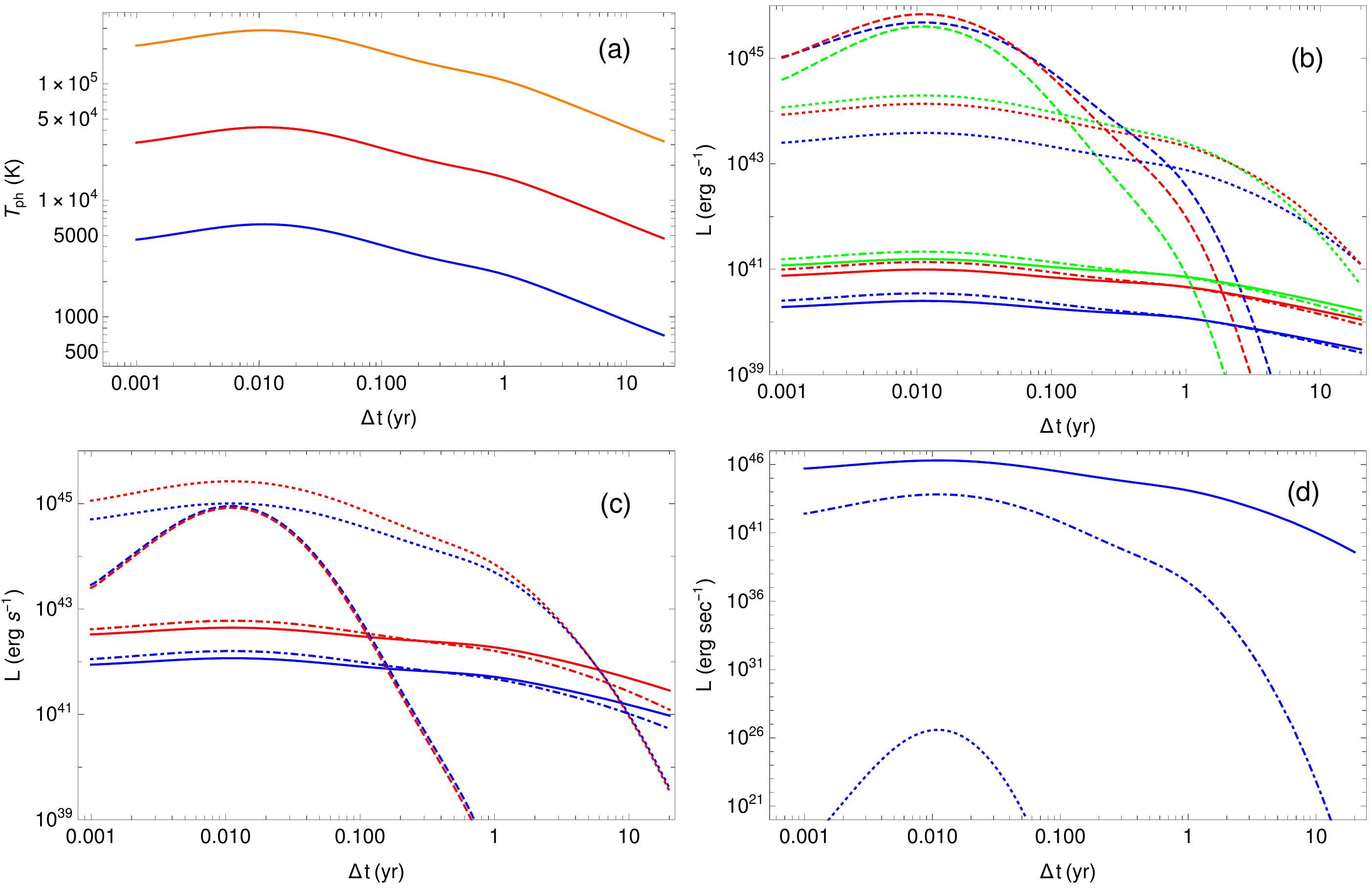}
\end{center}
\caption{(a) The temperature of the photosphere for $\rho_w = 10^{-8}~{\rm g~cm^{-3}}~{\rm (blue)},~10^{-10}~{\rm (red)}~{\rm g~cm^{-3}},~{\rm and}~10^{-12}~{\rm (orange)}~{\rm g~cm^{-3}}$. The reprocessed spectrum calculated using equation (\ref{lres}) for spectral bands V (5000-6000 \AA: Blue), B (3600 - 5000 \AA: Red) and U (3000-4000 \AA: Green) is shown in (b), for UVM2 (1660-2680 \AA: Blue) and UVW2 (1120 - 2640 \AA: Red) in (c) and X-ray (0.2 - 2 keV) in (d). The solid line shows the unprocessed spectral luminosity from the disc, dashed lines are for $\rho_w = 10^{-8}~{\rm g~cm^{-3}}$, dotted lines are for $\rho_w = 10^{-10}~{\rm g~cm^{-3}}$, and dot-dashed lines are for $\rho_w = 10^{-12}~{\rm g~cm^{-3}}$. The amount of reprocessing decreases with a decrease in density $\rho_w$ resulting in an increase in X-ray luminosity. The photosphere temperature is comparable to the $\sim ~{\rm few}~\times~10^{4}~{\rm K}$ from optical/UV observations. See section \ref{discuss} for more discussion. }
\label{rep}
\end{figure*}

Here, we show a simple reprocessing analysis of our disc luminosity with no corona ($b_2 =0$) by a static atmosphere (without bothering about their origin) as we do not have wind in our model. We assume a spherical atmosphere with the density profile given by  $\rho = \rho_{w} (r/ r_w)^{-2}$, where $r$ is the radial element and $\rho_w$ are the density at the inner radius of the atmosphere $r_w$. We assume the inner radius $r_w $ to be the circularization radius. Assuming a Thomson opacity $\kappa = 0.34~{\rm cm^{2}~ g^{-1}}$, the optical depth is given by $\tau_s \approx \kappa \rho_{w} r_w $ that is calculated assuming outer radius of the atmosphere $r_{\rm out,a} \gg r_w$. The temperature at the inner radius $r_w$ is then given by $T_{w} = [(3 \tau_s/4 \sigma) L_b/(4 \pi r_w^2 )]^{1/4}$ \citep{2016ApJ...827....3R}. If we assume the photon flow to be adiabatic until it reaches the photospheric radius $r_{\rm ph}$ such that $T \propto \rho^{1/3}$, the temperature of the photosphere is given by $T_{\rm ph} = T_w (r_{\rm ph}/r_w)^{-2/3}$. Following \citet{2009MNRAS.400.2070S}, we approximate the photosphere radius by $\kappa  r_{\rm ph} \rho(r_{\rm ph}) \simeq 1$ such that $r_{\rm ph} = \rho_w \kappa r_w^2$. Then, assuming a blackbody emission, the spectral luminosity is given by 

\begin{equation}
L = 4 \pi r_{\rm ph}^2 \int_{\nu_l}^{\nu_h} \frac{2 h }{c^2} \frac{\nu^3}{\exp\left(\frac{h \nu}{k_B T_{\rm ph}}\right) -1} \, \diff \nu.
\label{lres}
\end{equation}

\noindent We incident the bolometric luminosity obtained for case \Romannum{1} with $\mu = 1$ (A1) on the atmosphere shown by blue solid line in Fig. \ref{lbps}, and the emitted spectrum for two different values of $\rho_w = 10^{-8}~{\rm g~cm^{-3}}~{\rm and}~10^{-10}~{\rm g~cm^{-3}}$ is shown in Fig. \ref{rep}. As the photosphere radius $r_{\rm ph} \propto \rho_w$ and temperature $T_{\rm ph} \propto \rho_w^{-2/3}$, an increase in density increases the photosphere radius but decreases the temperature. The photosphere temperature is smaller than the disc single blackbody temperature and is comparable to the $\sim ~{\rm few}~\times~10^{4}~{\rm K}$ from optical/UV observations. The decrease in $\rho_w$ reduces the amount of reprocessing and thus the optical and UV luminosity decreases whereas the X-ray luminosity increases. The X-ray luminosity of the reprocessed spectrum is observationally negligible and thus implies that the observation dominates in UV and optical. TDEs such as PS1-10jh \citep{2012Natur.485..217G}, PS1-11af \citep{2014ApJ...780...44C} and others have been observed with weak to no X-ray observations. However, TDEs such as OGLE16aaa \citep{2020A&A...639A.100K} and ASASSN-15oi \citep{2017ApJ...851L..47G} have shown late time X-ray brightening with no significant observations at early time. This implies that the density of reprocessing medium declines with time which results in a reduction in reprocessing of disc luminosity. This simplistic static reprocessing model explains the observational temperature of optical and UV emissions and the dominance of these emissions over X-rays. In general, the reprocessing atmosphere may not be steady and evolves with time which will result in an evolving photosphere radius and the X-ray luminosity may dominate at the late times.

%%%%%%%%%%%%%%%%%%%%%%%%%%%%%%%%%%%%%%%%%%%%%%%%%%

% Don't change these lines
\bsp	% typesetting comment
\label{lastpage}
\end{document}